\documentclass{aa}  
\usepackage{graphicx}
\usepackage{txfonts}
\usepackage{natbib}
\bibpunct[]{(}{)}{;}{a}{}{,}

\begin{document}

\title{The dynamical state of the First Hydrostatic Core Candidate Cha-MMS1}
\author{A. E. Tsitali\inst{1}\thanks{Member of the International Max Planck Research School (IMPRS) for Astronomy and
Astrophysics at the Universities of Bonn and Cologne.} \and A. Belloche\inst{1} \and B. Commer\c{c}on\inst{2} \and K. M. Menten\inst{1}}

\institute{Max-Planck-Institut f\"{u}r Radioastronomie, Auf dem H\"{u}gel 69, 53121, Bonn, Germany
\and Laboratoire de radioastronomie, UMR 8112 du CNRS, \'{E}cole Normale Sup\'{e}rieure et Observatoire de Paris, 24 rue Lhomond, 75231, Paris Cedex 05, France}

\date{Received 31 January 2013; accepted 10 June 2013}

\abstract
% context heading (optional)
{First Hydrostatic Cores represent a theoretically predicted intermediate evolutionary link between the prestellar and protostellar phases. Studying observational characteristics of first core candidates is therefore vital for probing and understanding the earliest phases of star formation.} 
% aims heading (mandatory)
{We aim to determine the dynamical state of the First Hydrostatic Core candidate Cha-MMS1.}
% methods heading (mandatory)
{We observed Cha-MMS1 in various molecular transitions with the APEX 
and Mopra telescopes. Continuum data retrieved from the \emph{Spitzer Heritage Archive} were used to estimate the internal luminosity of the source. The molecular emission was modeled with a radiative transfer code to derive constraints on the kinematics of the envelope, which were then compared to predictions of magneto-hydrodynamic simulations.
}
% results heading (mandatory)
{We derive an internal luminosity of 0.08~L$_{\odot}$ --~0.18 L$_{\odot}$ for Cha-MMS1.
An average velocity gradient of $3.1\pm0.1$~km~s$^{-1}$~pc$^{-1}$ over $\sim0.08$~pc is found perpendicular to the filament in which Cha-MMS1 is embedded. The gradient is flatter in the outer parts and, surprisingly, also 
at the innermost $\sim2000$ AU to 4000~AU. The former features are consistent with solid-body rotation beyond 
4000~AU and slower, differential rotation beyond 8000~AU, but the origin of 
the flatter gradient in the innermost parts is unclear. The classical infall signature is detected in HCO$^+$~3--2 and CS~2--1. 
The radiative transfer modeling indicates a uniform infall velocity in the outer parts of the envelope. In the inner parts (at most 9000~AU), an infall velocity field scaling with $r^{-0.5}$ is consistent with the data but the shape of the profile is less well constrained and the velocity could also decrease toward the center.
The infall velocities are subsonic to transonic, 0.1~km~s$^{-1}$ --~0.2 km~s$^{-1}$ at $r\geq3300$ AU, and subsonic to supersonic, 0.04~km~s$^{-1}$ --~0.6 km~s$^{-1}$ at $r\leq3300$ AU.
Both the internal luminosity of Cha-MMS1 and the infall velocity field in its envelope are consistent with predictions of MHD simulations for the first core phase.
There is no evidence for a fast, large-scale outflow stemming from Cha-MMS1 but excess emission from the high-density tracers CS~5--4, CO~6--5, and CO~7--6 suggests the presence of higher-velocity material at the inner core.
}
% conclusions heading (optional), leave it empty if necessary 
{Its internal luminosity excludes Cha-MMS1 being a prestellar core. The kinematical properties of its envelope are consistent with Cha-MMS1 being a first hydrostatic core candidate or a very young Class 0 protostar.}


\maketitle

\section{Introduction}
\label{sec:intro}

Many advances have recently been made in the field of early low-mass star formation, spanning from the prestellar phase to the formation and evolution of Young Stellar Objects (YSOs), e.g., with the \textit{Spitzer} c2d Legacy Project \citep[`From Molecular Cores to Planet-Forming Disks',][]{evans03} and the \textit{Herschel} Gould Belt Survey \citep{andre10}. In particular, \textit{Herschel} \citep{pilbratt10} has provided valuable insight into the early star formation processes. Most dense starless cores in molecular clouds appear to be located along a complex network of long, thin filaments, suggesting that filament formation precedes the core formation process \citep{arzoumanian11,hill11,andre10,menshchikov10,molinari10}. The study of these early stages is necessary in order to address open questions such as the origin of the stellar Initial Mass Function (IMF), its relationship with the prestellar phase and the Core Mass Function \citep[CMF; e.g.,][]{andre09}, and the initial conditions needed for star formation to occur. Recent results from the \textit{Herschel} survey confirm the resemblance of the prestellar CMF to the stellar IMF in the Aquila and Polaris clouds \citep{koenyves10,andre10}. Such resemblance was already seen in various molecular clouds with ground-based single-dish telescopes, such as in the  Ophiuchus molecular cloud \citep{motte98}, or in the Pipe nebula 
\citep{rathborne09}.

It has recently become common to split the population of
starless cores in molecular clouds into two categories, the gravitationally 
bound and unbound cores. Prestellar cores represent the subset of starless 
cores that are self-gravitating and will thus very likely form stars 
\citep[e.g.,][]{difrancesco07,andre09}, while the gravitationally-unbound 
starless cores may be transient objects (``failed'' cores) or objects on the 
verge of becoming prestellar \citep[e.g.][]{belloche11b}. The 
gravitational collapse of a prestellar core leads to the formation of a 
stellar embryo, the protostar. This marks the beginning of the Class 0 phase,
during which the central object accretes mass from its protostellar envelope 
\citep[][]{andre00}. Theoretically, the early work of \citet{larson69} already
showed that the formation of the central protostar must be preceded by the 
formation of a larger, less dense, first hydrostatic core (hereafter FHSC).
The FHSC thus represents an intermediate evolutionary stage between the 
prestellar and protostellar phases. The detection of FHSCs is observationally 
very challenging because of their very short expected lifetime.

\subsection{FHSC: A Theoretical Background}
\label{sec:FCtheory}

The formation of the first hydrostatic core emerged from theory for the first time by \citet{larson69} but only a handful of objects have recently been observed and suggested as likely candidates. \citet{larson69} describes the process of forming a protostar from a parent molecular core using a spherical collapse model ignoring the presence of magnetic fields and rotation. The initial phase is characterised by an isothermal contraction of the molecular core.
When the central density exceeds 10$^{-13}$~g~cm$^{-3}$ the radiative cooling ceases to be efficient and an opaque, adiabatic core forms at the centre. The rise in temperature results in an increase of the thermal pressure, and finally, when the pressure balances the gravitational force the collapse ceases and the first hydrostatic core is formed. 
The initial central temperature of the FHSC is estimated to be around 170~K with an initial central density of 2$\times$10$^{-10}$~g~cm$^{-3}$. The so-called second, more compact (protostellar) core is formed after the dissociation of H$_2$ and subsequent collapse, when the central temperature and density reach 2~$\times$~10$^4$~K and 2~$\times$~10$^{-2}$~g~cm$^{-3}$, respectively \citep{larson69}. 

Various theoretical studies predict observational characteristics of the first core phase. Internal luminosities of up to $\sim0.1$ L$_\odot$ \citep{masunaga98, saigotomisaka11} or $\sim~0.25~L_{\odot}$ \citep{commercon12} have been predicted. First cores are characterised by radii and masses of the order of $\sim5$~AU -- 10 AU and 0.05 $M_{\odot}$ -- 0.1~$M_{\odot}$, respectively \citep{masunaga98, saigo08}. Their lifetimes range from a few 100 yr to a few 1000~yr, increasing with the rate of rotation. \citet{commercon12} derive lifetimes ranging from $\sim 1000$ yr to $> 4000$~yr for rotating, magnetised $1~M_\odot$ cores with 3D radiation-MHD simulations. The FHSC lifetime is shorter for higher levels of magnetisation due to the stronger magnetic braking that increases the mass accretion rate. These short lifetimes imply that first cores are rare and thus difficult to observe, although we note that \citet{tomida10} predict much longer lifetimes ($> 10^4$~yr) for first cores formed in very-low mass cloud cores (0.1~$M_{\odot}$).

Outflows at the first core phase are thought to be a significant observational signature characterising this evolutionary stage. \citet{machida08} used 3D resistive MHD simulations to study the driving mechanisms of outflows in the star formation process. Their predictions distinguish between an extended, slow molecular outflow driven by the first core and a highly collimated, fast jet later driven by the protostellar core, exhibiting typical velocities of $\sim3$ km s$^{-1}$ and 30~km~s$^{-1}$, respectively. The outflow driven by the first core is predicted to be extremely compact, spanning $\sim200$~AU --~800~AU in extent just before the start of the second collapse \citep{commercon12, commercon10,machida08}. The first core outflows are thought to result from the twisting of the magnetic field lines due to the rotation of the collapsing core, whose amplified toroidal component leads to the subsequent transfer of angular momentum to the gas outside of the core \citep{tomisakatomida11}.

\subsection{Cha-MMS1 and its evolutionary stage}

Chamaeleon-MMS1 (hereafter Cha-MMS1 for short) is a dense core embedded in a filament ($\sim0.5$ pc in length) within the Chamaeleon I molecular cloud \citep{belloche11a} at a distance of 150~pc \citep{whittet97,knudehog98}. Several previous studies of Cha-MMS1 suggest that it is an object at a very early evolutionary stage. 

\citet{reipurth96} discovered Cha-MMS1 in dust continuum emission at 
1.3~mm. They suggested that Cha-MMS1 is the driving source of the nearby 
HH~49/50 objects and identified it as a Class 0 protostar based on that 
association. During the Class 0 protostellar phase, the central object is deeply embedded within its collapsing envelope, which comprises more than half of the system's mass \citep{andre93}. \citet{lehtinen01} confirmed this classification based on a 
\textit{tentative} far-infrared detection, but \citet{lehtinen03} argued
that Cha-MMS1 possibly represents an evolutionary stage earlier than Class 0
based on its lack of thermal free-free emission at cm wavelengths.
Cha-MMS1 is embedded in a gravitationally-bound C$^{18}$O core 
\citep{haikala05}. Large deuterium fractionations of HCO$^{+}$ and 
N$_2$H$^{+}$ were derived, consistent with Cha-MMS1 being an evolved 
prestellar core or a young protostellar envelope \citep{belloche06}.

A faint \textit{Spitzer} 24~$\mu$m and 70~$\mu$m detection indicates the presence of a central object in Cha-MMS1, either a FHSC or a protostar \citep{belloche06}. \citet{belloche11a} derived a very low internal luminosity of $\sim0.015$~L$_\odot$ for this object based on the correlation between the 70 $\mu$m flux density and internal luminosity established by \citet{dunham08} for protostellar objects. As most Class 0 protostars feature an outflow \citep{andre00}, a search for a large-scale outflow driven by Cha-MMS1 was performed in CO 3--2 with APEX but none was found \citep{belloche06}. The non-detection of an outflow around Cha-MMS1 on scales of $\sim 10^4$~AU suggests that the central object may be less evolved than a Class 0 object. It could possess an outflow too compact to have been detected with the resolution of the previous studies, which would be in agreement with the predictions of the FHSC observational signatures. Furthermore, Cha-MMS1, the Class 0 protostar IRAM 04191+1522 (hereafter, IRAM 04191) and the Very Low Luminosity Object (VeLLO) L1521F \citep{crapsi04, bourke06} are located at approximately the same distance and a direct comparison of their \textit{Spitzer} fluxes supports the idea that Cha-MMS1 is less evolved \citep{belloche06}. Bearing all this in mind, Cha-MMS1 could be at the stage of the FHSC, inbetween the prestellar and Class 0 phases. However, its classification as such is very difficult to observationally confirm until a compact \citep[200~AU --~800~AU;][]{commercon10, commercon12}, slow outflow with velocities of the order of 2~km s$^{-1}$ --~4~km~s$^{-1}$ \citep{tomisaka02,machida08,commercon10, hennebellefromang08} is detected. We therefore merely consider it as a FHSC candidate.

\subsection{FHSC candidates}  

The detection of seven candidate first cores has been claimed so far: Cha-MMS1 \citep{belloche06, belloche11a}, L1448-IRS2E \citep{chen10}, Per-Bolo~58 \citep{enoch10, dunham11}, L1451-mm \citep{pineda11}, CB17-MMS \citep{chen12}, B1-bS, and B1-bN \citep{pezzuto12}. All but one (B1-bS) are so-called Very Low Luminosity Objects (VeLLOs; internal luminosity 
$L_{\rm int} < 0.1~ L_\odot$), in agreement with the range of luminosities predicted for FHSCs. With $L_{\rm bol}\sim0.49$~$L_\odot$, B1-bS might be too luminous for a FHSC.
L1448-IRS2E, Per-Bolo 58, and L1451-mm drive outflows that have been interferometrically detected.  L1448-IRS2E drives an outflow with velocities of $\sim25$~km~s$^{-1}$, one order of magnitude higher than predicted for a first core by MHD simulations \citep{machida08}. This suggests that it is at the more evolved second core stage and is likely ruled out as a first core candidate. Per-Bolo~58, L1451-mm, and CB17-MMS all have outflow velocities in agreement with theoretical predictions (see Sect.~\ref{sec:FCtheory}). However, the outflows of Per-Bolo~58 and CB17-MMS extend over 6000~AU --~8000 AU, with dynamical times $\sim$~10$^4$ yr, about one order of magnitude longer than the expected first core lifetime in a \emph{magnetised} collapsing dense core. Lifetimes of the order of $\sim$~4000~yr --~10000~yr are produced by non- or very-weakly magnetised simulations, but no outflow is produced at the FHSC stage in these cases \citep[e.g,][]{commercon12}. 
%\citet{price12} claim producing an extended, collimated, slow outflow in their MHD simulations at the FHSC stage, but issues related to their use of large sink particles cast doubts on their synthetic core being still at the FHSC stage \citep[see Sect.~5.3 of][ for more details]{commercon12}.
%In our current understanding, Per-Bolo 58 and CB17-MMS thus do not seem to be in the FHSC phase.
L1451-mm is therefore the only candidate driving an outflow with properties (maximum velocity 2.3~km~s$^{-1}$, dynamical time 1.6~$\times$~10$^{3}$ yr) consistent with current theoretical predictions at the first-core stage. There has been no outflow detection for the B1-bS and B1-bN condensations in Perseus, but the SED fitting of their \textit{Herschel} and \textit{Spitzer} (where applicable) fluxes seems to be consistent with the presence of a central object surrounded by a dusty envelope that is younger than the Class 0 phase \citep{pezzuto12}.  

The goal of this study is to set constraints on the kinematics of the envelope of Cha-MMS1 to test if it is consistent with Cha-MMS1 being at the FHSC stage. The structure of this paper is as follows. In Sect.~\ref{sec:observations} we summarise the observational details, we then present our results in Sect.~\ref{sec:results} and in Sect.~\ref{sec:mapyso} we perform radiative transfer modeling of the spectra towards Cha-MMS1. The discussion and conclusions follow in Sects.~\ref{sec:discussion} and ~\ref{sec:conclusions}, respectively. 

\section{Observations}
\label{sec:observations}

\begin{figure*}[htpb]
\begin{tabular}{rcc}
\includegraphics[width=55mm,angle=270]{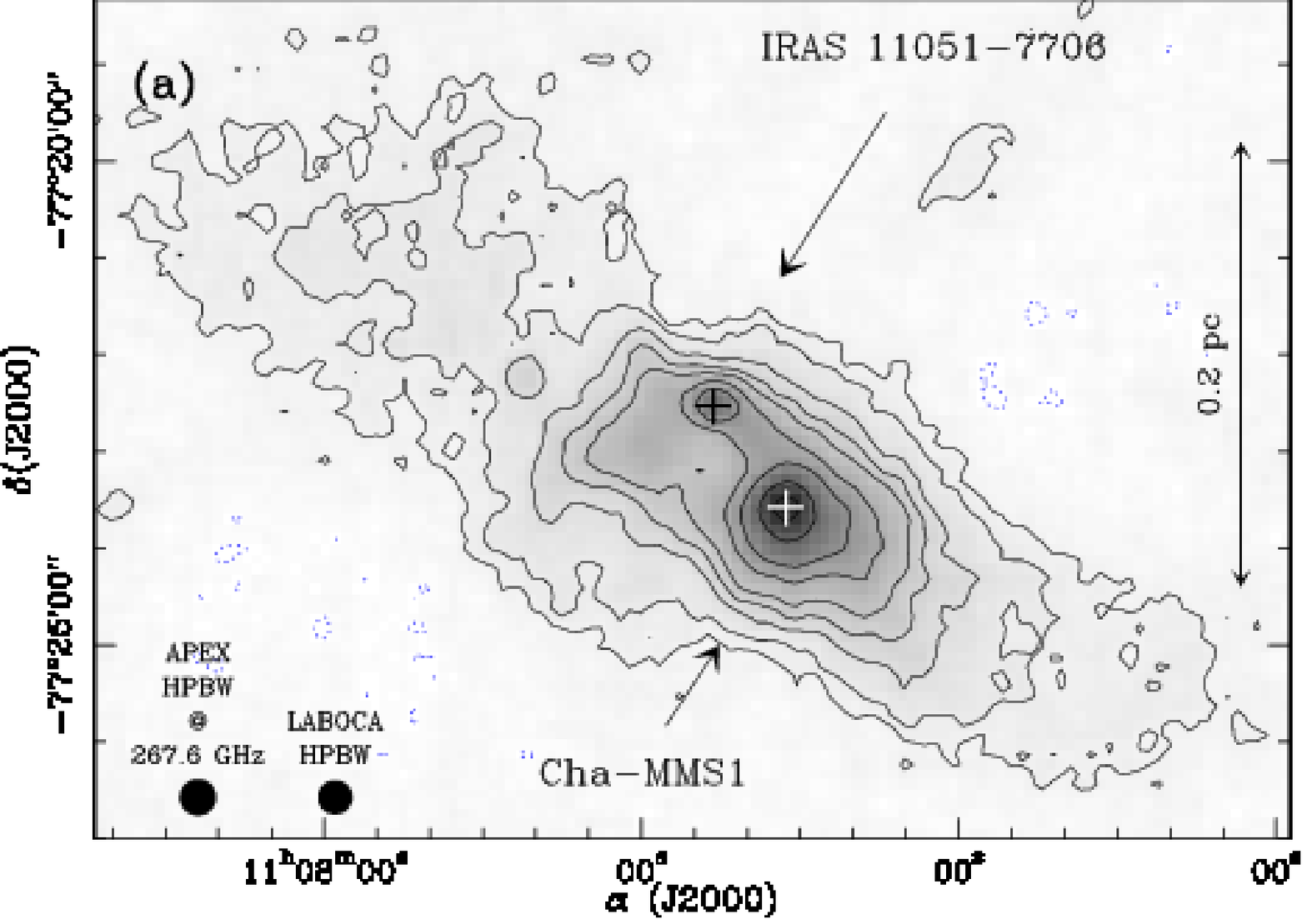} &
\includegraphics[width=55mm,angle=270]{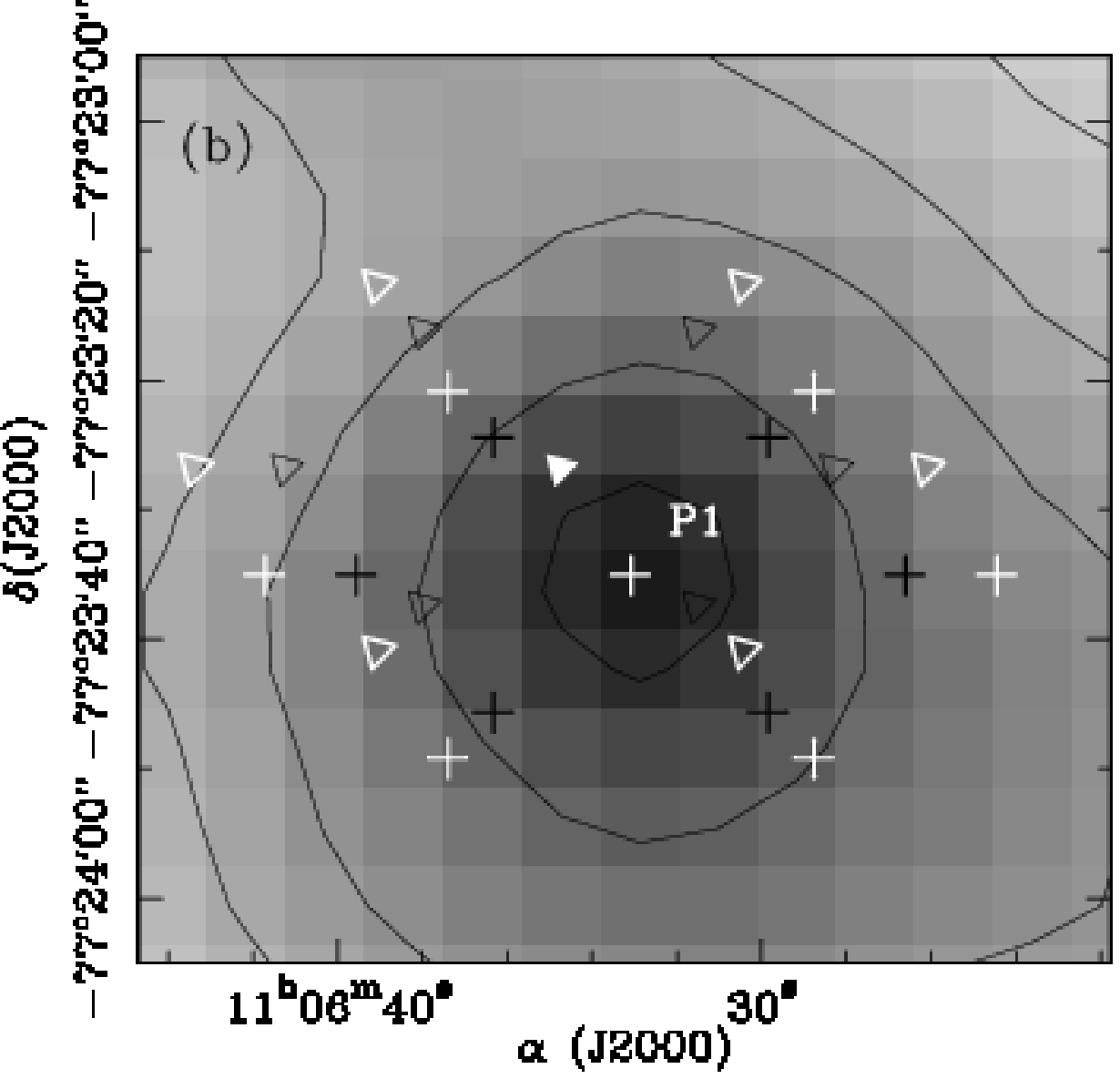} &
\includegraphics[width=55mm,angle=270]{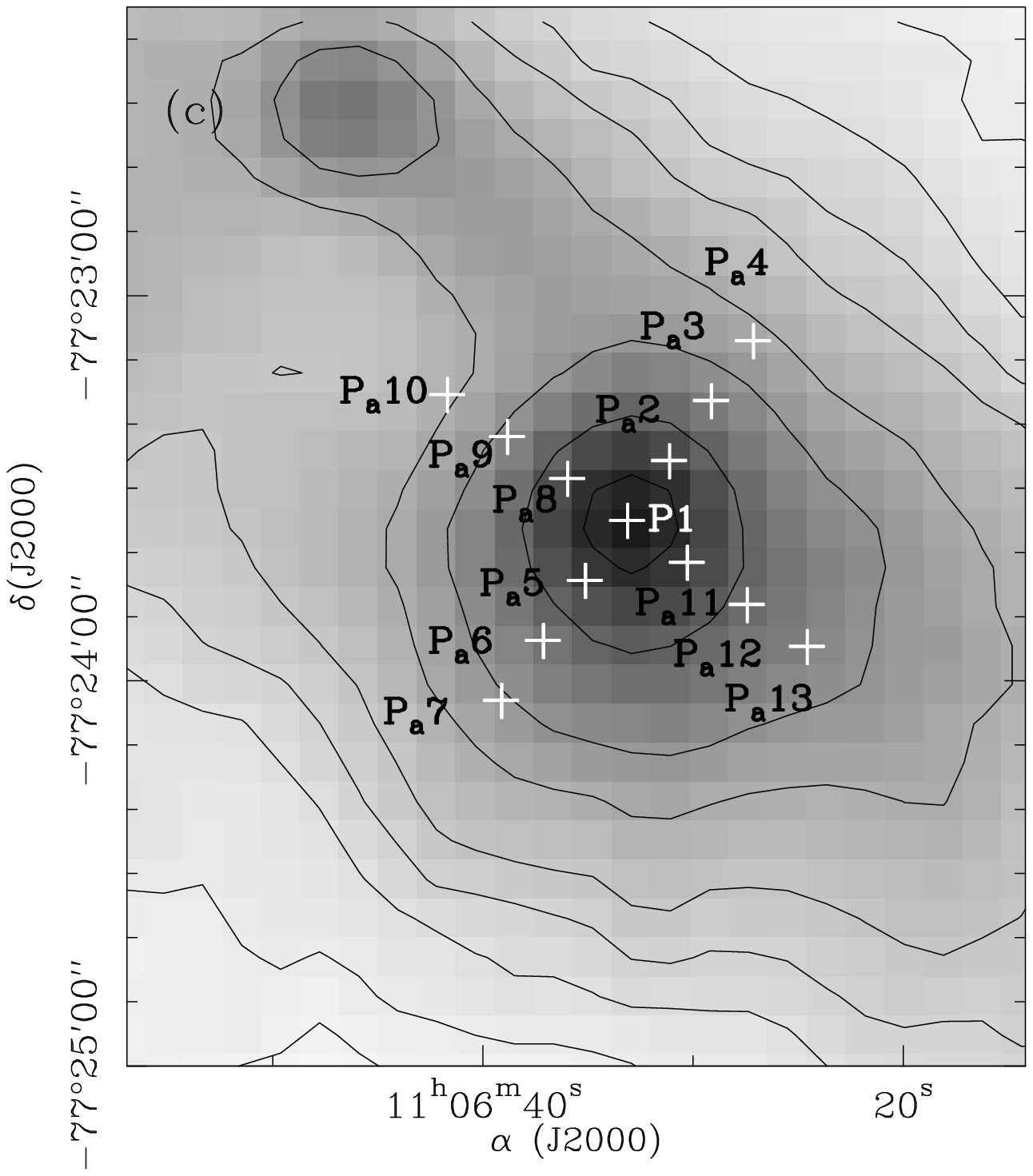} \\ 
\includegraphics[width=55mm,angle=270]{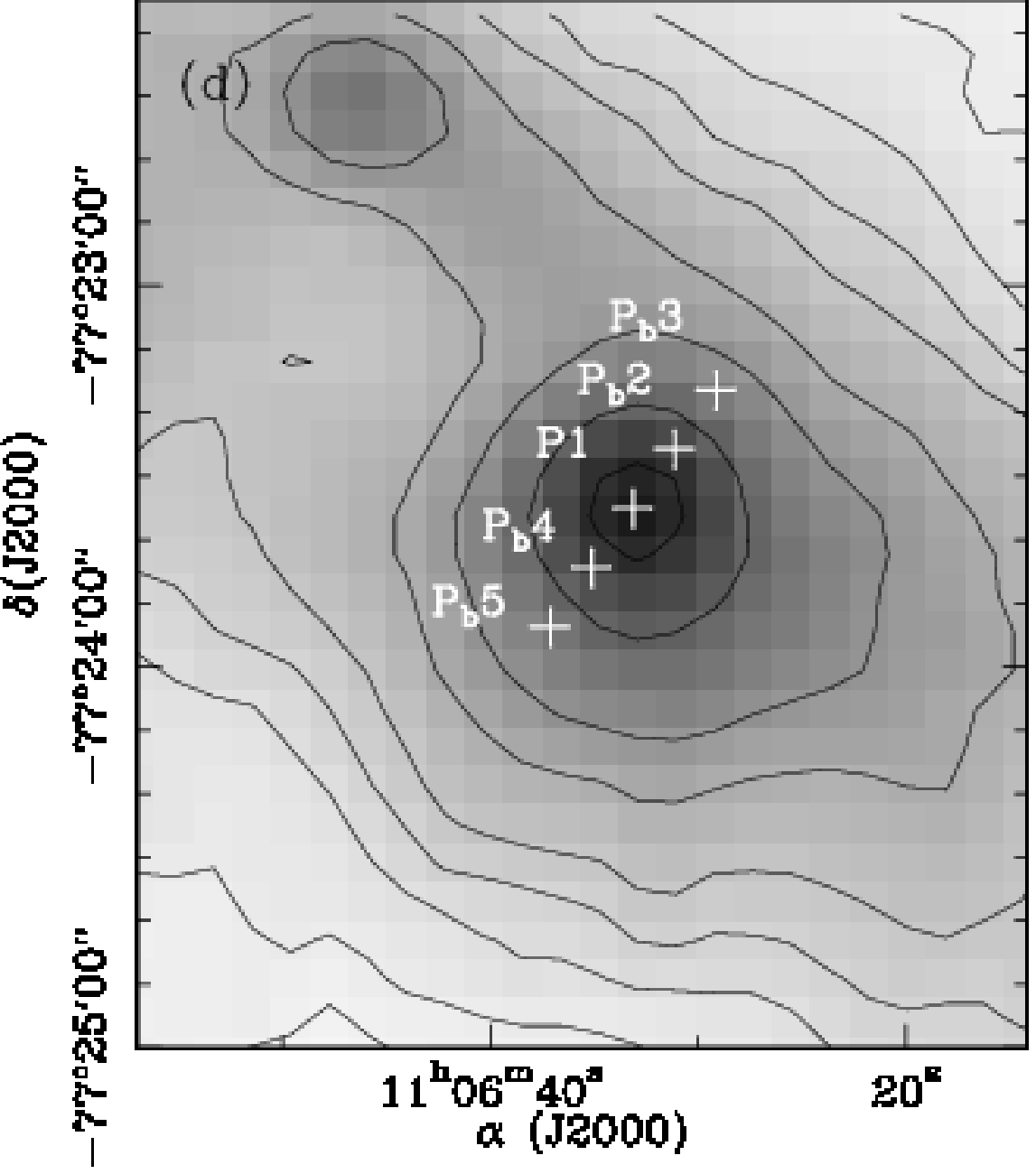} &
\includegraphics[width=55mm,angle=270]{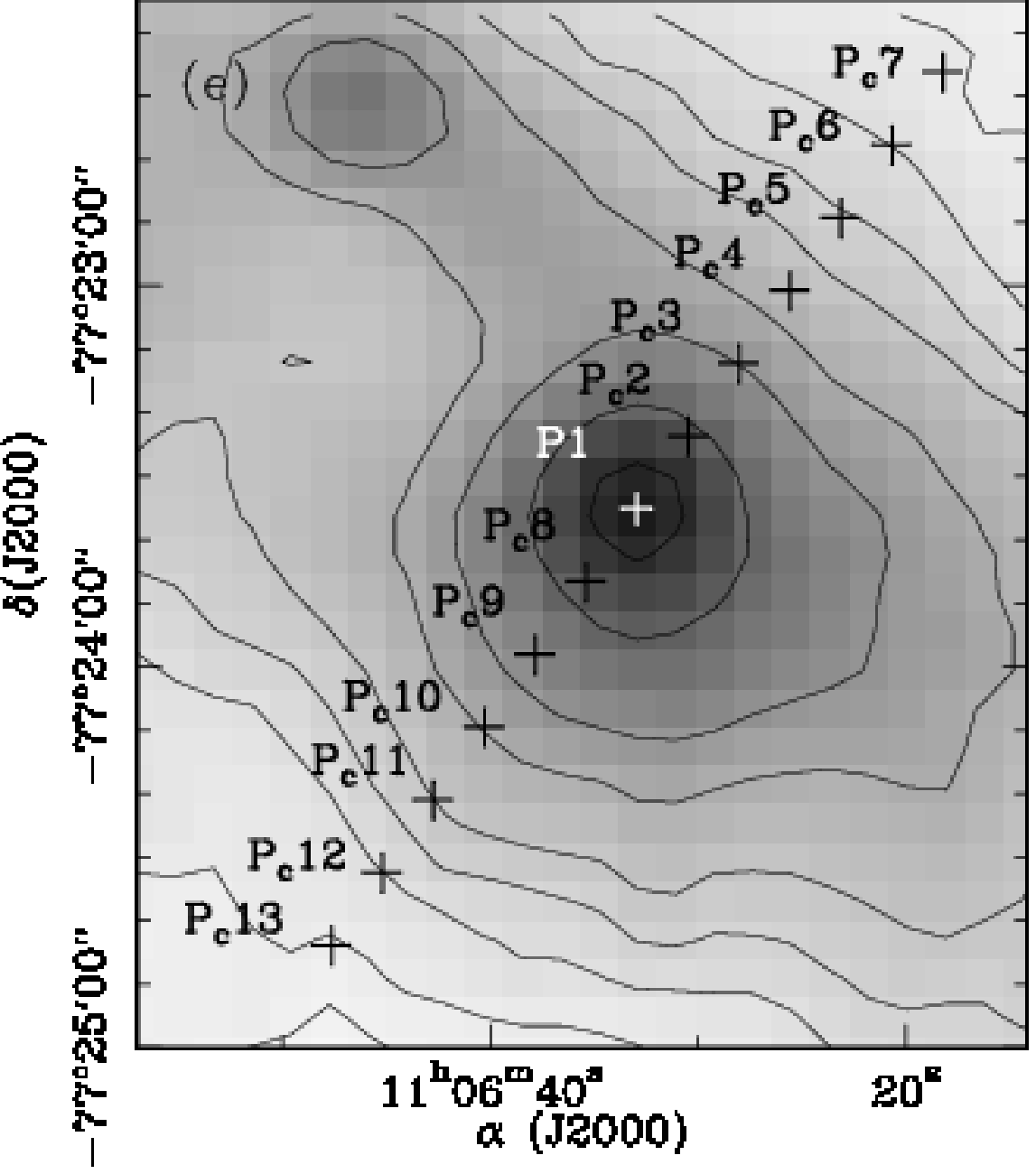} &
\includegraphics[width=55mm,angle=270]{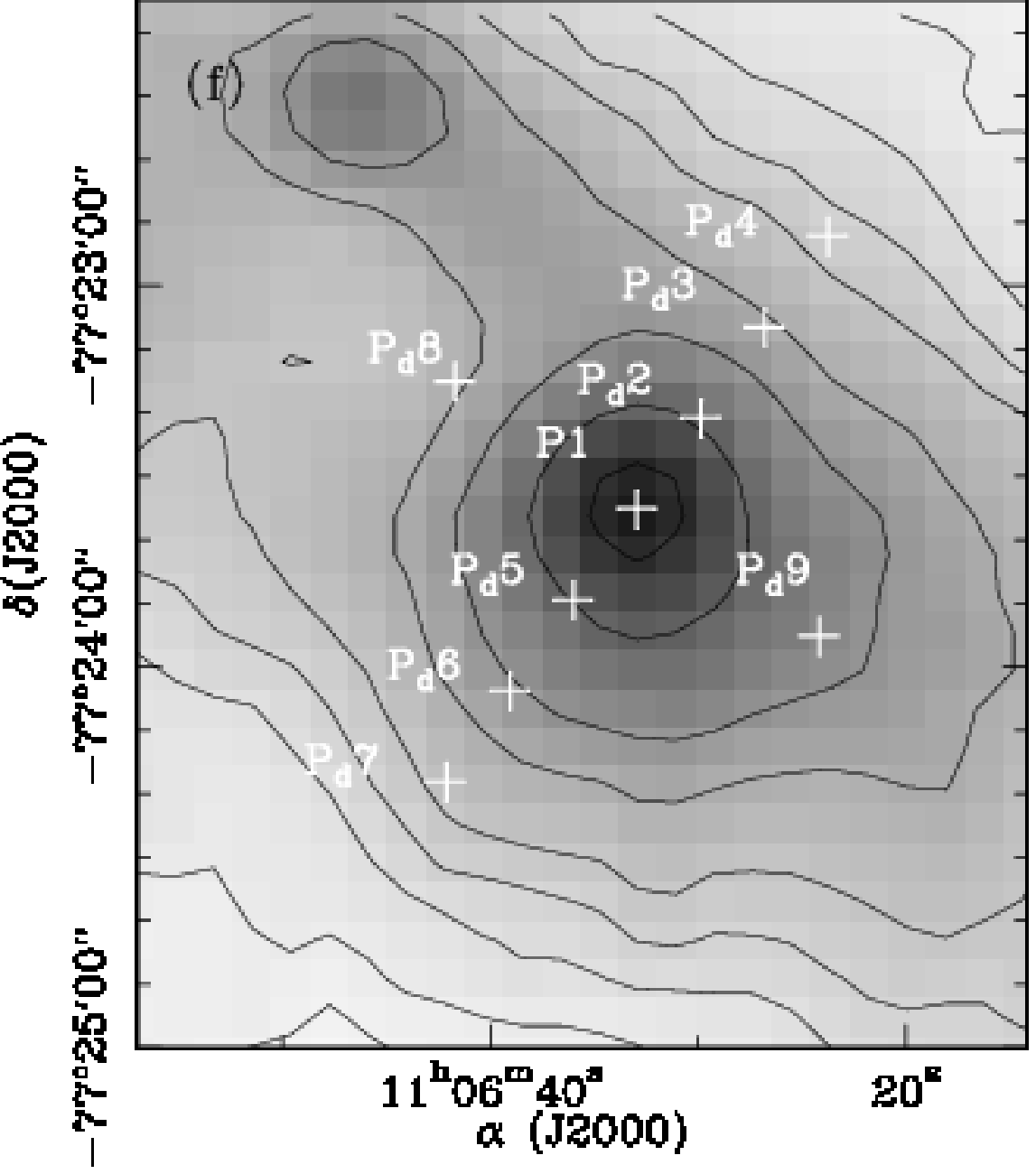} \\
\end{tabular}
 \caption[]{ (a) 870 $\mu$m map of the filament in which Cha-MMS1 is embedded, obtained with LABOCA as part of an unbiased survey of Chamaeleon~I \citep{belloche11a}. The contour levels correspond to $-a$, $a$, $2a$, $4a$, $6a$, $8a$, $12a$, $16a$, $24a$, $32a$, with $a = 36$~mJy/$21\arcsec$-beam ($3\sigma$). The white cross at $\alpha_{\textbf{J2000}}$=11$^{h}$06$^{m}$33$^{s}$.13, $\delta_{\textbf{J2000}}$=-77$^{\circ}$23$^{\prime}$35.1$^{\prime\prime}$ is the \textit{Spitzer} position of Cha-MMS1. The position of the nearby Class I object IRAS 11051-7706 is also shown with a black cross. 
(b)-(f) Zoom-in of the positions (P, white/black crosses/triangles) observed with APEX and Mopra. P$_1$ is the central position. Panel (b) refers to CO 6--5 (white crosses), $^{13}$CO 6--5 (white crosses), and CO 7--6 (black crosses), (c) to HCO$^+$~3--2 and H$^{13}$CO$^+$~3--2, (d) to CS~5--4, H$^{13}$CO$^+$~4--3, and CO~4--3, (e) to C$^{17}$O~2--1 and C$^{18}$O~2--1 (all with APEX), and (f) to the Mopra observations (see Table~\ref{table:mopra2011}). The spacing between adjacent positions is 11.4$^{\prime\prime}$ in panels (c) and (d), and 14$^{\prime\prime}$ in panel (e). In panel (f), it is 17.5$^{\prime\prime}$ for the positions perpendicular to the filament and 35$^{\prime\prime}$ for the positions along the filament. For panel (b) the pixel spacing of the CHAMP$^+$ array is 20$^{\prime\prime}$ for the white and 15$^{\prime\prime}$ for the black crosses relative to P1. The black (CO 7--6) and white (CO 6--5) open triangles mark the positions observed with CHAMP$^{+}$ centered on the offset position at (5.3$^{\prime\prime}$, 8.3$^{\prime\prime}$) relative to Cha-MMS1 (filled triangle). 
}
\label{fig:CHAmap}
\end{figure*}

We performed observations towards the dense core Cha-MMS1 at $\alpha_{2000}$=11$^{h}$06$^{m}$33$^{s}$.13, $\delta_{2000}$=-77$^{\circ}$23$^{\prime}$35.1$^{\prime\prime}$\footnote{Position from \textit{Spitzer} observations \citep{belloche11a}.} with the APEX and Mopra telescopes on the central core position as well as on offset positions close to the core along directions parallel and perpendicular to the filament in which it is embedded (see Fig.~\ref{fig:CHAmap}). The data were reduced with the CLASS software\footnote{see http://www.iram.fr/IRAMFR/GILDAS.}. 

\begin{table*}[!htpb]
\begin{center}
\caption{Parameters of 2010 APEX CHAMP$^+$ observations.} 
\vspace*{-1ex}
\hfill{}
\begin{tabular}{lcccccccccc} 
\hline
\hline
Transition & f\tablefootmark{a}            & ${\sigma_{\nu}}$\tablefootmark{b}  & $HPBW$\tablefootmark{c}            & $\delta f$\tablefootmark{d} & $\delta V$\tablefootmark{e}  &  ${N_{\mathrm{pos}}}$\tablefootmark{f} & ${F_{\mathrm{eff}}}$\tablefootmark{g}      & ${B_{\mathrm{eff}}}$\tablefootmark{g}        & $T_{\mathrm{sys}}$\tablefootmark{h}   & rms\tablefootmark{i}  \\
           & {\scriptsize (MHz)}  & {\scriptsize(kHz)} &  {\scriptsize($^{\prime\prime}$)}   & {\scriptsize (kHz)}      & {\scriptsize (km~s$^{-1}$)}  &    & {\scriptsize(\%)} &  {\scriptsize(\%)} & {\scriptsize(K)} & {\scriptsize(mK)}  \\ \hline
$^{13}$CO 6--5  & 661067.2766  & 0.5 & 9.2    & 183 & 0.083  & 7       & 95 & 36 & 1000-1200 & 76-85  \\
CO 6--5     & 691473.0763      & 0.5 & 8.8   & 183 & 0.079  & 14 & 95 & 36 & 900-1900 & 108-185 \\
CO 7--6     & 806651.8060      & 5.0 & 7.5   & 183 & 0.068  & 14 & 95 & 36 & 2200-6900 & 217-608 \\
\hline
\end{tabular}
\hfill{}
\label{table:apex2010}
\end{center}
\vspace*{-1ex}
\tablefoot{
\tablefoottext{a}{Rest frequency taken from the Cologne Database for Molecular Spectroscopy (CDMS, http://www.astro.uni-koeln.de/cdms).}
\tablefoottext{b}{Frequency uncertainty taken from the CDMS catalog.}
\tablefoottext{c}{Angular resolution.}
\tablefoottext{d}{Channel spacing in frequency.} 
\tablefoottext{e}{Channel spacing in velocity.} 
\tablefoottext{f}{Number of observed positions.}
\tablefoottext{g}{Forward and main-beam efficiencies from CHAMP$^+$ webpage.}
\tablefoottext{h}{System temperature.}
\tablefoottext{i}{rms sensitivity in $T_a^\star$ scale.}
}
%$^j$ Not including the observations at the reference position. [AB: not necessary to mention that]
\end{table*}

%%%%%%%%%%%%%%%%%%%%% 2011 APEX TABLE & FIGURE %%%%%%%%%%%%%%%%%%%%%%%%%%%%%%%%

\begin{table*}[!htpb]
\begin{center}
\caption{Parameters of 2011 APEX observations.} 
\vspace*{-1ex}
\hfill{}
\begin{tabular}{@{\extracolsep{-5pt}}lcccccccccccc} 
\hline
\hline
Transition & f\tablefootmark{a}         & ${\sigma_{\nu}}$\tablefootmark{b} & $HPBW$\tablefootmark{c}     & Receiver & Backend & $\delta f$\tablefootmark{d} & $\delta V$\tablefootmark{e} &  ${N_{\mathrm{pos}}}$\tablefootmark{f} & ${F_{\mathrm{eff}}}$\tablefootmark{g}      & ${B_{\mathrm{eff}}}$\tablefootmark{g}        & $T_{\mathrm{sys}}$\tablefootmark{h}   & rms\tablefootmark{i}  \\
           & {\scriptsize (MHz)} & {\scriptsize(kHz)} & {\scriptsize($^{\prime\prime}$)} &   &  & {\scriptsize (kHz)}   & {\scriptsize (km~s$^{-1}$)}  &    & {\scriptsize(\%)} &  {\scriptsize(\%)} & {\scriptsize(K)} & {\scriptsize(mK)}  \\ \hline
C$^{18}$O 2--1      & 219560.3541   & 1.5 & 27.7 & APEX-1 SSB & XFFTS2  & 76  & 0.104 & 13 & 95 & 75 & 232-240 & 136-195 \\
C$^{17}$O 2--1      & 224714.1870   & 80  & 27.1 & APEX-1 SSB & XFFTS2  & 76  & 0.101 & 13 & 95 & 75 & 200-217 &87-99 \\
CS 5--4            & 244935.5565   & 2.8 & 24.9 & APEX-1 SSB & XFFTS2  & 76  & 0.093 &  5 & 95 & 75 & 215-420 & 23-25 \\
H$^{13}$CO$^+$ 3--2 & 260255.3390   &  9.7  & 23.4 & APEX-1 SSB & XFFTS2  & 76  & 0.088 & 13 & 95 & 74 & 250-262 & 63-75 \\
HCO$^+$ 3--2       & 267557.6259   & 1.1 & 22.8 & APEX-2 SSB & FFTS1   & 122 & 0.137 & 13 & 95 & 74 & 200-306 & 120-152 \\
H$^{13}$CO$^+$ 4--3 & 346998.3440   & 11.9 & 17.5 & FLASH345 2SB & XFFTS & 76  & 0.066 &  5 & 95 & 73 & 280-305 & 48-53 \\
CO 4--3            & 461040.7682   & 0.5 & 13.2 & FLASH460 DSB & AFFTS & 183 & 0.119 &  5 & 95 & 60 & 1083-1288 & 234-360 \\
 \hline
\end{tabular}
\hfill{}
\label{table:apex2011}
\end{center}
\vspace*{-1ex}
\tablefoot{
\tablefoottext{a}{Rest frequency taken from the CDMS catalog. The frequencies given here for the H$^{13}$CO$^+$ 3--2 and 4--3 transitions do not account for their hyperfine structure. See the CDMS catalog for the specific frequencies corresponding to each hyperfine structure component of the transitions.} 
\tablefoottext{b}{Frequency uncertainty taken from the CDMS catalog.}
\tablefoottext{c}{Angular resolution.}
\tablefoottext{d}{Channel spacing in frequency.} 
\tablefoottext{e}{Channel spacing in velocity.} 
\tablefoottext{f}{Number of observed positions.}
\tablefoottext{g}{Forward and main-beam efficiencies.}
\tablefoottext{h}{System temperature.}
\tablefoottext{i}{rms sensitivity in $T_a^\star$ scale.}
}
\end{table*}

%%%%%%%%%%%%%%%    MOPRA TABLE & FIGURE  %%%%%%%%%%%%%%%%%%%%%%%%%%%%%%%%%%%%

\begin{table*}[!htpb]
\begin{center}
\caption{Parameters of Mopra observations} 
\vspace*{-1ex}
\hfill{}
\begin{tabular}{lccccccccccc} 
\hline
\hline
Transition & f\tablefootmark{a}               &  ${\sigma_{\nu}}$\tablefootmark{b}  & $HPBW$\tablefootmark{c}  & $\delta f$\tablefootmark{d}    & $\delta V$\tablefootmark{e}               & ${N_{\mathrm{pos}}}$\tablefootmark{f}  & ${B_{\mathrm{eff}}}$\tablefootmark{g}            & $T_{\mathrm{sys}}$\tablefootmark{h}   & rms\tablefootmark{i}  \\
           & {\scriptsize (MHz)} & {\scriptsize(kHz)}  & {\scriptsize($^{\prime\prime}$)} & {\scriptsize (kHz)} & {\scriptsize (km~s$^{-1}$)}  &                      &  {\scriptsize(\%)}  &  {\scriptsize(K)} &  {\scriptsize(mK)}    \\ \hline
c-C$_3$H$_2$ 3$_{2,2}$-3$_{1,3}$          & 84727.6909  & 3.4  & 40.6 & 34 & 0.12  &  1 & 34 & 185     & 18 \\
HC$^{18}$O$^+$ 1--0                      & 85162.2231 &  4.8  & 40.4 & 34 & 0.12  &  1 & 34 & 185     & 18 \\
H$^{13}$CO$^+$ 1$_{2,2}$--0$_{1,1}$ $^j$   & 86754.3004 &  3.9  & 39.7 & 34 & 0.12  &  1 & 34 & 165     & 17 \\
HN$^{13}$C 1$_{2,3,3}$--0$_{1,2,2}$ $^j$    & 87090.8297 &  3.8  & 39.5 & 34 & 0.12  &  1 & 34 & 165     & 14 \\
HNCO 4$_{0,4,5}$-3$_{0,3,4}$  $^j$         & 87925.2178 &  0.3  & 39.2 & 34 & 0.11  &  1 & 34 & 164     & 16 \\
HCO$^+$ 1--0                            & 89188.5247 &  4.1  & 38.6 & 34 & 0.11  &  1 & 34 & 147     & 17  \\
HC$_3$N 10$_{11}$--9$_{10}$ $^j$          & 90979.0024 &  1.0  & 37.9 & 34 & 0.11  &  9 & 34 & 208-230 & 17-32 \\
$^{13}$CS 2--1                           & 92494.3080 & 50.0  & 37.2 & 34 & 0.11  &  9 & 34 & 208-230 & 17-29 \\
N$_2$H$^+$ 1$_{2,3}$--0$_{1,2}$  $^j$      & 93173.7642 &  2.4  & 37.0 & 34 & 0.11  &  9 & 34 & 213-233 & 19-31 \\
C$^{34}$S 2--1                           & 96412.9495 &  2.2  & 35.7 & 34 & 0.10  &  9 & 34 & 216-235 & 19-35 \\
CH$_3$OH\emph{--E} 2$_{1,2}$-1$_{1,1}$              & 96739.362   & 5.0  & 35.6 & 34 & 0.10  &  9 & 34 & 229-251 & 20-33 \\
CH$_3$OH\emph{--A} 2$_{0,2}$-1$_{0,1}$              & 96741.375   & 5.0  & 35.6 & 34 & 0.10  &  9 & 34 & 229-251 & 20-33 \\
CH$_3$OH\emph{--E} 2$_{0,2}$-1$_{0,1}$              & 96744.550   & 5.0  & 35.6 & 34 & 0.10  &  9 & 34 & 229-251 & 20-33 \\
C$^{33}$S 2--1                           & 97172.0639 & 0.2  & 35.4 & 34 & 0.10  &  9 & 34 & 229-251 & 20-34 \\
CS 2--1                                 & 97980.9533 & 2.3  & 35.2 & 34 & 0.10  &  9 & 34 & 229-251 & 20-32 \\
 \hline
\end{tabular}
\hfill{}
\label{table:mopra2011}
\end{center}
\vspace*{-1ex}
\tablefoot{
\tablefoottext{a}{Rest frequency taken from the CDMS catalog.}
\tablefoottext{b}{Frequency uncertainty.}
\tablefoottext{c}{Angular resolution.}
\tablefoottext{d}{Channel spacing in frequency.}
\tablefoottext{e}{Channel spacing in velocity.}
\tablefoottext{f}{Number of observed positions.}
\tablefoottext{g}{Main beam efficiency.}
\tablefoottext{h}{Range of system temperature.}
\tablefoottext{i}{rms sensitivity in $T_a^\star$ scale.}
\tablefoottext{j}{Transition with hyperfine structure.}
}
\end{table*}

\subsection{2010 APEX Observations}
\label{sec:2010obs}

Observations with APEX{\footnote{The Atacama Pathfinder Experiment telescope (APEX) is a collaboration between the Max-Planck Institut f\"{u}r Radioastronomie, the European Southern Observatory, and the Onsala Space Observatory.}} using the CHAMP$^+$\footnote{see http://www3.mpifr-bonn.mpg.de/div/submmtech/heterodyne/ champplus/champmain.html.} heterodyne SSB receiver were carried out in 2010 July, in the following molecular transitions: $^{13}$CO~6--5, CO~6--5, and CO~7--6. 
CHAMP$^+$ is a 2 $\times$ 7 pixel array receiver connected to a Fast-Fourier-Transform spectrometer backend array (FFTS). It operates in two frequency bands simultaneously, around 690 GHz and 810 GHz. The channel spacing is 183 kHz. The corresponding velocity resolution for each transition is given in Table~\ref{table:apex2010}. All three transitions were observed with the central CHAMP$^+$ pixel pointed on the central core position at $\alpha_{J2000}$=11$^{h}$06$^{m}$33$^{s}$.13, $\delta_{J2000}$=-77$^{\circ}$23$^{\prime}$35.1$^{\prime\prime}$. In addition to the central position, CO~6--5 and CO~7--6 were also observed with the central CHAMP$^+$ pixel being centered at an offset position ($\Delta\alpha$,~$\Delta\delta$)~=~(5.3$^{\prime\prime}$,~8.3$^{\prime\prime}$) relative to the centre of Cha-MMS1 (see Fig.~\ref{fig:CHAmap}b).
The observations were done in position-switching mode with the reference position at ($\Delta\alpha$, $\Delta\delta$)  = (-600$^{\prime\prime}$, 4$^{\prime\prime}$). The reference position was checked to be free of emission with an rms 
sensitivity of 0.06 K and 0.24~K for the central pixel in CO 6--5 and CO 7--6, 
respectively, and for the spectral resolution given in Table~\ref{table:apex2010}. 

The observations were carried out on four different days, in the last two of which $^{13}$CO~6--5 was observed in parallel to CO~7--6. 
A comparison to the CO~7--6 spectra of the first two days of observation suggests that there is a pointing offset in the south-west direction parallel to the filament of up to $\sim 5\arcsec$ between the CO~6--5/CO~7--6 and $^{13}$CO~6--5/CO~7--6 datasets.
The forward and beam efficiencies used to convert antenna temperatures ${T_a}^{\star}$ into main-beam temperatures can be found in Table~\ref{table:apex2010} along with further information on line frequencies, system temperatures, and noise levels. The focus was optimised on Saturn and Mars and the pointing in CO 6--5 emission on the star IRAS 07454-7112. 

\subsection{2011 APEX Observations}

We carried out observations with the APEX telescope in 2011 April, June, and December in the following molecular transitions: CS 5--4, H$^{13}$CO$^{+}$~3--2, HCO$^{+}$~3--2, H$^{13}$CO$^{+}$~4--3, CO~4--3, C$^{18}$O~2--1, and C$^{17}$O~2--1. The observations were done in position-switching mode with the reference position at $\alpha_{J2000}$=11$^{h}$05$^{m}$23$^{s}$.7, $\delta_{J2000}$=-77$^{\circ}$11$^{\prime}$02.2$^{\prime\prime}$. The reference position is free of emission with an rms of 0.12 K and 0.33 K in HCO$^+$ 3--2 and CO 4--3, respectively, for the spectral resolution given in Table~\ref{table:apex2011}. The corresponding beamwidth, forward, and main beam efficiencies are listed in Table~\ref{table:apex2011}. The positions observed for each transition are shown in Figs.~\ref{fig:CHAmap}c to e, overlaid on the 870 $\mu$m map of the filament seen in Fig.~\ref{fig:CHAmap}a. All positions lie either perpendicular (position angle $-35^{\circ}$ East from North) or parallel to the filament. The transitions CS~5--4, H$^{13}$CO$^{+}$~4--3, CO~4--3, C$^{18}$O~2--1, and C$^{17}$O~2--1 were observed perpendicular to the filament only, while H$^{13}$CO$^{+}$~3--2 and HCO$^{+}$~3--2 were also observed parallel to it. Table~\ref{table:apex2011} gives information about the frontend and backend used for each transition, their respective spectral resolutions, the resultant system temperatures, and rms noise levels. The telescope pointing was checked every 1 h to 1.5~h and was performed on IRAS 07454-7112. The pointing accuracy is $\sim$2$^{\prime\prime}$ (rms). The focus was optimised on either Saturn, Jupiter, or Mars, and repeated approximately every 3~h. 

\subsection{Mopra Observations}

We performed observations with the Mopra telescope towards the central position of Cha-MMS1 along with eight other offset positions perpendicular and parallel to the filament (see Fig.~\ref{fig:CHAmap}f) in 2011 May in several molecular transitions using the zoom mode of the high resolution spectrometer MOPS.
The receiver was tuned at two different frequencies, 94554 MHz and 87190 MHz (only the central position was observed for the latter). Position switching observations were done with the reference position at $\alpha_{J2000}$=11$^{h}$05$^{m}$23$^{s}$.7, $\delta_{J2000}$=-77$^{\circ}$11$^{\prime}$02.2$^{\prime\prime}$. It was checked to be free of emission with an rms sensitivity ranging from 46~mK --~52 mK in all transitions for the spectral resolution given in Table~\ref{table:mopra2011}. The transitions that are used for the analysis in this paper are listed here: CS~2--1, C$^{34}$S~2--1, $^{13}$CS~2--1, C$^{33}$S~2--1, HC$_3$N~10--9, HN$^{13}$C~1--0, N$_2$H$^+$~1--0, HCO$^+$~1--0, H$^{13}$CO$^+$~1--0, HC$^{18}$O$^+$~1--0, HNCO~4--3, c-C$_3$H$_2$~3$_{2,2}$-3$_{1,3}$, CH$_3$OH\emph{--A}~2$_{0,2}$--1$_{0,1}$, and CH$_3$OH\emph{--E}~2$_{1,2}$--1$_{1,1}$. The final reduced dataset was obtained after averaging both polarisations. We did however notice some differences in integrated intensity for the two polarisations of up to $\sim$~10\% (see Appendix~\ref{sec:mopracal} for more details about this discrepancy). The channel spacing was 34 kHz.
The range of system temperatures $T_{\rm sys}$ for each transition is given in Table~\ref{table:mopra2011}. The beam efficiency used to convert antenna temperatures ${T_a}^{\star}$ into main-beam temperatures is 0.34. This value was derived from a detailed calibration analysis
(see Appendix~\ref{sec:mopracal}). The telescope pointing was checked 
approximately every hour on U~Men for Cha-MMS1, and AH~Sc 
and IK~Tau for the calibration sources Oph~A~SM1N and IRAM~04191, 
respectively. The transitions, their rest frequencies, and the 
number of observed positions are listed in Table~\ref{table:mopra2011}. 

\subsection{Spitzer Archive Data}

We used MIPS1 24~$\mu$m and MIPS2 70~$\mu$m continuum data taken from the \textit{Spitzer Heritage Archive}\footnote{see http://irsa.ipac.caltech.edu/data/SPITZER/docs/ \hbox{spitzerdataarchives/.}} (AORkeys: 19978496, 3962112, 19979264).

\subsection{CO 3--2 Data}

Cha-MMS1 was observed in CO 3--2 in 2005 with the APEX telescope, and the data were presented in \citet{belloche06}. We used the CO 3--2 data along the direction perpendicular to the filament in conjuction with the other CO transitions and isotopologues when modeling the spectra in Sect.~\ref{sec:mapyso}.

\section{Results}
\label{sec:results}

\subsection{Internal luminosity derivation}
\label{sec:intlum}

We performed aperture photometry on MIPS1 24~$\mu$m and MIPS2~70 $\mu$m \emph{Spitzer} data, and derived flux densities for the dense core Cha-MMS1. We used the IDL procedure aper.pro\footnote{From the IDL Astronomy User's Library (http://idlastro.gsfc.nasa.gov/contents.html).} with the following aperture and background inner and outer radii: 16$^{\prime\prime}$ (18$^{\prime\prime}$ - 39$^{\prime\prime}$) and 13$^{\prime\prime}$ (20$^{\prime\prime}$ - 32$^{\prime\prime}$) for the 70 $\mu$m and 24~$\mu$m data, respectively. 
Fine-scale aperture corrections of 2.16 for MIPS1 and 1.17 for MIPS2 were taken from the \emph{MIPS Instrument Handbook}\footnote{see http://irsa.ipac.caltech.edu/data/SPITZER/docs/mips/ \hbox{mipsinstrumenthandbook/1/.}}. The flux densities derived before and after correction are given in Table~\ref{table:fluxes}. In the following, we use the average value of the two independent, aperture-corrected 70~$\mu$m flux density measurements to estimate the internal luminosity of Cha-MMS1. We determine the internal luminosity of Cha-MMS1 via two methods. 

\subsubsection{Method 1}

\citet{dunham08} calculated the internal luminosity of low-luminosity
protostars based on a parametric model consisting of a protostellar envelope,
a disk, and an outflow cone, coupled to a 2D radiative transfer code. They derived the following empirical relation between the internal luminosity of a protostar and its observed 70~$\mu$m flux:
\begin{equation}
L_{int} = 3.3 \times 10^8F_{70}^{0.94} L_{\odot},
\end{equation}
where F$_{70}$ is normalised to 140~pc and is in cgs units (cm$^{-2}$~s$^{-1}$). 
With this equation, we derive an internal luminosity of 0.025~$\pm$~0.003~$L_{\odot}$ 
after correction for the distance of Cha-MMS1.

\subsubsection{Method 2}
\label{sec:intlum_method2}

\citet{commercon12} recently presented the evolution of the 24 $\mu$m and 70 $\mu$m flux densities in the course of the first core lifetime as well as the time evolution of the FHSC internal luminosity via 3D radiation-magnetohydrodynamic (hereafter, RMHD) simulations of a 1 $M_{\odot}$ dense core collapse. A 3D RMHD simulation for the case of a 5 $M_{\odot}$ dense core collapse was also computed (see Sect.~\ref{sec:discussion_models}). Both models have a strong initial magnetisation level \citep[MU2 model;][]{commercon12}.

We looked for a correspondence between Cha-MMS1's 24 $\mu$m and 70 $\mu$m flux densities and the model flux density predictions of FHSC obtained for the 1 $M_{\odot}$ and 5 $M_{\odot}$ dense cores. In the case of the 1 $M_{\odot}$ model, we find consistent flux densities within a factor of $\sim2$ for inclinations to the line-of-sight $45^{\circ} < i < 60^{\circ}$, and a first core age of 850~yr. In this case, the internal luminosity prediction is $\sim0.08$ \textbf{$L_{\odot}$} -- 0.13 $L_{\odot}$, at least three times larger than the internal luminosity derived using the relation by \citet{dunham08}.   

In the 5 $M_{\odot}$ case at inclinations $30^{\circ} < i < 45^{\circ}$ and for a first core age of $\sim2680$ yr, the observed and predicted 24 $\mu$m flux densities are consistent within a factor of $\sim2.5$, and we thus obtain an internal luminosity estimate of $\sim0.13$ $L_{\odot}$ -- 0.18 $L_{\odot}$. 

In the framework of this MHD model, the overall range is $\sim0.08$ $L_{\odot}$ -- 0.18 $L_{\odot}$ for inclinations of $30^{\circ} \le i < 60^{\circ}$. The internal luminosity derived from the empirical relation of \citet{dunham08} is therefore lower by a factor of $\sim3$~--~7 compared to the predictions of 3D RMHD simulations. 

We adopt an internal luminosity of $\sim0.1$ L$_{\odot}$ as an approximation, which is within the luminosity range we derived based on the RMHD simulations. We use this value for the radiative transfer modeling that follows in Sect.~\ref{sec:mapyso}. Even if we were to adopt the upper limit of 0.18 $L_{\odot}$, the temperature profile of the inner envelope would not significantly change (Equation~\ref{eq:tdust}, Sect.~\ref{sec:input}). 

\begin{table}
\begin{center}
\caption{Flux densities of Cha-MMS1 from aperture photometry.} 
\vspace*{-1ex}
\begin{tabular}{@{\extracolsep{-8pt}}ccccc} 
\hline
\hline
Instrument & $\lambda$\tablefootmark{a}        & AORkey\tablefootmark{b}    & $F_\lambda$\tablefootmark{c}                & $F_\lambda^{\rm corr}$\tablefootmark{d}    \\
           & {\scriptsize($\mu$m)} &           & {\scriptsize(mJy)}  & {\scriptsize(mJy)}   \\ \hline
MIPS2      & 70                    & 19978496  & 139 $\pm$ 32        & 300 $\pm$ 70   \\
           &                       & 3962112   & 184 $\pm$ 25        & 397 $\pm$ 54    \\
           &                       & \textit{average} &              & 349 $\pm$ 44    \\
MIPS1      & 24                    & 19978496  & 2.82 $\pm$ 0.64     & 3.3 $\pm$ 0.7    \\
           &                       & 3962112   & 2.32 $\pm$ 0.60     & 2.7 $\pm$ 0.8    \\
           &                       & 19979264  & 2.50 $\pm$ 0.63     & 2.9 $\pm$ 0.7   \\
           &                       & \textit{average} &              & 3.0 $\pm$ 0.4    \\\hline
\end{tabular}
\label{table:fluxes}
\end{center}
\vspace*{-1ex}
\tablefoot{
\tablefoottext{a}{Wavelength.}
\tablefoottext{b}{AORkey of \textit{Spitzer} observations.}
\tablefoottext{c}{Flux density from aperture photometry.}
\tablefoottext{d}{Flux density after fine-scale correction.}
}
\end{table}

\subsection{Spectra towards Cha-MMS1}
\label{sec:vlsrvelocities}

Figures ~\ref{fig:central_mopra} and ~\ref{fig:central_apex} show the spectra of the transitions observed with APEX and Mopra towards the central position of Cha-MMS1. Apart from $^{13}$CS 2--1 and C$^{33}$S 2--1 for which we can only draw upper limits, most transitions are detected. Tables~\ref{table:mopra_vlsr} to ~\ref{table:apex_vlsr3} give the centroid velocities derived for these transitions after performing gaussian or hyperfine-structure fits in CLASS to the observed spectra (``GAUSS'' and ``HFS'' methods). Only the spectra that have either a gaussian shape or a well-defined hyperfine structure and no self-absorption features can be fitted in this way. Tables ~\ref{table:mopra_vlsr} to ~\ref{table:apex_vlsr3} list the systemic velocities of groups of transitions observed at the same offset positions (see Fig~\ref{fig:CHAmap}). 
The centroid velocities of transitions that were only observed at the central position of the core are given in Table~\ref{table:mopra_vlsr_s2}. 

\subsubsection{Issues with the systemic velocity}
\label{sec:vlsrvelocities_issue}

It is apparent from Fig.~\ref{fig:central_apex} that all the APEX spectra are redshifted by 0.1~km~s$^{-1}$ compared to the systemic velocity derived from a hyperfine-structure fit to the N$_2$H$^+$ 1--0 multiplet observed with Mopra. There is therefore an issue with one of the two datasets. As the shift is seen for both the high and low-density APEX tracers (e.g., H$^{13}$CO$^+$~4--3 and C$^{18}$O~2--1), it is not likely to be an intrinsic characteristic of the source, but rather an instrumental  effect. 

We compared the spectra of the calibration source IRAM 04191, observed with both APEX, Mopra, and with the IRAM 30-m telescope, to pinpoint the source of the systemic velocity inconsistency. In addition, as we have Mopra observations of the central position of Cha-MMS1 from both 2010 and 2012 (as part of a survey targeting starless cores in Cha I and III, Tsitali et al. in prep.), we compare them to the 2011 data that we present in this paper. The N$_2$H$^+$~1--0 IRAM 04191 spectra are consistent with each other and consequently, we cannot draw any conclusions about the velocity shift. However, the N$_2$H$^+$~1--0 central spectrum of Cha-MMS1 observed with Mopra in 2010 and 2012 gives a velocity estimate consistent with the APEX data, i.e. 4.4~kms$^{-1}$, but inconsistent with the 2011 Mopra data, i.e. 4.3~kms$^{-1}$. Hence, we assume that the Mopra 2010, 2012, and APEX 2011 datasets are correct and apply a correction of 0.1~kms$^{-1}$ to the systemic velocity derived from the 2011 Mopra N$_2$H$^+$~1--0 transition whenever we use it along with the 2011 APEX spectra. We will explicitly mention it in the text whenever this correction is applied.

\onlfig{
\begin{figure*}
\begin{center}
\begin{tabular}{c}
\includegraphics[width=140mm,angle=270]{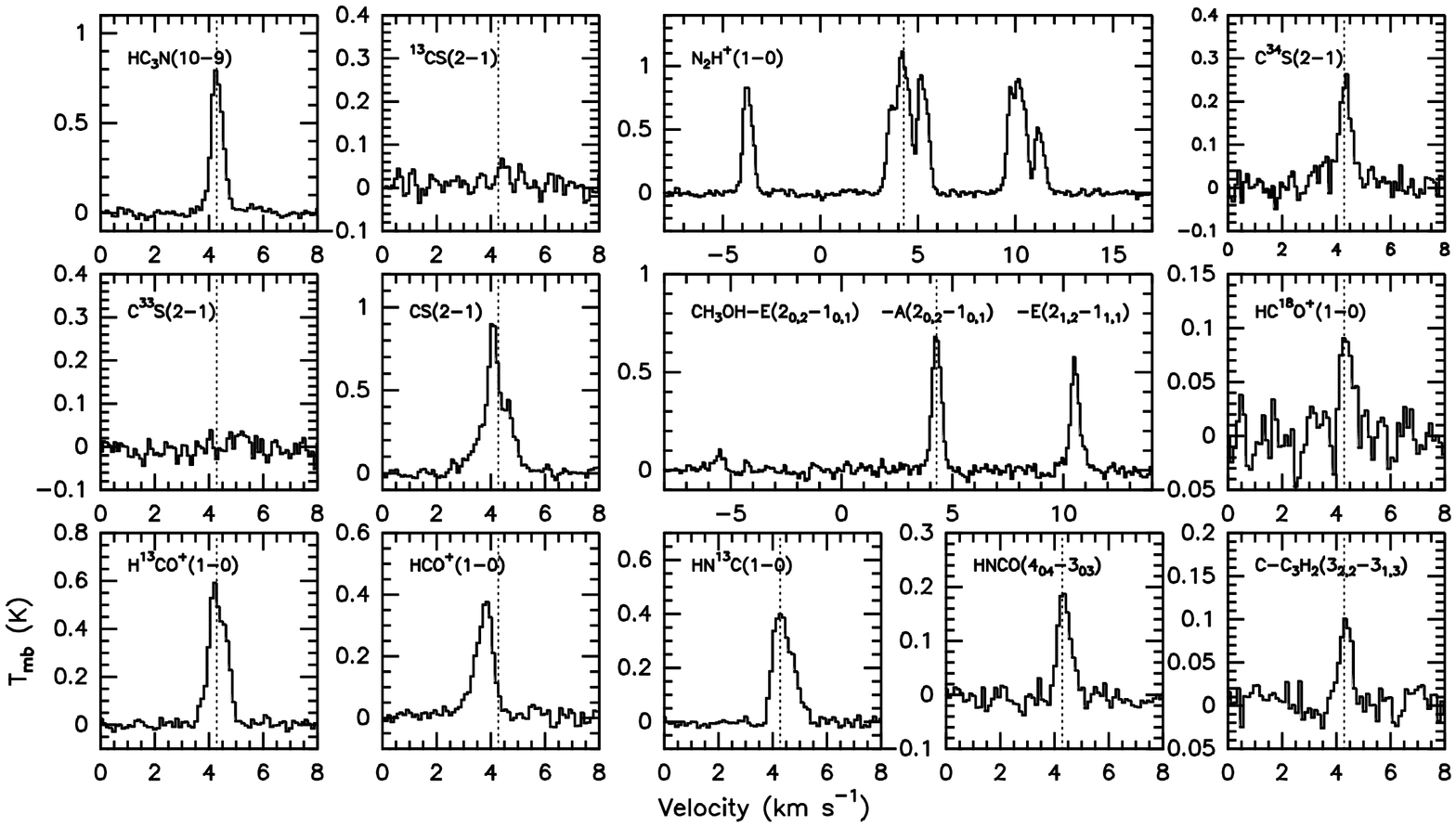} 
\vspace*{-10ex}
\end{tabular}
 \caption[]{Transitions observed with Mopra towards the central position of Cha-MMS1, in main-beam brightness temperature scale. The dotted line shows the systemic velocity of Cha-MMS1, derived from a seven component hyperfine fit to the N$_2$H$^+$ 1--0 multiplet. \label{fig:central_mopra}}
\end{center}
\end{figure*}
}

\onlfig{
\begin{figure*}
\begin{center}
\begin{tabular}{c}
\includegraphics[width=150mm,angle=270]{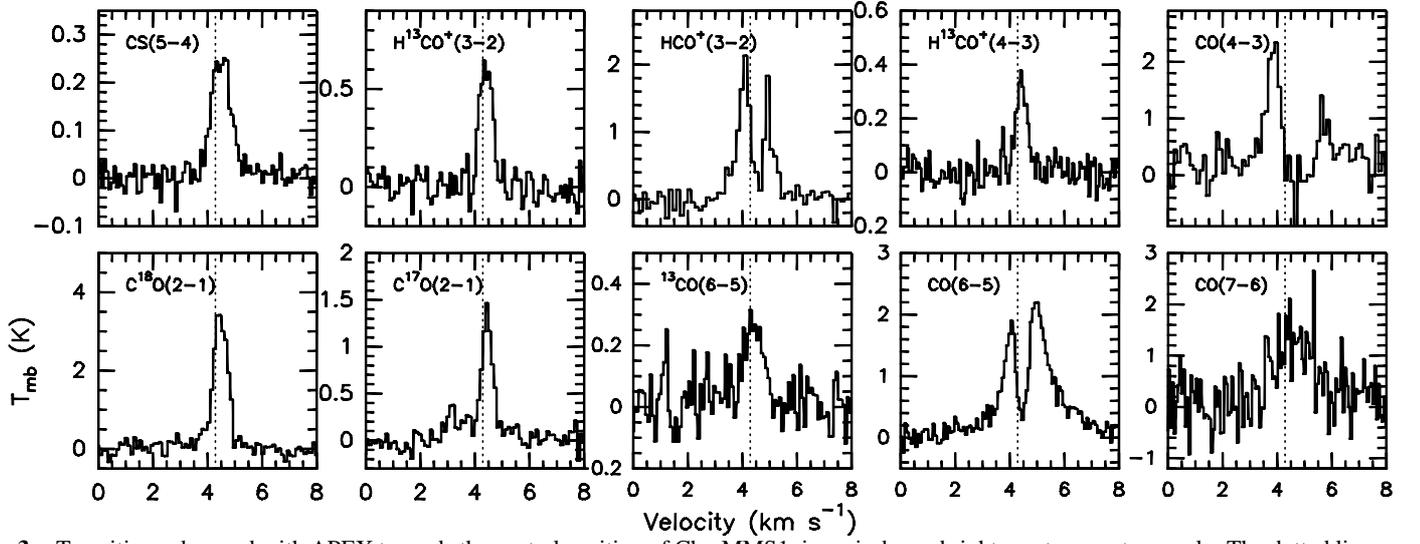} 
\vspace*{-38.5ex}
\end{tabular}
 \caption[]{Transitions observed with APEX towards the central position of Cha-MMS1, in main-beam brightness temperature scale. The dotted line shows the systemic velocity of Cha-MMS1, derived from a seven component hyperfine fit to the N$_2$H$^+$ 1--0 multiplet observed with Mopra \textit{without} correcting for the +0.1~km~s$^{-1}$ velocity shift (see Sect.~\ref{sec:vlsrvelocities_issue}). \label{fig:central_apex}}
\end{center}
\end{figure*}
}

%%%%%%%%%%%% Mopra Table VLSR %%%%%%%%%%%%%%%%

\onltab{
\begin{table*}%[!htpb]
\begin{center}
\caption{Mopra centroid velocities at positions perpendicular and parallel to the filament.} 
\vspace*{-1ex}
\begin{tabular}{ccccccccc} 
\hline
\hline 
Position\tablefootmark{a} & $\Delta\alpha$\tablefootmark{b} & $\Delta\delta$\tablefootmark{b}   &  Spacing\tablefootmark{c}  &               &                   & $V_{\rm LSR}$\tablefootmark{d} {\scriptsize(km~s$^{-1}$)} &              & \\ 
              &  {\scriptsize($^{\prime\prime}$)}    & {\scriptsize($^{\prime\prime}$)}   &   {\scriptsize($^{\prime\prime}$)}    & C$^{34}$S 2--1 & HC$_3$N 10--9\tablefootmark{e} & N$_2$H$^+$ 1--0\tablefootmark{e} & CH$_3$OH\emph{--A} 2$_{0,2}$--1$_{0,1}$ & CH$_3$OH\emph{--E} 2$_{1,2}$--1$_{1,1}$ \\ \hline 

P1       & 0 & 0 & 0     & 4.33$\pm$0.02 & 4.27$\pm$0.004 & 4.299$\pm$0.002 & 4.29$\pm$0.01   & 4.29$\pm$ 0.01   \\ 
P$_{d}$2 & -10.0 & 14.3 & 17.5  & 4.30$\pm$0.02 & 4.29$\pm$0.004 & 4.279$\pm$0.002 & 4.29$\pm$0.01   & 4.28$\pm$ 0.01   \\ 
P$_{d}$3 & -20.1 & 28.7 & 35.0  & 4.26$\pm$0.02 & 4.23$\pm$0.004 & 4.339$\pm$0.002 & 4.25$\pm$0.01   & 4.25$\pm$ 0.01   \\
P$_{d}$4 & -30.1 & 43.0 & 52.5  & 4.21$\pm$0.02 & 4.12$\pm$0.004 & 4.168$\pm$0.002 & 4.21$\pm$0.01   & 4.21$\pm$ 0.01   \\
P$_{d}$5 & 10.0  & -14.3 & 17.5  & 4.35$\pm$0.02 & 4.34$\pm$0.004 & 4.346$\pm$0.003 & 4.33$\pm$0.01   & 4.34$\pm$ 0.01   \\
P$_{d}$6 & 20.1  & -28.7 & 35.0  & 4.39$\pm$0.02 & 4.41$\pm$0.004 & 4.403$\pm$0.004 & 4.37$\pm$0.01   & 4.36$\pm$ 0.01   \\
P$_{d}$7 & 30.1  & -43.0 & 52.5  & 4.42$\pm$0.02 & 4.47$\pm$0.006  & 4.435$\pm$0.005 & 4.41$\pm$0.01   & 4.38$\pm$ 0.01   \\
P$_{d}$8 & 20.1  & 28.7  & 35.0  & 4.39$\pm$0.02 & 4.33$\pm$0.006  & 4.294$\pm$0.004 & 4.29$\pm$0.02   & 4.35$\pm$ 0.02   \\
P$_{d}$9 & -20.1 & -28.7 & 35.0  & 4.35$\pm$0.02 & 4.36$\pm$0.001  & 4.286$\pm$0.004 & 4.36$\pm$0.01   & 4.35$\pm$ 0.01   \\ \hline
\end{tabular}
\label{table:mopra_vlsr}
\end{center}
\vspace*{-1ex}
\tablefoot{
\tablefoottext{a}{These positions are shown in Fig.~\ref{fig:CHAmap}f.}
\tablefoottext{b}{J2000 equatorial offset relative to the central position P1.}
\tablefoottext{c}{Angular distance to P1.} 
\tablefoottext{d}{The correction of 0.1~km~s$^{-1}$ discussed in Sect.~\ref{sec:vlsrvelocities_issue} has not been applied.}
\tablefoottext{e}{HC$_3$N 10--9 and N$_2$H$^+$ 1--0 both have a hyperfine structure and were therefore fitted using the method ``HFS'' with CLASS.}
}  
\end{table*}
}

%%%%%%%%%%%% Mopra Table VLSR Setup 2 %%%%%%%%%%%%%%%%
\onltab{
\begin{table}
\begin{center}
\caption{Mopra centroid velocities toward the central position, P1.} 
\vspace*{-1ex}
\begin{tabular}{lc} 
\hline
\hline 
 Transition\tablefootmark{a} & $V_{\rm LSR}$\tablefootmark{b} \\ 
            & {\scriptsize(km~s$^{-1}$)} \\ \hline
 HN$^{13}$C 1--0                  & 4.37$\pm$0.01 \\ 
 HNCO 4$_{0,4}$--3$_{0,3}$         & 4.36$\pm$0.02 \\ 
 \textbf{c}-C$_3$H$_2$ 3$_{2,2}$-3$_{1,3}$  & 4.35$\pm$0.03 \\ 
 HC$^{18}$O$^+$ 1--0              & 4.39$\pm$0.03 \\ 
 H$^{13}$CO$^+$ 1--0              & 4.32$\pm$0.01 \\ \hline
\end{tabular}
\label{table:mopra_vlsr_s2}
\end{center}
\vspace*{-1ex}
\tablefoot{
\tablefoottext{a}{Only the central position was observed for each of these transitions.}
\tablefoottext{b}{The correction of 0.1~km~s$^{-1}$ discussed in Sect.~\ref{sec:vlsrvelocities_issue} has not been applied.}
}
\end{table}
}

%%%%%%%%%%%% APEX Table VLSR 1 %%%%%%%%%%%%%%
\onltab{
\begin{table}[!ht]
\begin{center}
\caption{CS 5--4 and H$^{13}$CO$^+$ 4--3 centroid velocities (APEX) at positions perpendicular to the filament.} 
\vspace*{-1ex}
\begin{tabular}{@{\extracolsep{-4pt}}cccccc} 
\hline
\hline 
Position\tablefootmark{a} & $\Delta\alpha$\tablefootmark{b}  & $\Delta\delta$\tablefootmark{b} & Spacing\tablefootmark{c} & \multicolumn{2}{c}{$V_{\rm LSR}$ {\scriptsize(km~s$^{-1}$)}}  \\ 
             & {\scriptsize($^{\prime\prime}$)} & {\scriptsize($^{\prime\prime}$)} & {\scriptsize($^{\prime\prime}$)} & CS 5--4  & H$^{13}$CO$^+$ 4--3  \\\hline 
P1       & 0      & 0      & 0      & 4.50$\pm$0.02 & 4.43$\pm$0.02 \\ 
P$_b$2   & -6.5  & 9.3   & 11.4 & 4.42$\pm$0.02 & 4.41$\pm$0.02 \\ 
P$_b$3   & -13.1 & 18.7  & 22.8 & 4.40$\pm$0.01 & 4.44$\pm$0.06 \\ 
P$_b$4   & 6.5   & -9.3  & 11.4 & 4.43$\pm$0.02 & 4.42$\pm$0.04 \\ 
P$_b$5   & 13.1  & -18.7 & 22.8 & 4.37$\pm$0.02 & 4.65$\pm$0.31 \\ \hline
\end{tabular}
\label{table:apex_vlsr1}
\end{center}
\vspace*{-1ex}
\tablefoot{
\tablefoottext{a}{These positions are shown in Fig.~\ref{fig:CHAmap}d.}
\tablefoottext{b}{J2000 equatorial offset relative to the central position.}  
\tablefoottext{c}{Angular distance to the central position.}
}
\end{table}
}

%%%%%%%%%%%% APEX Table VLSR 2 %%%%%%%%%%%%%%
\onltab{
\begin{table}
\begin{center}
\caption{H$^{13}$CO$^+$ 3--2 centroid velocities (APEX) at positions parallel and perpendicular to the filament.} 
\vspace*{-1ex}
\begin{tabular}{ccccc} 
\hline
\hline 
Position\tablefootmark{a} & $\Delta\alpha$\tablefootmark{b} & $\Delta\delta$\tablefootmark{b} & Spacing\tablefootmark{c} & $V_{\rm LSR}$ {\scriptsize(km~s$^{-1}$)}  \\ 
         & {\scriptsize($^{\prime\prime}$)} & {\scriptsize($^{\prime\prime}$)} & {\scriptsize($^{\prime\prime}$)} & H$^{13}$CO$^+$ 3--2 \\\hline 
P1      & 0       & 0      & 0    & 4.41$\pm$0.02 \\ 
\multicolumn{4}{l}{perpendicular to the filament} & \\
P$_a$2  & -6.5   & 9.3   & 11.4 & 4.40$\pm$0.01 \\ 
P$_a$3  & -13.1  & 18.7  & 22.8 & 4.41$\pm$0.02 \\ 
P$_a$4  & -19.6  & 28.0  & 34.2 & 4.38$\pm$0.02 \\ 
P$_a$5  & 6.5    & -9.3  & 11.4 & 4.36$\pm$0.04 \\ 
P$_a$6  & 13.1   & -18.7 & 22.8 & 4.44$\pm$0.03 \\ 
\multicolumn{4}{l}{parallel to the filament} & \\
P$_a$8  & 9.3    & 6.5   & 11.4 & 4.35$\pm$0.02 \\ 
P$_a$9  & 18.7   & 13.1  & 22.8 & 4.32$\pm$0.04 \\ 
P$_a$10 & 28.0   & 19.6  & 34.2 & 4.41$\pm$0.03 \\ 
P$_a$11 & -9.3   & -6.5  & 11.4 & 4.41$\pm$0.02 \\ 
P$_a$12 & -18.7  & -13.1 & 22.8 & 4.44$\pm$0.02 \\ 
P$_a$13 & -28.0  & -19.6 & 34.2 & 4.43$\pm$0.02 \\ \hline
\end{tabular}
\label{table:apex_vlsr2}
\end{center}
\vspace*{-1ex}
\tablefoot{
\tablefoottext{a}{These positions are shown in Fig.~\ref{fig:CHAmap}c.} 
\tablefoottext{b,c}{Same as Table~\ref{table:apex_vlsr1}.}
}
\end{table}
}

%%%%%%%%%%%% APEX Table VLSR 3 %%%%%%%%%%%%%%
\onltab{
\begin{table}
\caption{C$^{17}$O 2--1 and C$^{18}$O 2--1 centroid velocities (APEX) at positions perpendicular to the filament.} 
\vspace*{-1ex}
\begin{tabular}{cccccc} 
\hline
\hline 
Position\tablefootmark{a} & $\Delta\alpha$\tablefootmark{b} & $\Delta\delta$\tablefootmark{b} & Spacing\tablefootmark{c} & \multicolumn{2}{c}{$V_{\rm LSR}$ {\scriptsize(km~s$^{-1}$)}} \\ 
  & {\scriptsize($^{\prime\prime}$)} & {\scriptsize($^{\prime\prime}$)} & {\scriptsize($^{\prime\prime}$)} & C$^{17}$O 2--1 & C$^{18}$O 2--1 \\\hline 
P1     & 0   & 0     & 0    & 4.45$\pm$0.01 & 4.47$\pm$0.01 \\ 
P$_c$2 & -8  & 11.5  & 14.0 & 4.42$\pm$0.02 & 4.45$\pm$0.01 \\ 
P$_c$3 & -16 & 23    & 28.0 & 4.37$\pm$0.02 & 4.40$\pm$0.01 \\ 
P$_c$4 & -24 & 34.5  & 42.0 & 4.30$\pm$0.03 & 4.32$\pm$0.01 \\ 
P$_c$5 & -32 & 46    & 56.0 & 4.23$\pm$0.02 & 4.28$\pm$0.01 \\ 
P$_c$6 & -40 & 57.5  & 70.0 & 4.22$\pm$0.02 & 4.27$\pm$0.01 \\ 
P$_c$7 & -48 & 69    & 84.1 & 4.38$\pm$0.03 & 4.33$\pm$0.02 \\ 
P$_c$8 & 8   & -11.5 & 14.0 & 4.44$\pm$0.02 & 4.48$\pm$0.01 \\ 
P$_c$9 & 16  & -23   & 28.0 & 4.47$\pm$0.03 & 4.51$\pm$0.01 \\ 
P$_c$10 & 24  & -34.5 & 42.0 & 4.53$\pm$0.03 & 4.52$\pm$0.01 \\ 
P$_c$11 & 32  & -46   & 56.0 & 4.56$\pm$0.03 & 4.55$\pm$0.01 \\ 
P$_c$12 & 40  & -57.5 & 70.0 & 4.55$\pm$0.03 & 4.51$\pm$0.01 \\ 
P$_c$13 & 48  & -69   & 84.1 & 4.50$\pm$0.04 & 4.52$\pm$0.02 \\ \hline
\end{tabular}
\label{table:apex_vlsr3}
\tablefoot{
\tablefoottext{a}{These positions are shown in Fig.~\ref{fig:CHAmap}e.} 
\tablefoottext{b,c}{Same as Table~\ref{table:apex_vlsr1}.}
}
\end{table}
}

 %%%%%%%%%%%%%%%%%%%%%%%%%%%%%%%%%%%%%

\subsection{Rotation}
\label{sec:rotation}

\begin{figure*}%[htpb]
\begin{center}
\begin{tabular}{cc}
\includegraphics[width=90mm,angle=0]{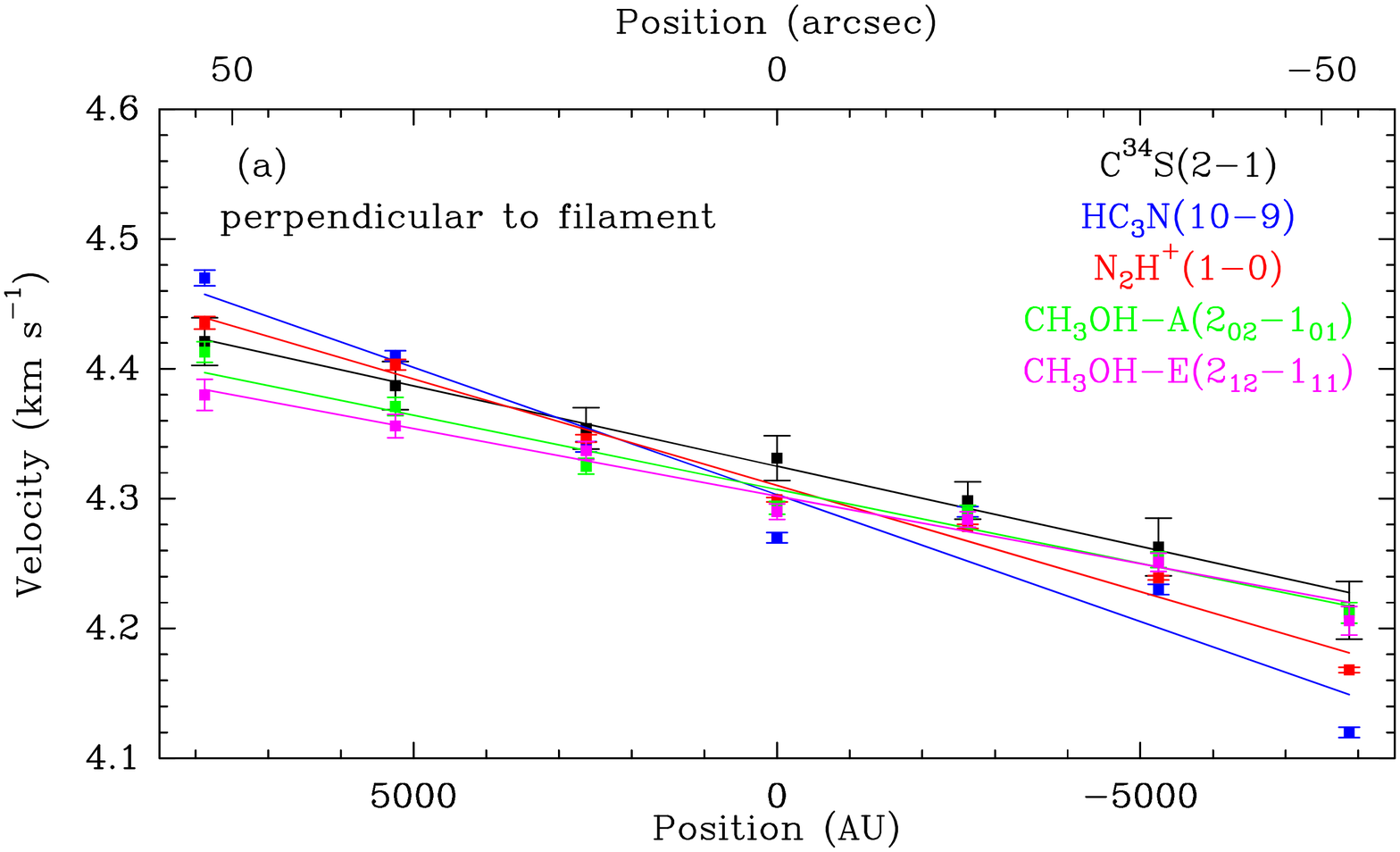} & 
\includegraphics[width=90mm,angle=0]{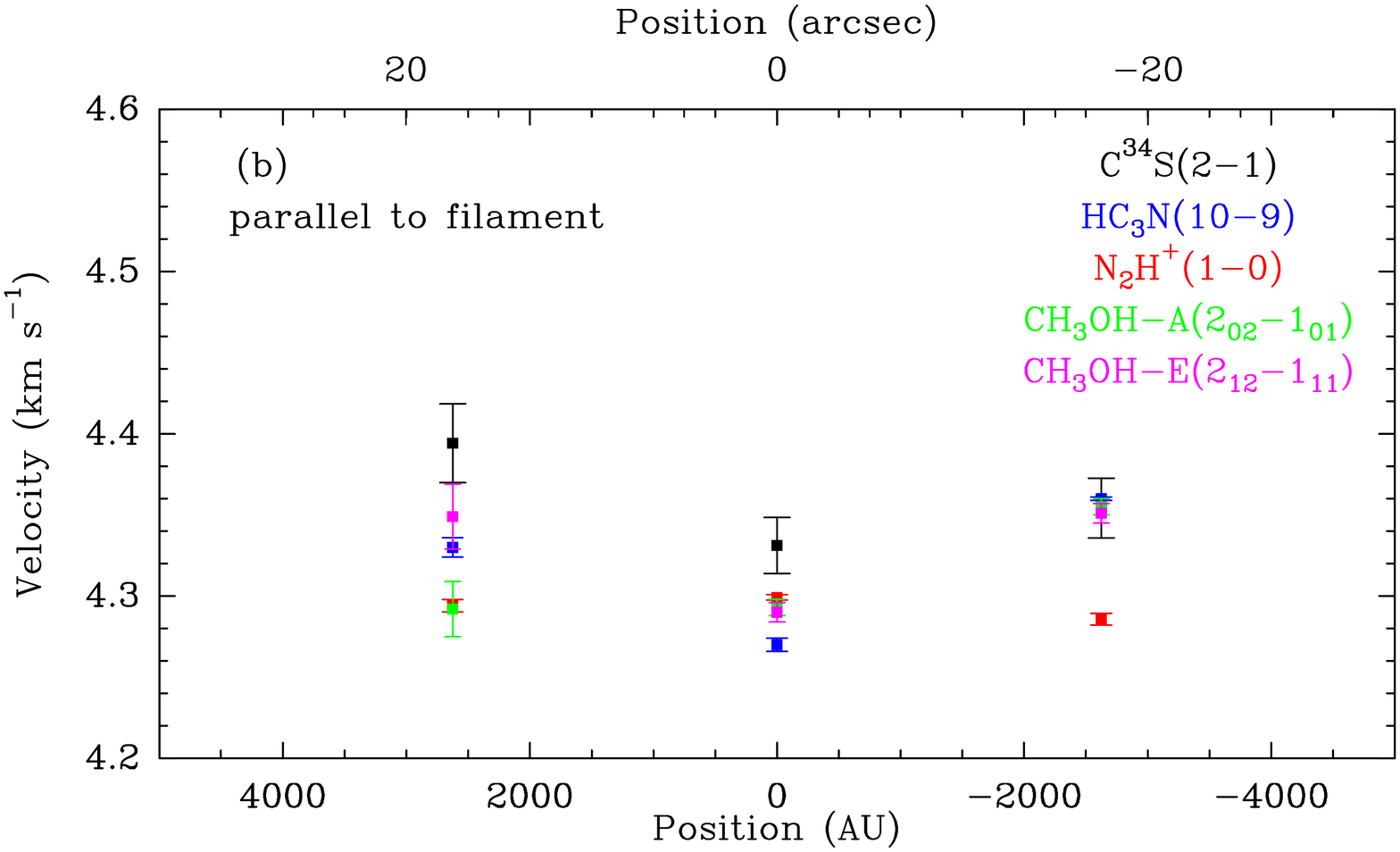} \\
\end{tabular}
\end{center}
\caption[]{Position-velocity diagrams of the Mopra C$^{34}$S 2--1 (black), HC$_3$N 10--9 (blue), N$_2$H$^+$ 1--0 (red), CH$_3$OH\emph{--A} 2$_{0,2}$--1$_{0,1}$ (green), and CH$_3$OH\emph{--E} 2$_{1,2}$--1$_{1,1}$ (pink) transitions, (a) perpendicular and (b) parallel to the filament. The error bars represent standard deviations (1$\sigma$). Linear fits to the velocities for each transition are shown with the same colour in panel (a).  \label{fig:pvmopra} }
\end{figure*}

\begin{figure*}%[!ht]
\begin{center}
\begin{tabular}{ccc}
\includegraphics[width=92mm,angle=0]{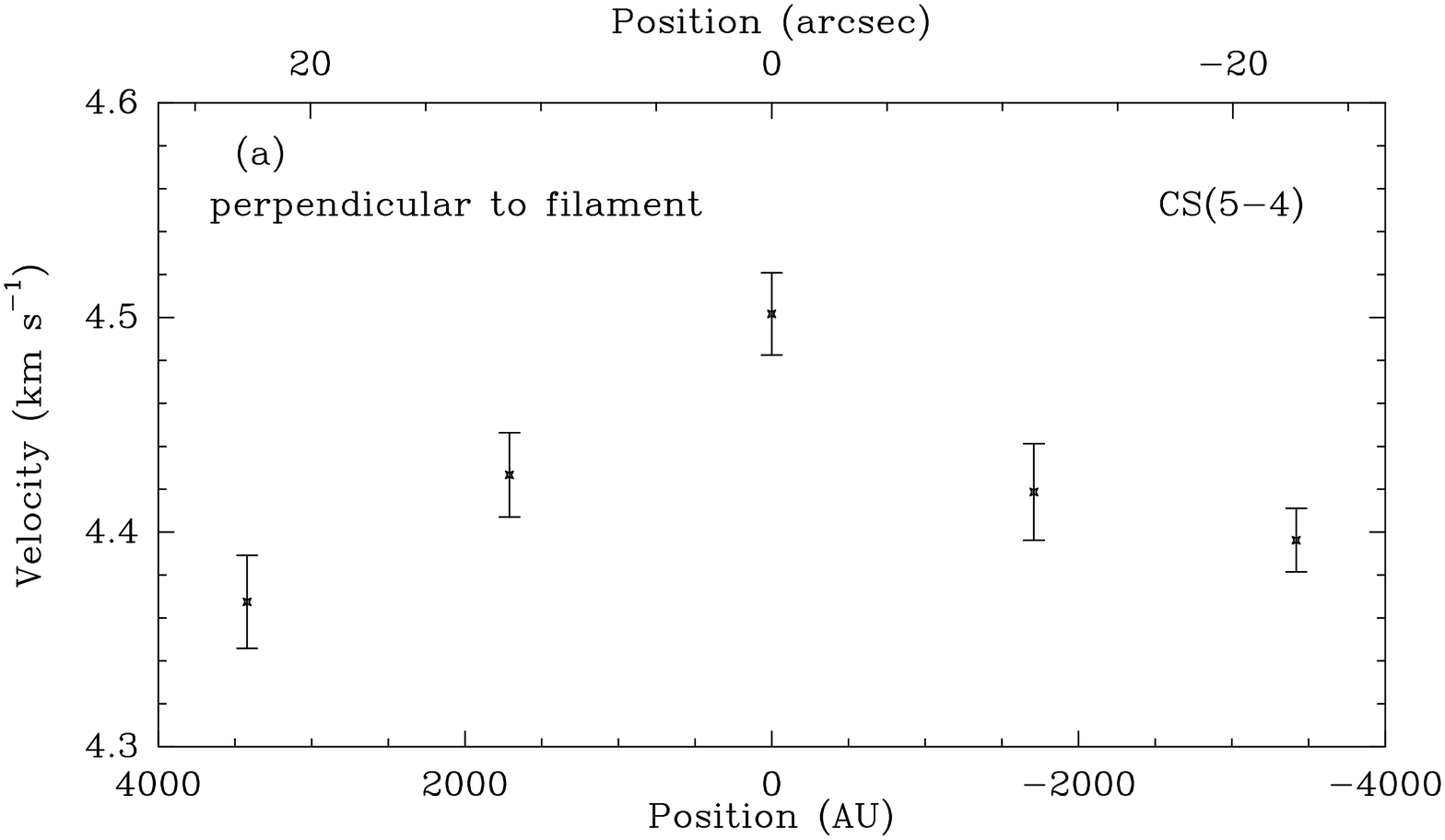} &
\includegraphics[width=90mm,angle=0]{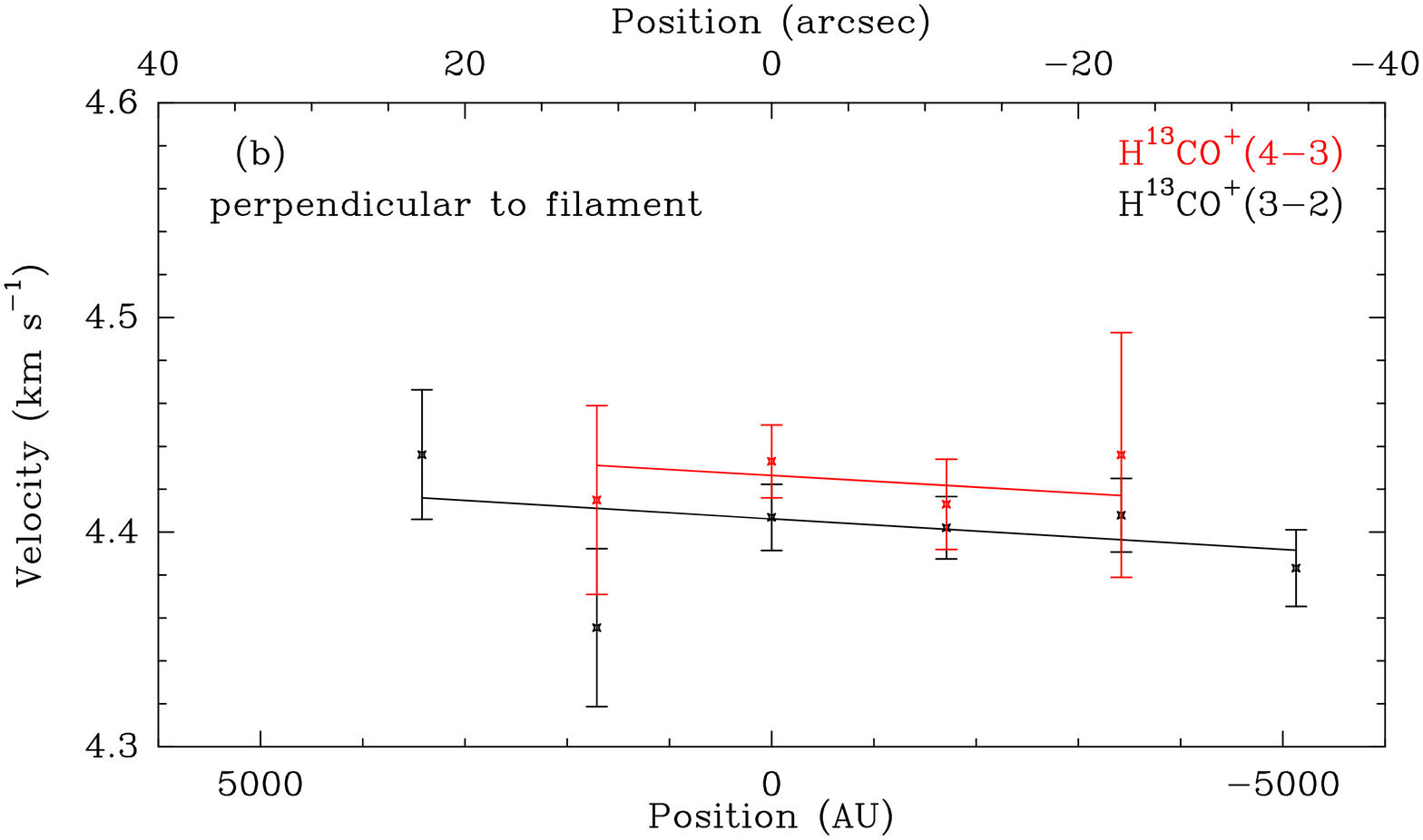} \\
\includegraphics[width=90mm,angle=0]{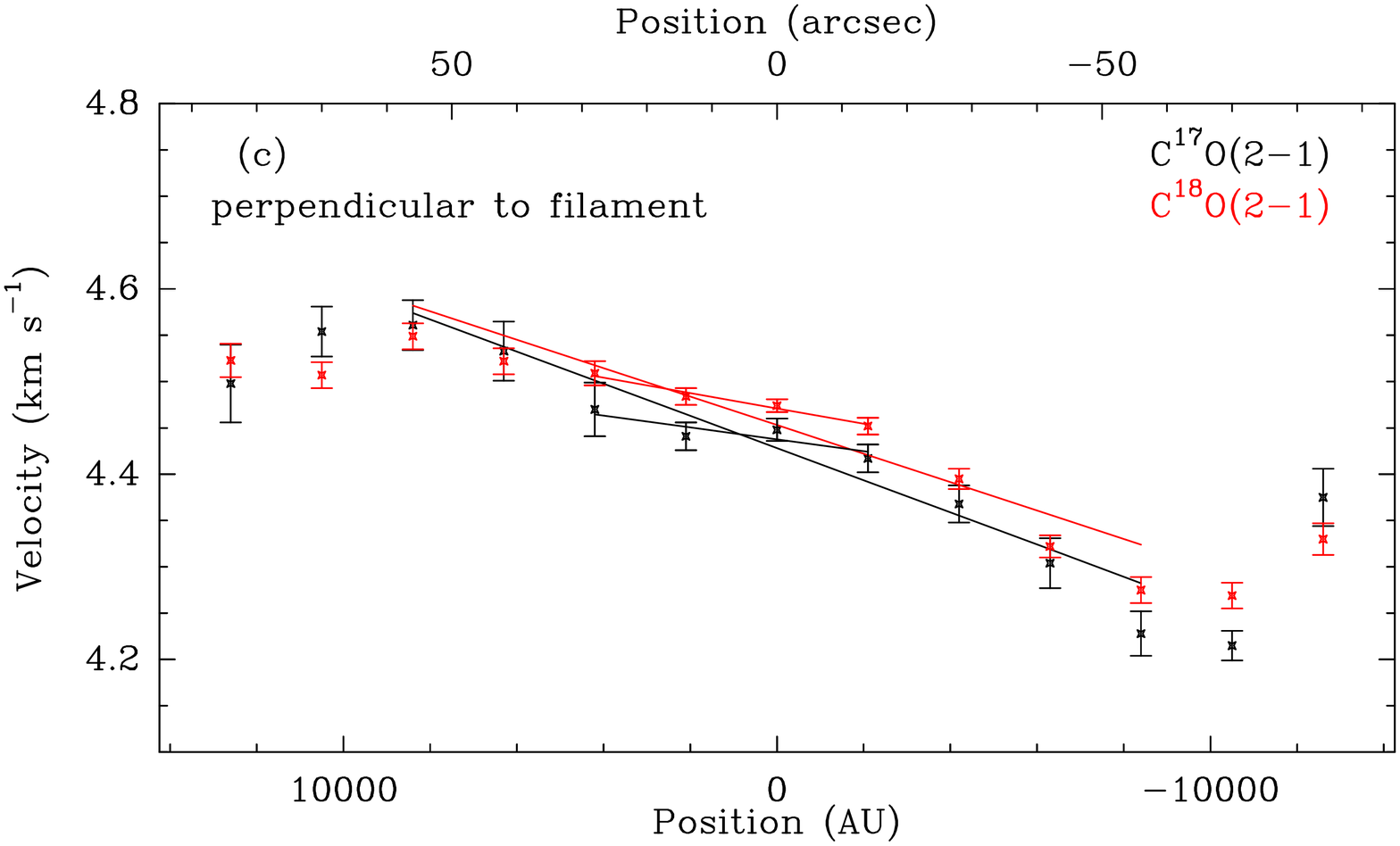} &
\includegraphics[width=90mm,angle=0]{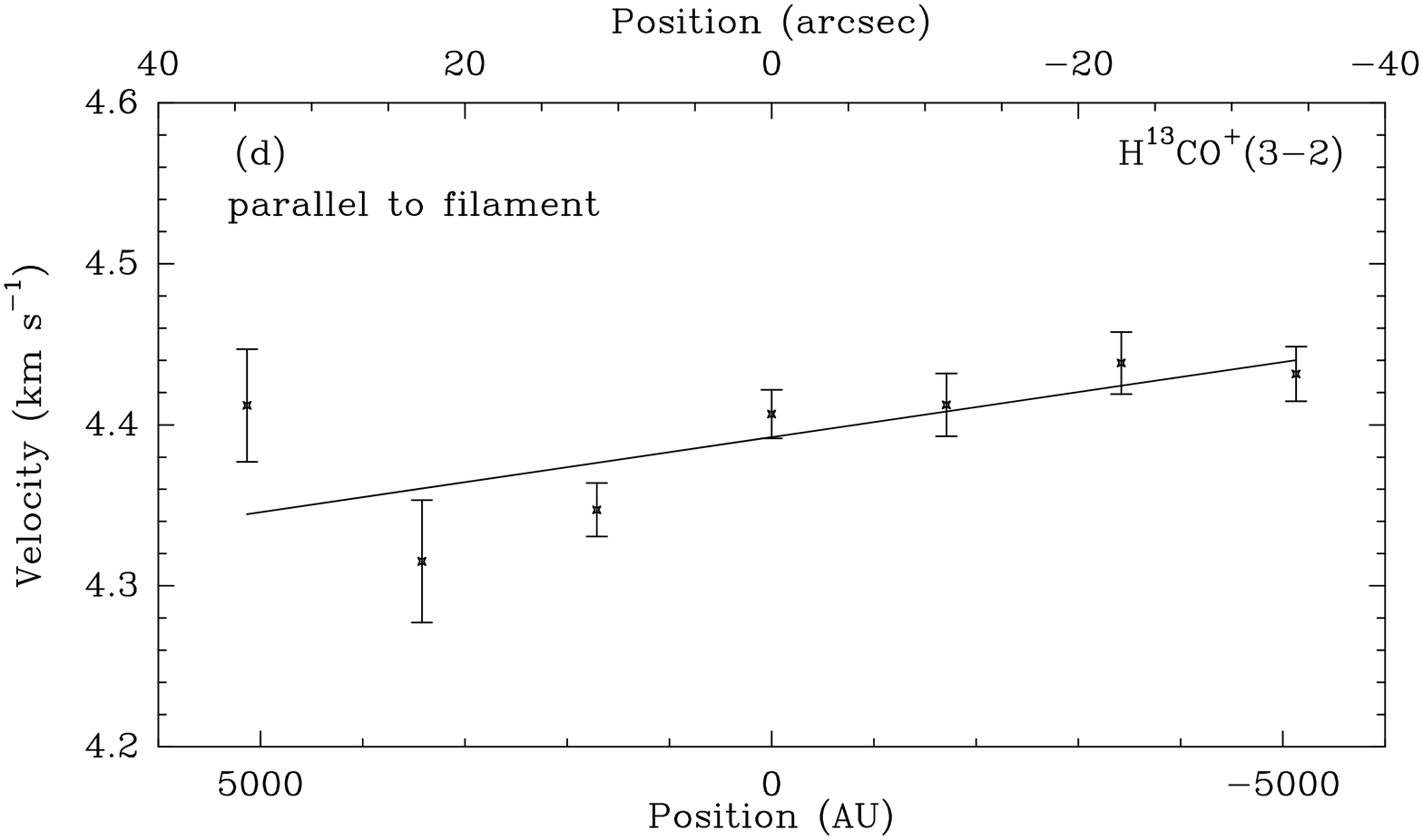} \\
\end{tabular}
\end{center}
 \caption[]{Position-velocity diagrams of the APEX CS 5--4, H$^{13}$CO$^+$ 3--2, H$^{13}$CO$^+$ 4--3, C$^{17}$O 2--1, and C$^{18}$O 2--1 transitions perpendicular (a, b, c) and parallel (d) to the filament. Linear fits are shown as straight lines. For C$^{17}$O 2--1 and C$^{18}$O 2--1 in (c) the two outer points from either side of the curve were excluded from the fits. Fits to the innermost positions of C$^{17}$O 2--1 and C$^{18}$O 2--1 are also displayed. \label{fig:pvapex} }
\end{figure*}

We constructed position-velocity (P-V) diagrams for the Mopra C$^{34}$S 2--1, HC$_3$N 10--9, N$_2$H$^+$ 1--0, CH$_3$OH\emph{--A} 2$_{0,2}$--1$_{0,1}$, and CH$_3$OH\emph{--E} 2$_{1,2}$--1$_{1,1}$ transitions and the APEX CS 5--4, H$^{13}$CO$^{+}$ 3--2, H$^{13}$CO$^{+}$ 4--3, C$^{17}$O 2--1, and C$^{18}$O 2--1 transitions (Figs.~\ref{fig:pvmopra} and ~\ref{fig:pvapex}) based on the centroid velocities measured in Sect.~\ref{sec:vlsrvelocities}.  

%%parallel to filament

We performed linear fits to these P-V diagrams to search for velocity gradients. The results are listed in Table~\ref{table:velgrad} and shown in Figs.~\ref{fig:pvmopra} and \ref{fig:pvapex}. Combining all tracers, there is no clear velocity gradient parallel to the filament ($\leq 2$~km~s$^{-1}$~pc$^{-1}$, see Figs.~\ref{fig:pvmopra}b and \ref{fig:pvapex}d).

%%perpendicular to filament

The P-V diagrams for the direction perpendicular to the filament are given in Figs.~\ref{fig:pvmopra}a and \ref{fig:pvapex}a--c. 
There is a clear velocity gradient along this direction with 
an amplitude of $\sim 2$ km s$^{-1}$~--~4.5~km~s$^{-1}$~pc$^{-1}$ up to 
$\sim8000$~AU. 
The average velocity gradient is $\sim 3.1 \pm 0.1$~km~s$^{-1}$~pc$^{-1}$. However, the C$^{17}$O 2--1 and C$^{18}$O 2--1 curves are significantly flatter at the inner, $r \le 4000$~AU radii, compared to the range 4000~AU --~8000~AU (Fig.~\ref{fig:pvapex}c), with velocity gradients $\leq 2$~km~s$^{-1}$~pc$^{-1}$. The H$^{13}$CO$^+$ 4--3 and H$^{13}$CO$^+$ 3--2 P-V curves are also consistent with no gradient for the inner $\sim 4000$~AU. 

The P-V curves of C$^{17}$O 2--1 and C$^{18}$O 2--1 depart from a straight line for radii larger than $\sim8000$ AU. The weighted average velocity gradient for the two transitions between the two outermost positions (at $\pm$ 12500~AU) is $\sim1.5 \pm 0.2$~km~s$^{-1}$~pc$^{-1}$, i.e. smaller by a factor of $\sim2$ than within 8000~AU. Such an ''S'' shape was reported by \citet{belloche02} for the Class 0 protostar IRAM 04191 and was interpreted as an indication of differential rotation in the envelope beyond a certain radius. If the velocity gradient of  Cha-MMS1 perpendicular to the filament is due to rotation then the bulk of the envelope is roughly in solid-body rotation between $\sim 4000$ AU and 8000~AU, and the outermost parts of the filament are rotating more slowly. Correcting for an inclination of 60$^\circ$~--~30$^\circ$ (Sect.~\ref{sec:intlum}), the average angular velocity for radii between 4000 AU and 8000~AU is $\Omega \sim 3.6$ km s$^{-1}$ pc$^{-1}$~--~6.2~km~s$^{-1}$~pc$^{-1}$, respectively. At 12500~AU, and for the same inclinations, we obtain $\Omega \sim 1.8$ km s$^{-1}$ pc$^{-1}$~--~3.0~km~s$^{-1}$~pc$^{-1}$.

The P-V diagram of CS 5--4 (Fig.~\ref{fig:pvapex}a) shows a centrally
peaked shape. 
The profile is not well resolved but the centroid velocity at the 
central position is significantly higher than at one beam spacing on each 
side. This velocity shift is due to the presence of an excess of redshifted emission toward the central position which is not reproduced by our radiative transfer modeling (see Sect.~\ref{sec:mapyso} and Fig.~\ref{fig:CSmodel}).
This issue is discussed in Sect.~\ref{sec:discussion_outflow}.

%%%%%%%%%%%%%%% TABLE 4 %%%%%%%%%%%%%%
%% Velocity Gradients

\begin{table}%[htpb]
\begin{center}
\caption{Velocity gradients perpendicular and parallel to the filament.} 
\vspace*{1ex}
\begin{tabular}{@{\extracolsep{-8pt}}lllll} 
\hline\hline
           & \multicolumn{2}{c}{Perpendicular} & \multicolumn{2}{c}{Parallel} \\
           & \multicolumn{2}{c}{\underline{to filament}} & \multicolumn{2}{c}{\underline{to filament}} \\
Transition & $\bigtriangledown$v\tablefootmark{a} & Extent\tablefootmark{b}          &  $\bigtriangledown$v\tablefootmark{a} & Extent\tablefootmark{b}  \\ 
            & {\scriptsize(km s$^{-1}$ pc$^{-1}$)} &  {\scriptsize(AU)} & {\scriptsize(km s$^{-1}$ pc$^{-1}$)} &  {\scriptsize(AU)} \\ \hline
C$^{34}$S 2--1                   & 2.6$\pm$0.1    & 15750  & 0.6$\pm$1.0   & 5250 \\
HC$_3$N 10--9                   & 4.0$\pm$0.5    & 15750   & -1.6$\pm$1.5 & 5250 \\
N$_2$H$^+$ 1--0                & 3.4$\pm$0.2    & 15750   & 0.1$\pm$0.1   & 5250  \\
CH$_3$OH\emph{--A} 2$_{0,2}$--1$_{0,1}$    & 2.4$\pm$0.2     & 15750   & -2.0$\pm$0.7  & 5250   \\
CH$_3$OH\emph{--E} 2$_{1,2}$--1$_{1,1}$    & 2.2$\pm$0.2     & 15750   & -1.5$\pm$1.4  & 5250   \\
H$^{13}$CO$^+$ 3--2             & 0.6$\pm$0.6     & 8550    & -1.9$\pm$0.8  & 10260  \\
H$^{13}$CO$^+$ 4--3             & 0.6$\pm$1.1     & 5130    & -             & - \\
C$^{17}$O 2--1                  & 3.6$\pm$0.5     & 16800   & -             & - \\
C$^{17}$O 2--1$^c$              & 1.3$\pm$0.7     & 6300    & -             & - \\
C$^{18}$O 2--1                  & 3.2$\pm$0.5     & 16800   & -             & - \\
C$^{18}$O 2--1$^c$              & 1.7$\pm$0.2     & 6300    & -             & - \\
\hline
\end{tabular}
\label{table:velgrad}
\end{center}
\tablefoot{
\tablefoottext{a}{The velocity gradients were estimated from linear fits to the position-velocity curves in Figs.~\ref{fig:pvmopra} and ~\ref{fig:pvapex}.}
\tablefoottext{b}{Total extent over which a linear fit to the data was performed.} 
\tablefoottext{c}{Gradients corresponding to the fits limited to the inner parts in Fig.~\ref{fig:pvapex}c.}
}
%\label{table:velgrad}
\end{table}

%%%%%%%%%%%%%%%%%%%%%%%%%%%%%%%%%%%%%

%%%%%% Figure 1, P-V diagram %%%%%%%%%%%%%%%
%\vspace*{-4ex}
%%%%%%%%%%%%%%%%%%%%%%%%%%%%%%%%%%

%%%%%% Figure, P-V diagram %%%%%%%%%%%%%%%
%\vspace*{-4ex}

%%%%%%%%%%%%%%%%%%%%%%%%%%%%%%%%

\subsection{Turbulence}
\label{sec:turbulence}

\begin{figure*}%[htpb]
\begin{tabular}{ccc}
\includegraphics[width=0.25\hsize,angle=270]{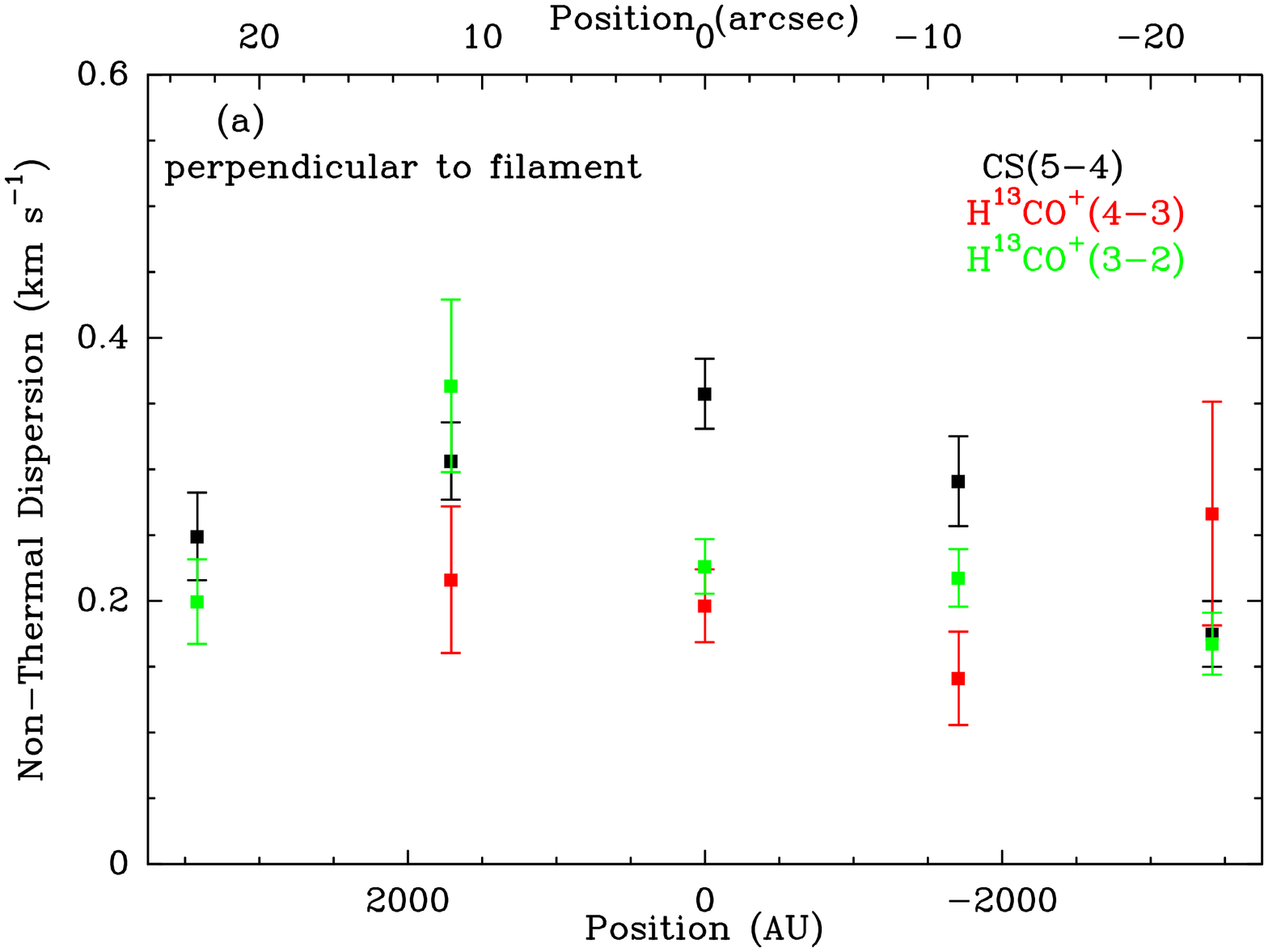} &
\includegraphics[width=0.25\hsize,angle=270]{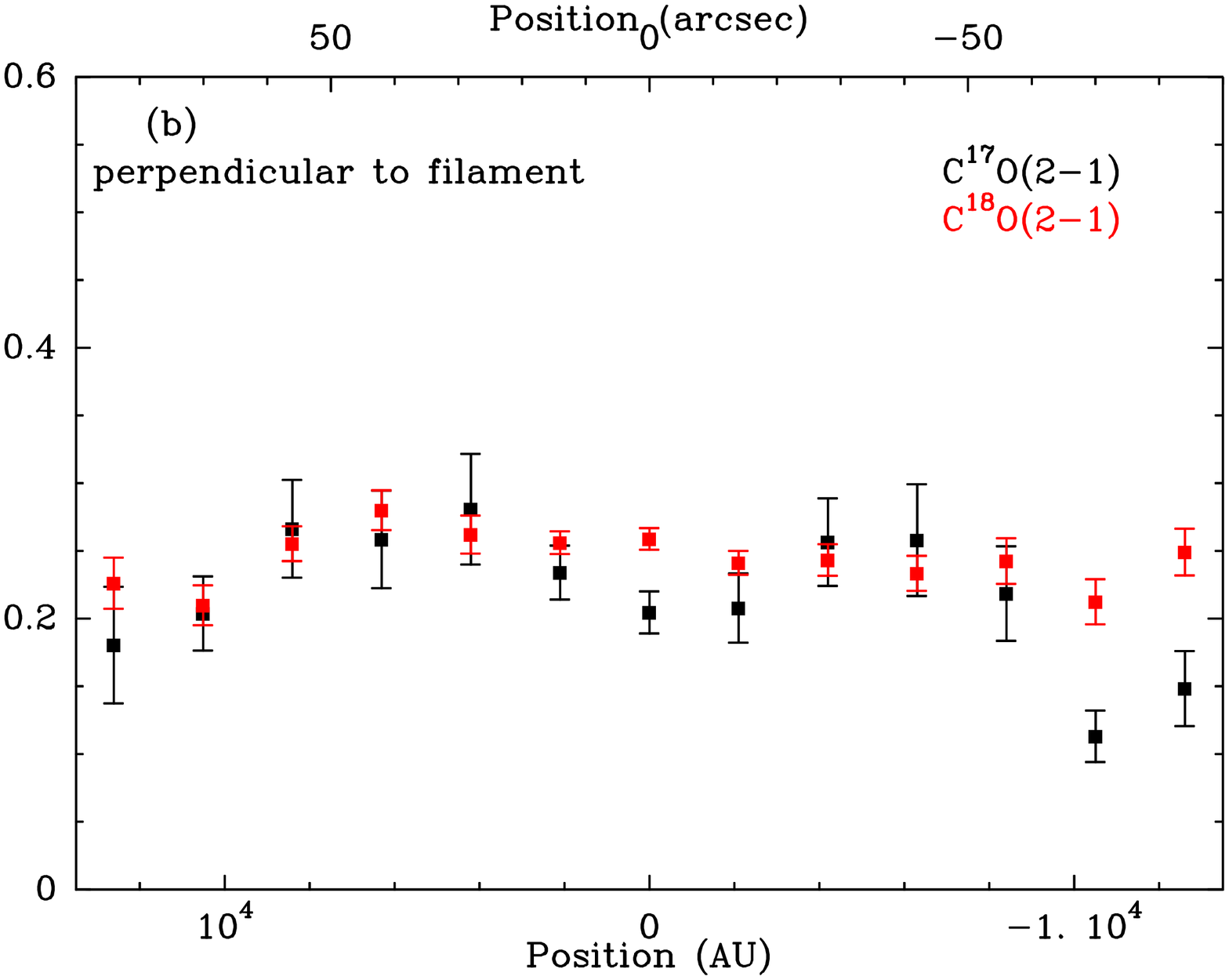} &
\includegraphics[width=0.25\hsize,angle=270]{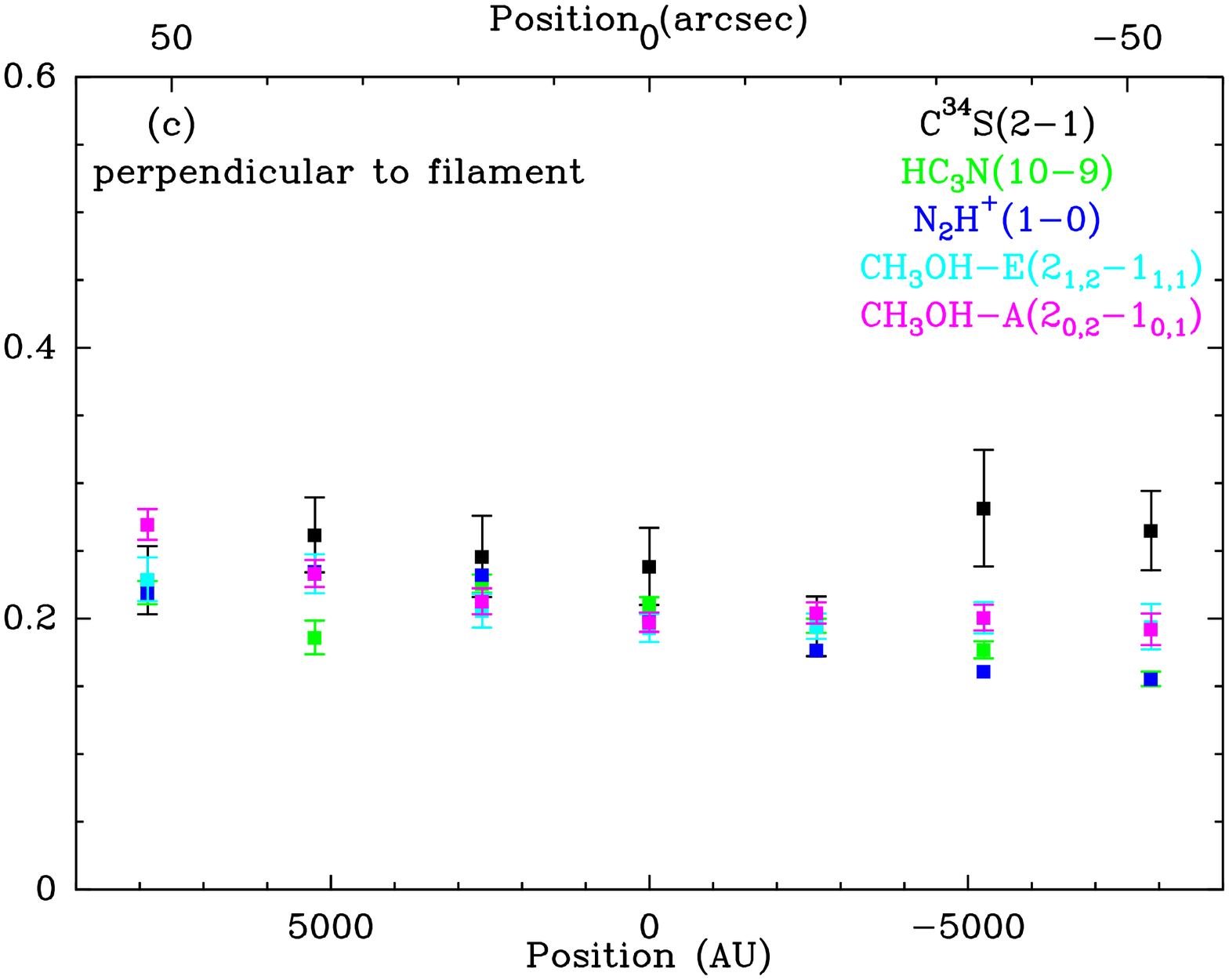} \\
\end{tabular}
\caption[]{\small Non-thermal velocity dispersion, $\sigma_{\rm nth}$, versus position perpendicular to the filament for (a) CS 5--4 (black), H$^{13}$CO 4--3 (red), and H$^{13}$CO$^+$ 3--2 (green), (b) C$^{17}$O 2--1 and C$^{18}$O 2--1, all observed with the APEX telescope, and (c) C$^{34}$S 2--1 (black), HC$_3$N 10--9 (green), N$_2$H$^+$ 1--0 (dark blue), CH$_3$OH\emph{--E} 2$_{1,2}$-1$_{1,1}$ (light blue), and CH$_3$OH\emph{--A} 2$_{0,2}$-1$_{0,1}$ (pink) observed with Mopra. The errorbars are standard deviations. The thermal dispersion was calculated assuming a temperature of $T$~=~9~K.}
\label{fig:posfwhm}
\end{figure*}

The spatial variation of the non-thermal velocity dispersion is shown in Fig.~\ref{fig:posfwhm} for several transitions. The ($FWHM$) linewidths were estimated using the ``GAUSS'' and ``HFS'' fitting methods in CLASS (see Table~\ref{table:fwhm}), as mentioned in Sect.~\ref{sec:vlsrvelocities}. We compute the thermal velocity dispersion of each molecule as follows,
\begin{equation} \label{eq:sigmath}
\sigma_{\rm th} = \sqrt{\frac{k_BT}{\mu_M m_H}},  
\end{equation}
where $\mu_{\rm M}$ is the molecular weight of the molecule, $k_{\rm B}$ the Boltzmann's constant, $m_{\rm H}$ the hydrogen mass, and $T$ the kinetic temperature that we assume to be 9 K (see Sect.~\ref{sec:mapyso}). The non-thermal linewidths are computed as follows,
\begin{equation} \label{eq:sigmanth}
\sigma_{\rm nth} = \sqrt{{\sigma_{\rm obs}}^2 - {\sigma_{\rm th}}^2},
\end{equation}
while their respective uncertainties, denoted as $\delta \sigma$, are calculated using:
\small
\begin{equation}
\delta \sigma_{\rm nth} = \sqrt{\Bigg({\frac{\partial \sigma_{\rm nth}}{\partial \sigma_{\rm observed}}}\Bigg)^2{\delta \sigma_{\rm obs}}^2 + {\Bigg({\frac{\partial \sigma_{\rm nth}}{\partial \sigma_{\rm th}}}\Bigg)^2}{\delta \sigma_{\rm th}}^2}. 
\end{equation}
\normalsize
If we assume that $\delta \sigma_{th} = 0$, the above relation gives,
\begin{equation}
\delta \sigma_{\rm nth} = \frac{\sigma_{\rm obs}\times \delta \sigma_{\rm obs}}{\sigma_{\rm nth}}.
\end{equation}
We compare the non-thermal velocity dispersion to the thermal dispersion of the mean particle, with $\mu = 2.37$:
\begin{equation} \label{eq:sigmath_mean}
\sigma_{\rm th,mean} = \sqrt{\frac{k_BT}{\mu m_H}}. 
\end{equation}
For $T=9$~K, $\sigma_{\rm th,mean}$ $\sim$ 0.18~km~s$^{-1}$, or $FWHM_{\rm th,mean} \sim 0.42$~km~s$^{-1}$. Table~\ref{table:fwhm} lists the observed linewidths and the derived thermal and non-thermal velocity dispersions for all transitions that have a low optical depth. 

From Table~\ref{table:fwhm} we infer that the non-thermal velocity dispersion is comparable to the mean thermal velocity dispersion. Therefore, there is almost an equipartition between thermal and non-thermal motions, provided that our asumption of $T=9$~K is valid. 

\begin{figure*}%[htpb]
\begin{tabular}{cc}
\includegraphics[width=65mm,angle=270]{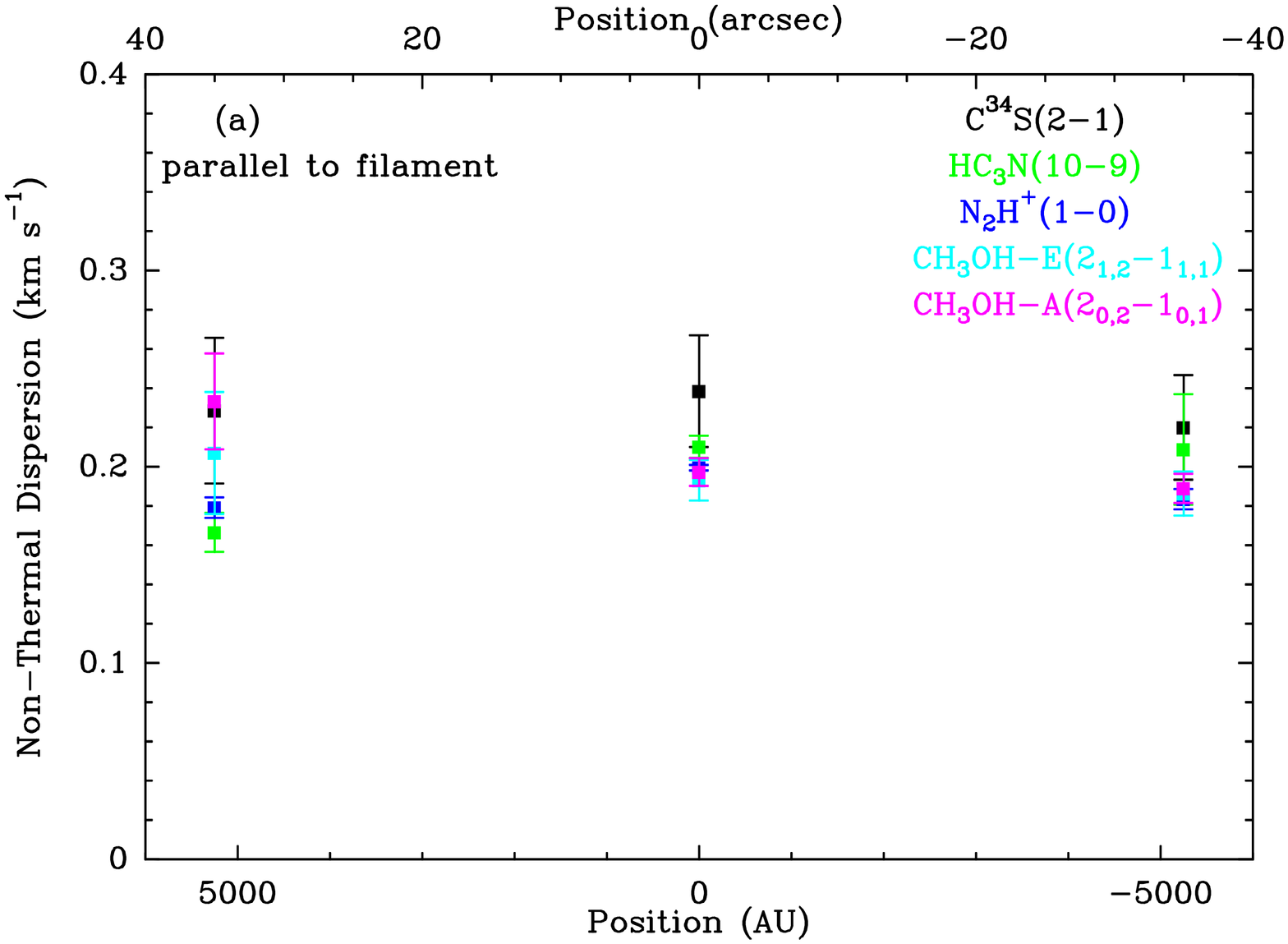} &
\includegraphics[width=65mm,angle=270]{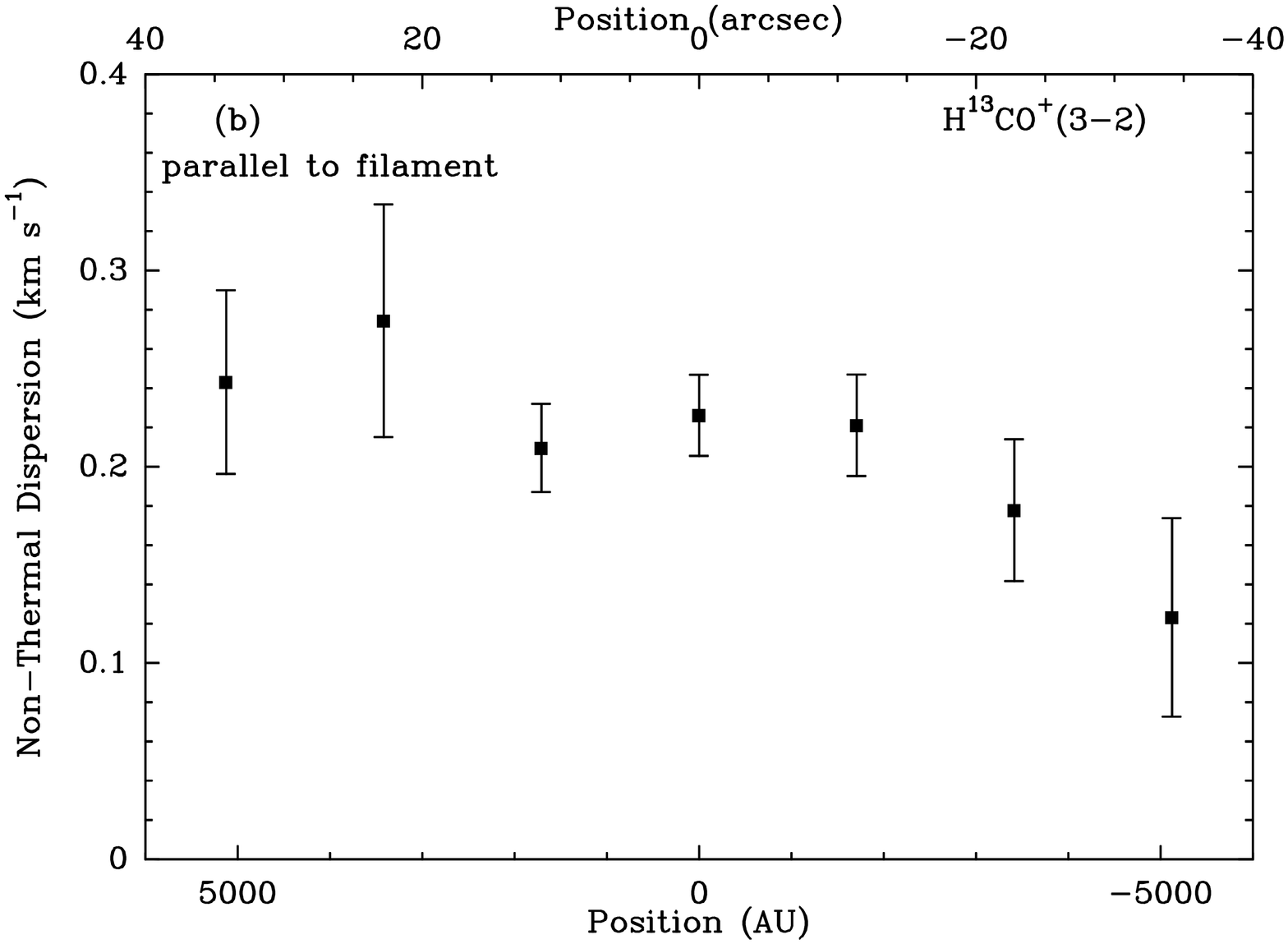} \\
\end{tabular}
\caption[]{\small Non-thermal velocity dispersion, $\sigma_{\rm nth}$, versus position parallel to the filament for 
(a) C$^{34}$S 2--1 (black), HC$_3$N 10--9 (green), N$_2$H$^+$ 1--0 (dark blue), CH$_3$OH\emph{--E} 2$_{1,2}$-1$_{1,1}$ (light blue), and CH$_3$OH\emph{--A} 2$_{0,2}$-1$_{0,1}$ (pink) observed with Mopra, and
(b) H$^{13}$CO$^+$ 3--2 observed with the APEX telescope. The errorbars are standard deviations. The thermal dispersion was calculated assuming a temperature of T~=~9~K.} \label{fig:posfwhm_p8p9}
\end{figure*}

The non-thermal dispersion does not vary significantly along or across the filament (Figs.~\ref{fig:posfwhm} and ~\ref{fig:posfwhm_p8p9}). One exception is, however, CS 5--4. Its non-thermal linewidth peaks at the centre and decreases at the outer parts. This is due to the excess redshifted emission observed at the central position (see Sect.~\ref{sec:rotation}, and Sects.~\ref{sec:mapyso} and ~\ref{sec:discussion_outflow} for further discussion).

Average values for the linewidths and non-thermal velocity dispersions over all positions are given in Table~\ref{table:avgfwhm}. The non-thermal dispersions, $\sigma_{\rm nth}$, have typical values of 0.2 km~s$^{-1}$, comparable to the mean thermal dispersion. Uniform non-thermal dispersions at scales $\sim0.1$ pc of the same order as the thermal dispersions have also been previously observed in other dense cores \citep[e.g.,][]{tafalla04, barranco98}. 

%%%%%%%%%%%%%%%%%% Table 5 %%%%%%%%%%%%%%%

\onltab{
\begin{table}
\caption{Observed linewidth, thermal, and non-thermal velocity dispersions toward the central position.} 
\vspace*{-5ex}
\begin{tabular}{@{\extracolsep{-8pt}}lllll}\\ [2ex] \hline\hline 
Line & $FWHM$\tablefootmark{a} & $\sigma_{\rm th}$\tablefootmark{b} & $\sigma_{\rm nth}$\tablefootmark{c} & $\frac{\sigma_{\rm nth}}{\sigma_{\rm th,mean}}$\tablefootmark{d} \\ 
& {\scriptsize(km s$^{-1}$)} & {\scriptsize(km s$^{-1}$)} & {\scriptsize(km s$^{-1}$)} & \\ \hline
C$^{34}$S 2--1                     & 0.57$\pm$0.05  & 0.04 & 0.24$\pm$0.02   & 1.33$\pm$0.11 \\ 
HC$_3$N 10--9 \tablefootmark{e}   & 0.50$\pm$0.009  & 0.04 & 0.21$\pm$0.006 & 1.17$\pm$0.03  \\ 
N$_2$H$^{+}$ 1--0 \tablefootmark{e}  & 0.48$\pm$0.002 & 0.05 & 0.20$\pm$0.001  & 1.11$\pm$0.006 \\ 
CH$_3$OH\emph{--A} 2$_{0,2}$--1$_{0,1}$        & 0.48$\pm$0.01  & 0.05 & 0.20$\pm$0.007 & 1.11$\pm$0.04 \\ 
CH$_3$OH\emph{--E} 2$_{1,2}$--1$_{1,1}$        & 0.47$\pm$0.02  & 0.05 & 0.19$\pm$0.01 & 1.06$\pm$0.06 \\ 
H$^{13}$CO$^+$ 3--2                & 0.55$\pm$0.03  & 0.05 & 0.23$\pm$0.02   & 1.28$\pm$0.11 \\ 
H$^{13}$CO$^+$ 4--3                & 0.48$\pm$0.04  & 0.05 & 0.20$\pm$0.03   & 1.11$\pm$0.17 \\ 
C$^{17}$O 2--1                     & 0.50$\pm$0.04  & 0.05 & 0.20$\pm$0.02  & 1.11$\pm$0.11  \\ 
C$^{18}$O 2--1                     & 0.62$\pm$0.02  & 0.05 & 0.26$\pm$0.008 & 1.44$\pm$0.04 \\ 
CS 5--4                           & 0.85$\pm$0.04  & 0.04 & 0.36$\pm$0.03  & 2.00$\pm$0.17 \\
HN$^{13}$C 1--0                    & 0.42$\pm$0.03  & 0.05 & 0.17$\pm$0.01  & 0.94$\pm$0.06 \\ 
HNCO 4$_{0,4}$--3$_{0,3}$           & 0.54$\pm$0.03  & 0.04 & 0.22$\pm$0.01  & 1.22$\pm$0.06 \\ 
HC$^{18}$O$^+$ 1--0                & 0.50$\pm$0.06  & 0.05 & 0.21$\pm$0.03  & 1.17$\pm$0.17  \\ 
c-C$_3$H$_2$ 3$_{2,2}$-3$_{1,3}$     & 0.52$\pm$0.08  & 0.04 & 0.22$\pm$0.03  & 1.22$\pm$0.17 \\ 
\hline
\end{tabular}\\[1ex] 
\label{table:fwhm} 
\tablefoot{
\tablefoottext{a}{Observed linewidth deduced from gaussian or hyperfine fits to the spectra.}
\tablefoottext{b}{Thermal dispersion computed with Equation~\ref{eq:sigmath}.} 
\tablefoottext{c}{Non-thermal dispersion computed using Equation~\ref{eq:sigmanth}. } 
\tablefoottext{d}{Ratio of non-thermal to \emph{mean} thermal dispersion, with $\mu$$=$2.37.}
\tablefoottext{e}{These transitions have a hyperfine structure and were fitted with the ``HFS'' fitting method in CLASS. The other transitions were fitted using the ``GAUSS'' method.}
}  
\end{table} 
}

\begin{table}
\caption{Linewidths and non-thermal velocity dispersions averaged over all positions.} 
\vspace*{-5ex}
\begin{tabular}{llll}\\ [2ex] \hline\hline 
Line &  $FWHM$  &      $\sigma_{\rm nth}$  & $\frac{\sigma_{\rm nth}}{\sigma_{\rm th,mean}}$ \\ 
     & {\scriptsize(km s$^{-1}$)} &  {\scriptsize(km s$^{-1}$)} & \\ \hline
C$^{34}$S 2--1                & 0.57$\pm$0.009     & 0.24$\pm$0.002 & 1.33$\pm$0.17 \\ 
HC$_3$N 10--9                 & 0.47$\pm$0.003    & 0.20$\pm$0.001 & 1.09$\pm$0.05 \\ 
N$_2$H$^{+}$ 1--0             & 0.47$\pm$0.005    & 0.19$\pm$0.003 & 1.08$\pm$0.02 \\ 
CH$_3$OH\emph{--A} 2$_{0,2}$--1$_{0,1}$  & 0.51$\pm$0.003     & 0.21$\pm$0.001 & 1.18$\pm$0.06 \\ 
CH$_3$OH\emph{--E} 2$_{1,2}$--1$_{1,1}$  & 0.49$\pm$0.004     & 0.20$\pm$0.001 & 1.13$\pm$0.08 \\
H$^{13}$CO$^+$ 3--2           & 0.51$\pm$0.01     & 0.21$\pm$0.004 & 1.18$\pm$0.27 \\ 
H$^{13}$CO$^+$ 4--3           & 0.50$\pm$0.02     & 0.20$\pm$0.005 & 1.14$\pm$0.28 \\ 
C$^{17}$O 2--1                & 0.53$\pm$0.01     & 0.22$\pm$0.002 & 1.21$\pm$0.17 \\
C$^{18}$O 2--1                & 0.59$\pm$0.005     & 0.24$\pm$0.001 & 1.35$\pm$0.08 \\ 
CS 5--4 \tablefootmark{a}    & 0.61$\pm$0.02     & 0.26$\pm$0.004 & 1.42$\pm$0.17 \\ 
\hline
\end{tabular}\\[1ex] 
\label{table:avgfwhm} 
\tablefoot{
\tablefoottext{a}{The linewidth and non-thermal velocity dispersion of the \emph{central} CS 5--4 position was not taken into account for computing the average value, due to the spectrum's broadened shape. See Sect.~\ref{sec:discussion_outflow} for further discussion.}
}
\end{table}

%%%%%%%%%%%%%%%%%%%%%%%%%%%%%%%%%%%%%%%%%

%%%% Figures 3 & 4, FWHM vs POS %%%%%%%%%%

%%%%%%%%%%%%%%%%%%%%%%%%%%%%%%%

\subsection{Infall Signature}
\label{sec:infallsig}

\begin{figure}%[htpb]
\centerline{\resizebox{1.0\hsize}{!}{\includegraphics[angle=270]{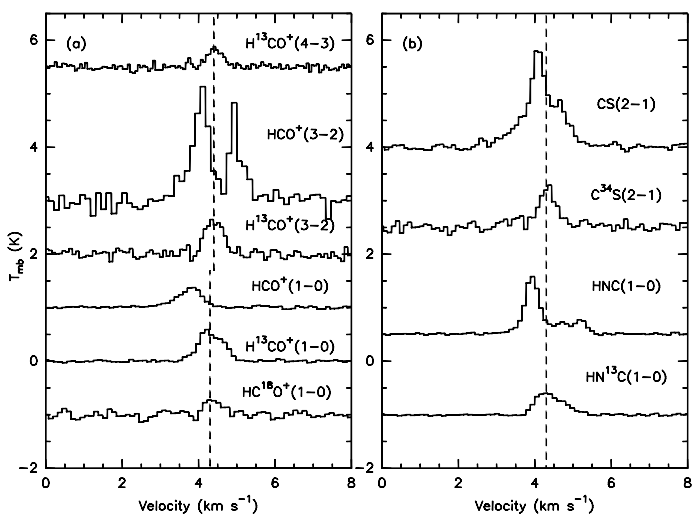}}}
\caption[]{Spectra obtained toward the central position of Cha-MMS1 
in the optically thick CS 2--1, HCO$^+$ 3--2, HCO$^+$ 1--0, and 
HNC 1--0 transitions and the low optical depth C$^{34}$S 2--1,  
H$^{13}$CO$^+$ 1--0, H$^{13}$CO$^+$ 3--2, H$^{13}$CO$^+$ 4--3, HC$^{18}$O$^+$ 1--0, and HN$^{13}$C 1--0 transitions. The dashed line corresponds to the systemic velocity derived from a seven-component hyperfine fit to the N$_2$H$^+$ 1--0 multiplet (4.3 km~s$^{-1}$). It is corrected to 4.4~km~s$^{-1}$ for the APEX transitions HCO$^+$ 3--2, H$^{13}$CO$^+$ 3--2, and H$^{13}$CO$^+$ 4--3 (see Sect.~\ref{sec:vlsrvelocities_issue}). \label{fig:infallsig}}
\end{figure}

Cha-MMS1 exhibits the classical signature of infall in various transitions 
(Fig.~\ref{fig:infallsig}). The infall signature manifests itself as a 
self-absorbed asymmetric, optically thick line with the blue peak being 
stronger than the red one, and an optically thin line peaking 
in-between these two peaks. This profile is indicative of inward 
motions as long as the excitation temperature increases towards the centre 
\citep[e.g.,][]{walker86, zhou92}.

Figure~\ref{fig:infallsig} shows that the absorption dips of CS 2--1 and HCO$^+$ 3--2 are redshifted with respect to the systemic velocity, shown as a dashed line. The systemic velocity was estimated by a seven component hyperfine-structure fit to the N$_2$H$^+$ 1--0 multiplet, using the HFS method in CLASS, giving a value of 4.299~$\pm$~0.002~km~s$^{-1}$. As discussed in Sect.~\ref{sec:vlsrvelocities_issue}, we apply a correction of 0.1~km~s$^{-1}$ to this systemic velocity to compare it to the APEX spectra. The self-absorption dip of the optically thick CS 2--1 transition has a velocity of 4.50~$\pm$~0.05~km~s$^{-1}$, which gives a velocity shift of 0.20~$\pm$~0.05~km~s$^{-1}$. Using the corrected value of the systemic velocity we derive a velocity shift of 0.18~$\pm$~0.05~km~s$^{-1}$ for the optically thick HCO$^+$ 3--2 line, whose absorption dip has a velocity of 4.58~$\pm$~0.06~km~s$^{-1}$. Since the self-absorption dips are produced by the outer parts of the envelope where the opacity of the CS and HCO$^+$ lines becomes unity, this observed redshift points to the fact that the outer layers of Cha-MMS1 undergo inward motions with velocities on the order of 0.2~km~s$^{-1}$.

%%%%%%%%%%%%%%%%%%%%%%%%%%%%%%%%%
%%%%%%%%%%%%%%%%%%%%%%%%%%%%%%%

\subsection{Overview of CHAMP$^+$ data}
\label{sec:champ+} 

\begin{figure*}%[htpb]
\begin{tabular}{cccc}\hspace{-3 em}
\includegraphics[width=48mm,angle=0]{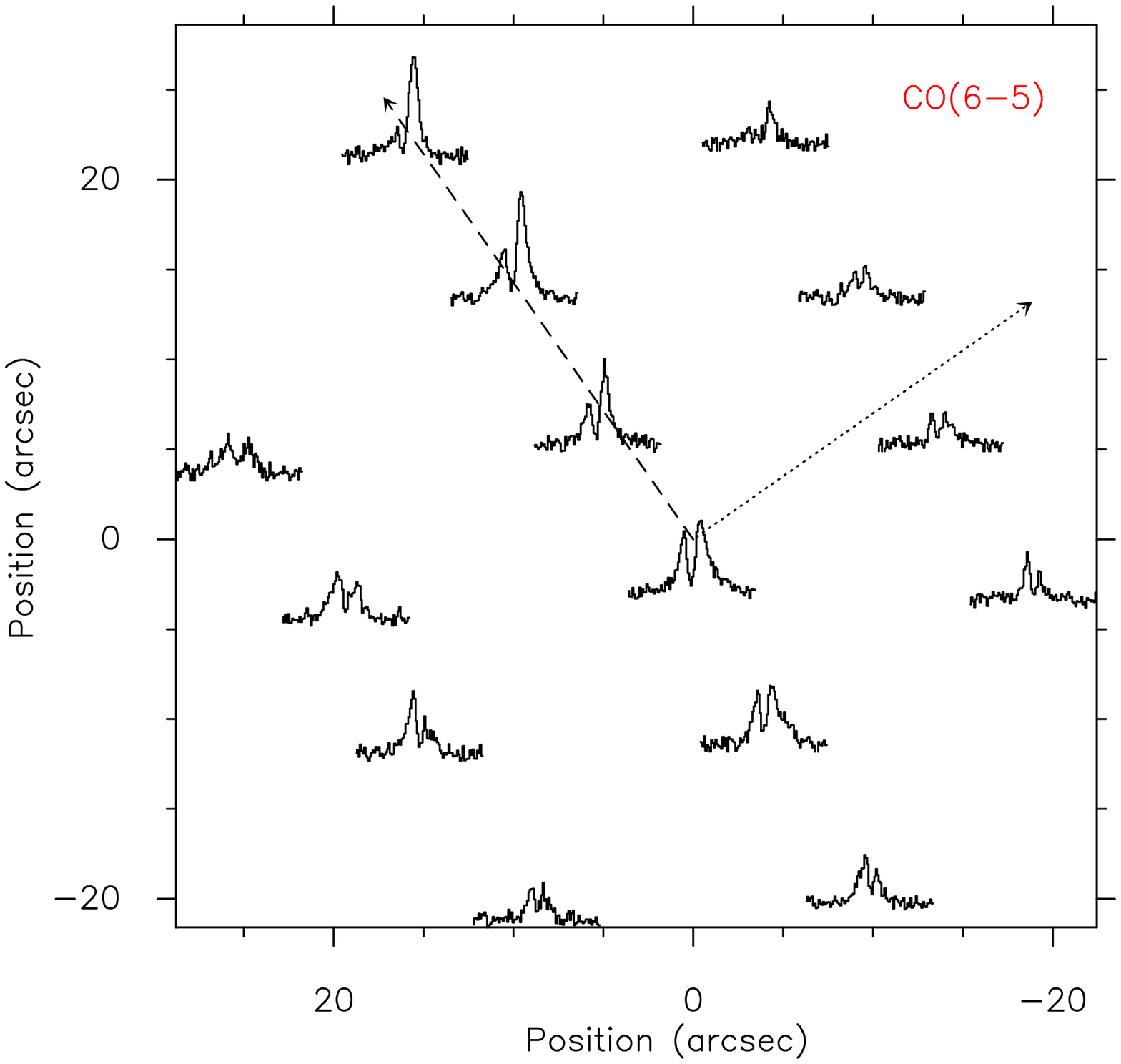}\hspace{-1. em} &
\includegraphics[width=44mm,angle=0]{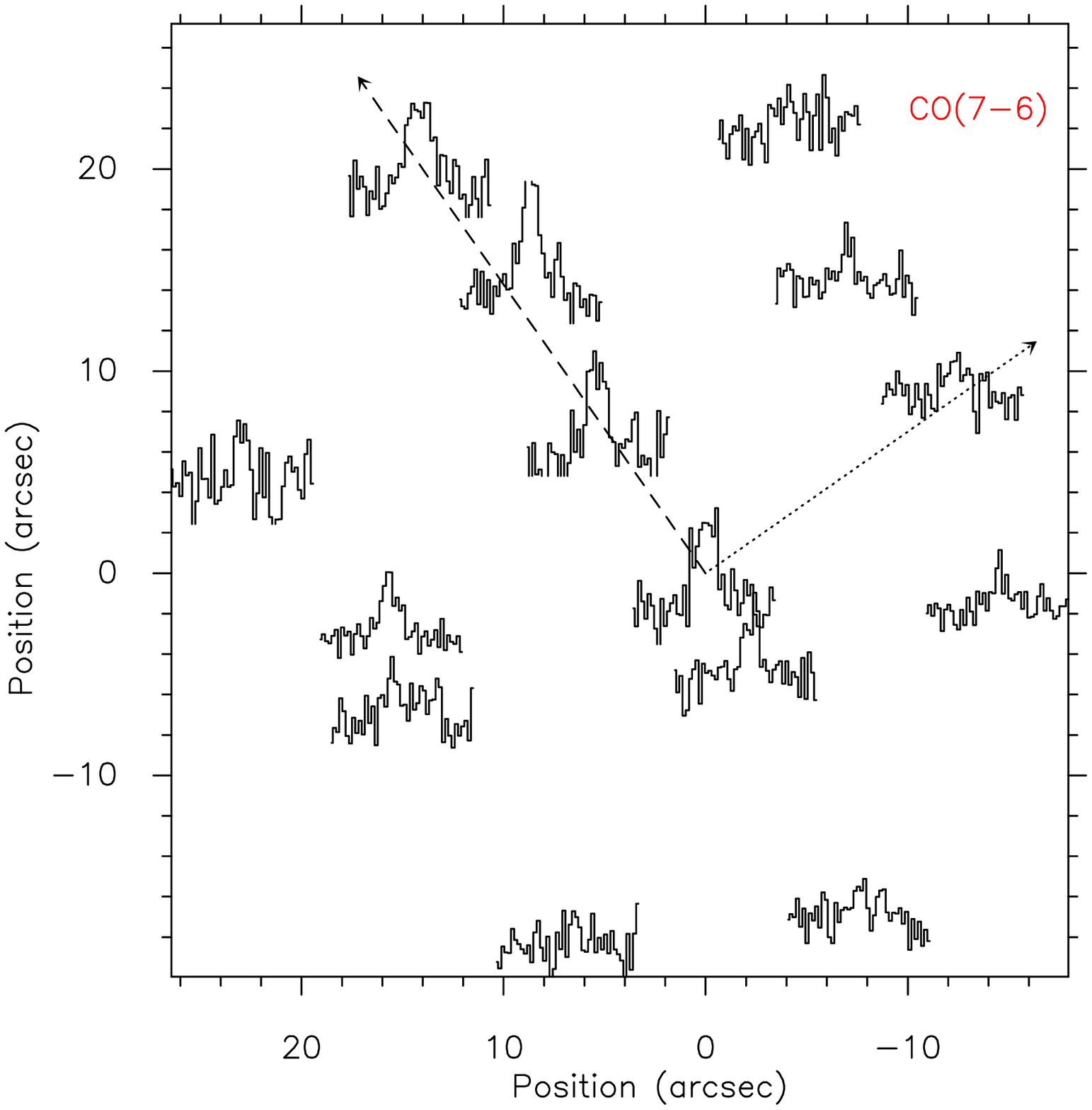}\hspace{-1. em} &
\includegraphics[width=50mm,angle=0]{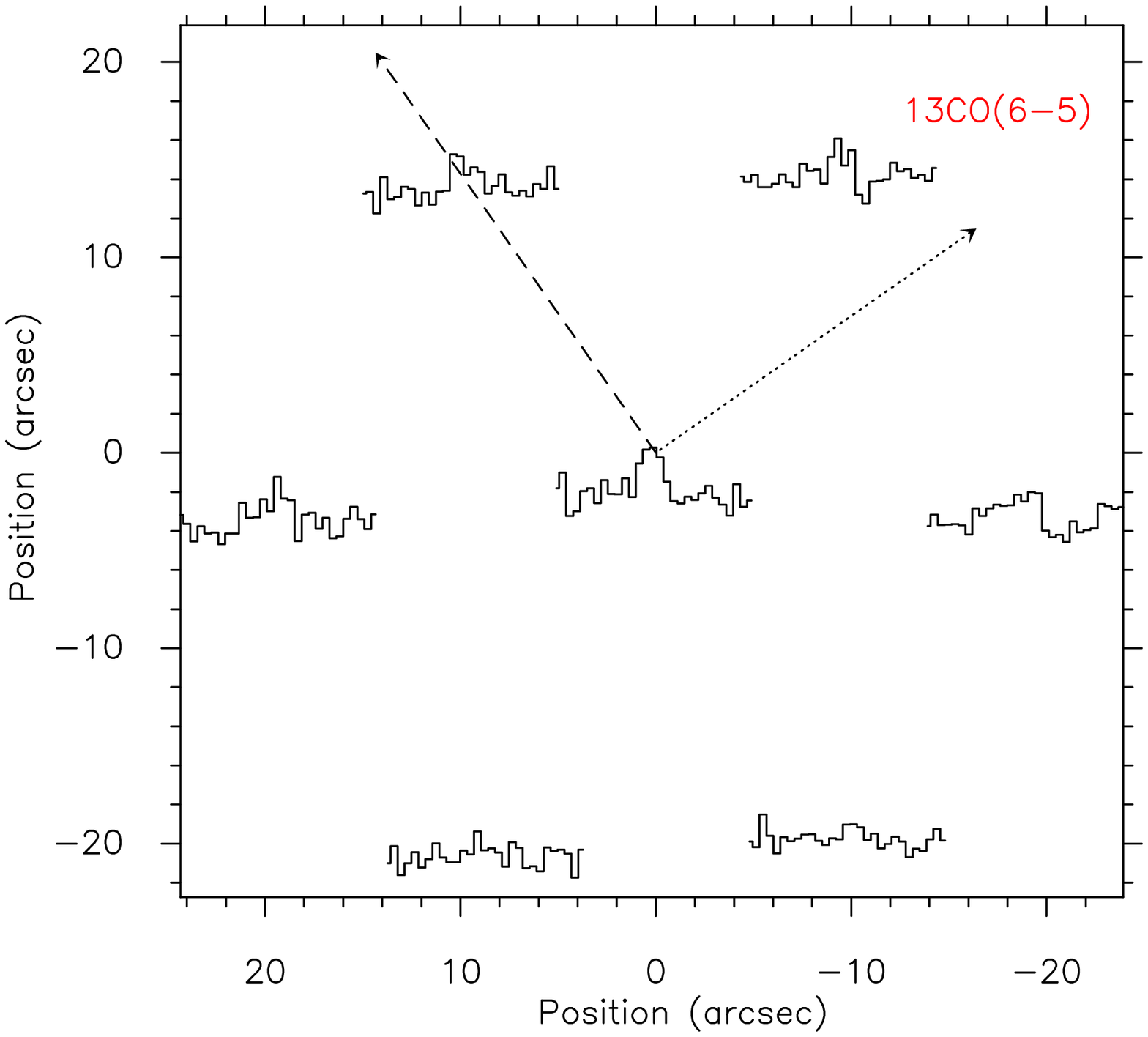} &
%\vspace*{-5 mm}
\includegraphics[angle=0, width=35mm]{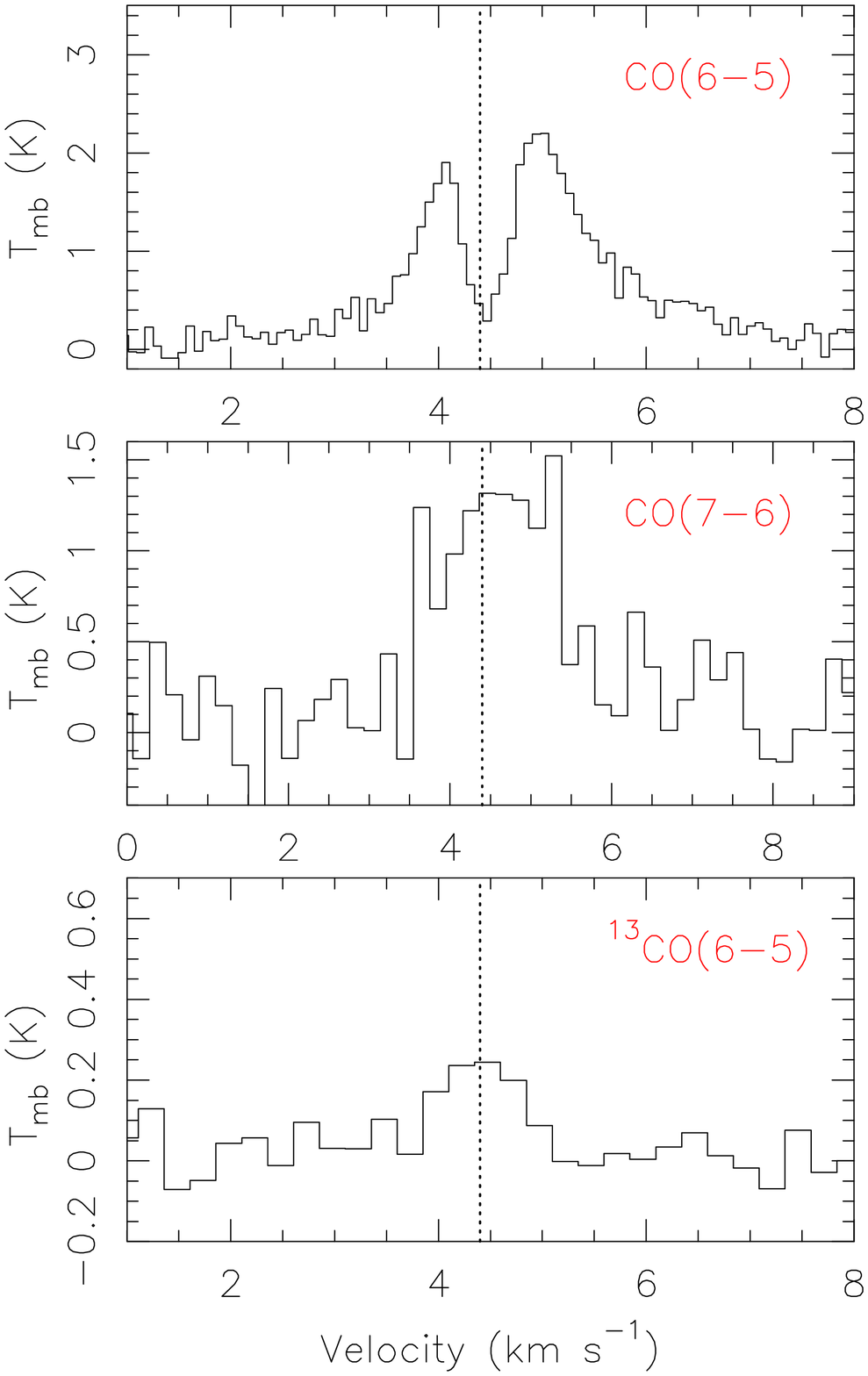} \\
%\vfill
\end{tabular}
\caption[]{CO 6--5, CO 7--6, and $^{13}$CO 6--5 spectra observed with the APEX telescope. The central position of Cha-MMS1 is at (0,0). The directions parallel and perpendicular to the filament are shown as dashed and dotted arrows, respectively. The central spectra of the three transitions are plotted at the rightmost panel in scale of the main-beam brightness temperature. The dotted line shows the systemic velocity of Cha-MMS1 derived from a seven-component hyperfine fit to the N$_2$H$^+$ 1--0 multiplet and corrected for the 0.1~km~s$^{-1}$ velocity shift (see Sect.~\ref{sec:vlsrvelocities_issue}). \label{fig:champspectra}}
\end{figure*}

We probed Cha-MMS1 in the CO 6--5, CO 7--6, and $^{13}$CO 6--5 molecular transitions with the APEX CHAMP$^+$ 2$\times$7-pixel heterodyne receiver array in order to search for emission indicative of outflowing material. Figure~\ref{fig:champspectra} shows position maps of all spectra obtained for each transition. 
The CO~6--5 and 7--6 spectra along the filament have slightly higher intensities toward the north-east as opposed to the south-west. This could be due to contamination from the outflow of the Class I object IRAS 11051-7706 lying close to Cha-MMS1 at the north-east direction \citep{belloche06} and it is further discussed in Sect.~\ref{sec:discussion_outflow}.

%%%%%%%%%%%%%% CHAMP+ SPECTRA%%%%%%%%%%%%%%%%

%%%%%%%%%%%%%%%%%%%%%%%%%%%%%%%%%%%%%%%%%%%%

\section{Radiative Transfer Modeling}
\label{sec:mapyso}

We used the Monte Carlo radiative transfer code MAPYSO \citep{blinder97, belloche02}, which assumes spherical symmetry, to derive kinematic constraints on the dense core Cha-MMS1 by modeling the observed spectra. We modeled the emission of three sets of molecules, namely CS, HCO$^+$, and CO with their respective isotopologues. 

We model all molecular transitions with the \emph{same} density, temperature, and turbulence distributions, as described below. These distributions are shown in Figs.~\ref{fig:ALLprofiles}a, c, and e, respectively. Given the shape of the continuum emission, we assume that Cha-MMS1 is embedded in a filament and that the physical structure perpendicular to the filament in the plane of the sky is similar to the structure along the line-of-sight. We thus model the spectra taken along the direction perpendicular to the filament in addition to the central spectra. We perform the fit optimisation by eye, by focusing on three main features of the spectra: the peak temperature, the position of the self-absorption dip (when there is one), and the linewidth of each spectrum. 

\subsection{Input Parameters}
\label{sec:input}

\subsubsection{Temperature profile}

Using the internal luminosity derived in Sect.~\ref{sec:intlum} we can constrain the inner dust temperature profile of the source, 
which we assume to be dominated by the central heating. 
Following \citet{terebey93} and \citet{motteandre01} we assume that in the inner part the dust temperature behaves as:  
\begin{equation}
T_{\rm dust}( r ) = 38~{\rm K} \times \left(\frac{r}{100~\mathrm{AU}}\right)^{-q}\times \left(\frac{L_{\rm int}}{1~L_{\odot}}\right)^{q/2},
\label{eq:tdust}
\end{equation}
with
\begin{equation}
q = \frac{2}{4+\beta}.
\end{equation}
The constant $\beta$ depends on the dust properties and the values 2, 1.5, and 1 are often used for molecular clouds, protostellar envelopes, and protostellar disks, respectively. Because this source is possibly at a very early evolutionary stage, perhaps before the protostellar phase, we adopt an intermediate value between the first two, 1.85, which gives $q$~$\sim$~0.34.

We assume that the gas and dust are well coupled for densities above $\sim10^{5}$~cm$^{-3}$ \citep{lesaffre05}, and we use Equation~\ref{eq:tdust} to define the kinetic temperature in the inner part. We assume a uniform temperature in the outer parts. We find from the modeling that a uniform kinetic temperature of 9~K produces spectra in agreement with the observations. More specifically, the optically thick CS 2--1 line sets the major constraints on the outer kinetic temperature, as this transition shows a strong asymmetry in the blue-red peak strengths (infall signature) with the blue peak being stronger by about a factor of $\sim$~2. Its absorption dip and asymmetry require uniform outer temperatures to be well fitted. The radius at which the temperature drops to 9~K is fixed by the central heating (Equation~\ref{eq:tdust}).

\subsubsection{Density profile}

We use a power-law density profile with an external radius of 60000 AU. We adopt a spherically symmetric $r^{-2}$ density profile for the envelope, as is expected from models of spherical gravitational collapse of nonsingular isothermal spheres during the core formation either without the effect of magnetic fields \citep[e.g.,][]{masunaga98, fosterchevalier93, bodenheimer68, larson69, penston69} or from models of axisymmetric, isothermal cloud contraction with magnetic fields \citep[e.g.,][]{tassis07, fiedler93}. Very recently, a 3D radiation hydrodynamic simulation of the collapse of an axisymmetric cloud core towards the formation of a first core led to a $r^{-2}$ density distribution for the first core for $r > 100$ AU \citep{furuya12}. Since we do not resolve the inner $r < 700$~AU of Cha-MMS1, a simple, $r^{-2}$ density profile is probably adequate in describing the envelope of Cha-MMS1. 

Such density profiles have also been observed in dense cores. Density distributions close to a $r^{-2}$ profile were derived for the starless cores L1498 and L1517B in Taurus \citep{tafalla04}. \citet{bacmann00} and \citet{alves01} also concluded that low-mass prestellar cores are well described by a density profile following an $r^{-2}$ dependence, excluding however the sharp edges and flattened centre. Sharp density edges have been observed at the edges of starless cores, with exponents as steep as $r^{-3.5}$ \citep{nielbock12}. For simplicity, we do not account for steep outer density slopes.
 
The mass of Cha-MMS1 was derived from the LABOCA 870~$\mu$m continuum map of the Chamaeleon I cloud. The flux density measured within a radius of 3750 AU gives a mass of $\sim$1.44~$M_{\odot}$ for the Cha-MMS1 core \citep{belloche11a}. We use this value to scale our input density profile.  

\subsubsection{Inner turbulent broadening and isotopic ratios}

In Sect.~\ref{sec:turbulence}, we found that the non-thermal dispersion for all the observed transitions shows no significant spatial variations for radii up to $\sim12500$ AU ($\sim0.06$ pc). We adopt a \emph{uniform} turbulent broadening up to this radius and keep the shape of the outer profile for $12500$ AU $<r< 60000$~AU ($\sim 0.3$~pc) as a free parameter. The non-thermal dispersion appears to be uniform at scales $\sim 0.1$~pc within the interiors of dense cores \citep[e.g.,][]{barranco98, goodman98} with increasing dispersion at larger scales \citep[e.g.,][]{goodman98} that follows the Larson scaling law \citep{larson81}.  

We assume the local ISM abundance isotopic ratios: $^{12}$C/$^{13}$C~$\sim$~77 \citep{wilsonrood94}, $^{32}$S/$^{34}$S~$\sim$~22 \citep{frerking80}, $^{16}$O/$^{18}$O~$\sim$~560 \citep{wilsonrood94}, and $^{18}$O/$^{17}$O~$\sim$~4.11 \citep{wouterloot05}.

\subsection{CS Modeling}
\label{sec:CSmodeling}

\begin{figure*}%[h]
\begin{tabular}{c}
\centerline{\resizebox{1.00\hsize}{!}{\includegraphics[angle=270]{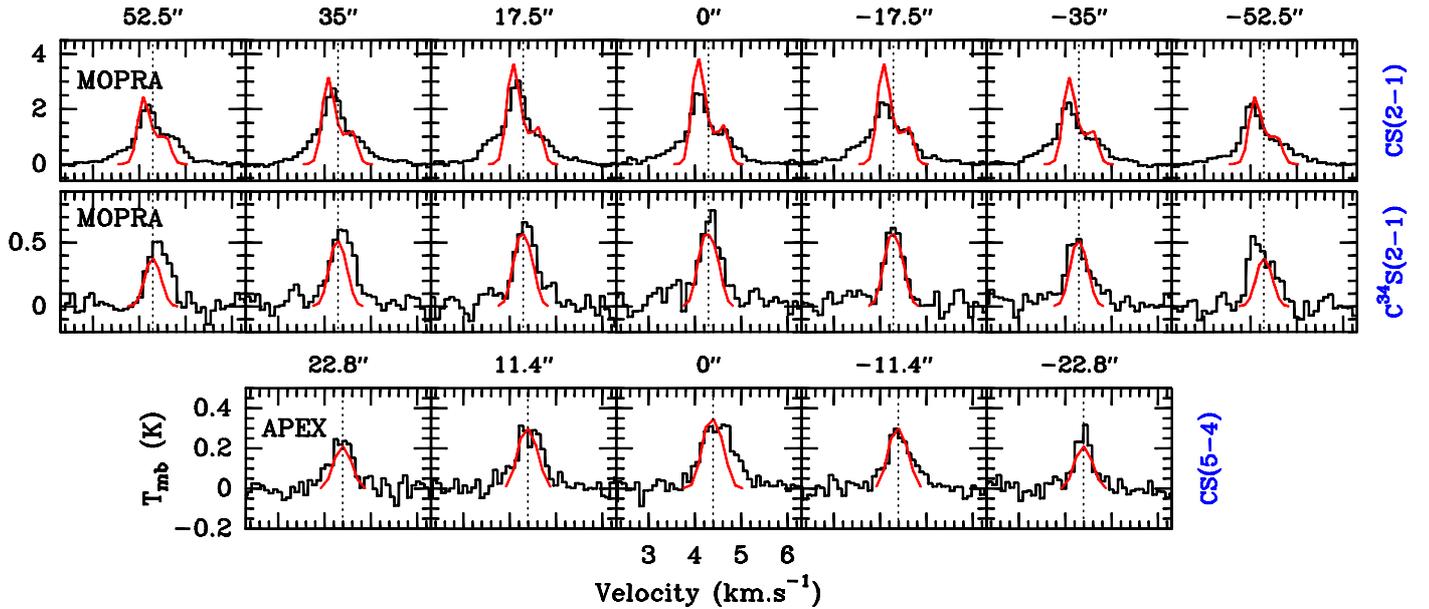}}}
\end{tabular}
\caption[]{\small Best fit model 'MCS' for the CS and C$^{34}$S molecular transitions. The black lines represent the observed spectra while the red spectra are generated by the MAPYSO radiative transfer code. The dotted line shows the systemic velocity of Cha-MMS1 derived from a seven-component hyperfine fit to the N$_2$H$^+$ 1--0 multiplet. For the APEX spectra, a correction of 0.1~km~s$^{-1}$ was added (see Sect.~\ref{sec:vlsrvelocities_issue}). The angular separation of the different positions with respect to the central spectrum is shown on top of the respective spectra. The model assumes spherical symmetry and therefore the model spectra at symmetric positions are identical. The telescope used for conducting the obervations is shown at the leftmost box of each row. The spectra from left to right correspond to the south-east to north-west direction perpendicular to the filament. }
\label{fig:CSmodel}
\end{figure*}

\begin{figure*}
\centerline{\resizebox{1.00\hsize}{!}{\includegraphics[angle=270]{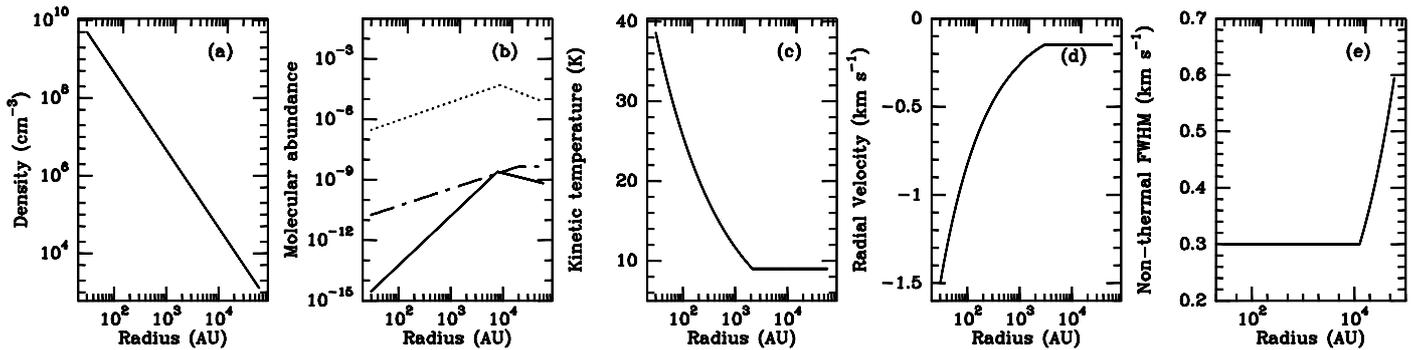}}} 
\caption{Input profiles for the best-fit models 'MCS', 'MHCOP', and 
'MCO': (a) density, (b) CS (solid), HCO$^+$ (dash-dot), and CO (dot) 
abundance, (c) kinetic temperature, (d) radial velocity, and (e) turbulent 
linewidth. }
\label{fig:ALLprofiles}
\end{figure*}

The following transitions of CS and its isotopologues were modeled: CS~2--1 (Mopra), C$^{34}$S~2--1 (Mopra), and CS~5--4 (APEX). $^{13}$CS~2--1 and C$^{33}$S~2--1 were only used as upper limits due to their non-detections and are not presented here. 
We perform the modeling of the spectra for the direction perpendicular to the filament, along which five positions were observed for CS~5--4 and seven for the other transitions. Figure~\ref{fig:CSmodel} shows one of the ``best fit'' models (hereafter 'MCS' model) for the CS molecular transitions.  
The distributions of density, abundance, kinetic temperature, radial 
velocity, and turbulent broadening characterising 'MCS' can be seen in 
Fig.~\ref{fig:ALLprofiles}.

%%%%%%%%%%%%%%% M669 Model %%%%%%%%%%%%%%

% \vspace*{-25ex}

\onlfig{
\begin{figure}
\centerline{\resizebox{1.0\hsize}{!}{\includegraphics[angle=0]{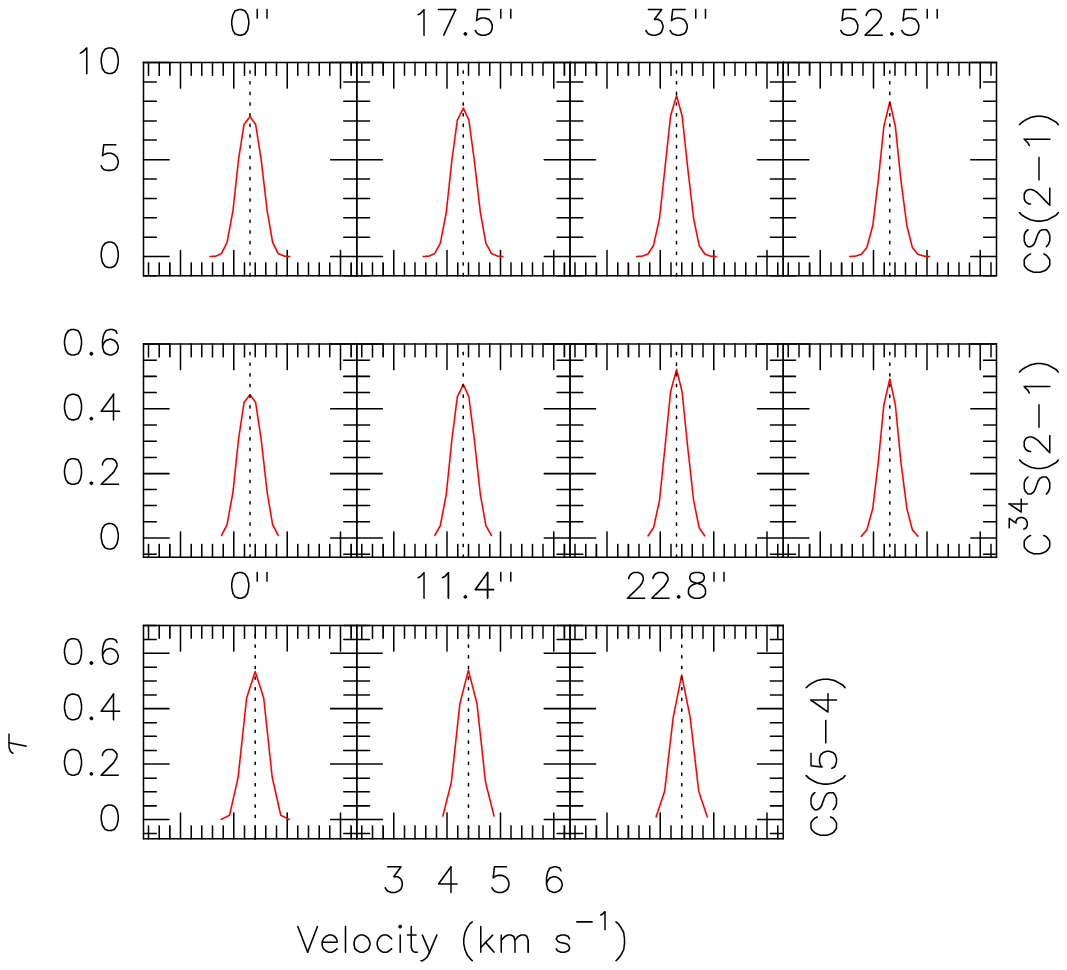}}} 
\caption[]{Transition opacities for the best fit model 'MCS'. The dotted line shows the systemic velocity of Cha-MMS1. For the APEX spectra, a correction of 0.1~km~s$^{-1}$ was added (see Sect.~\ref{sec:vlsrvelocities_issue}). 
}
\label{fig:CSopac}
\end{figure}
}

\onlfig{
\begin{figure}
\centerline{\resizebox{1.0\hsize}{!}{\includegraphics[angle=270]{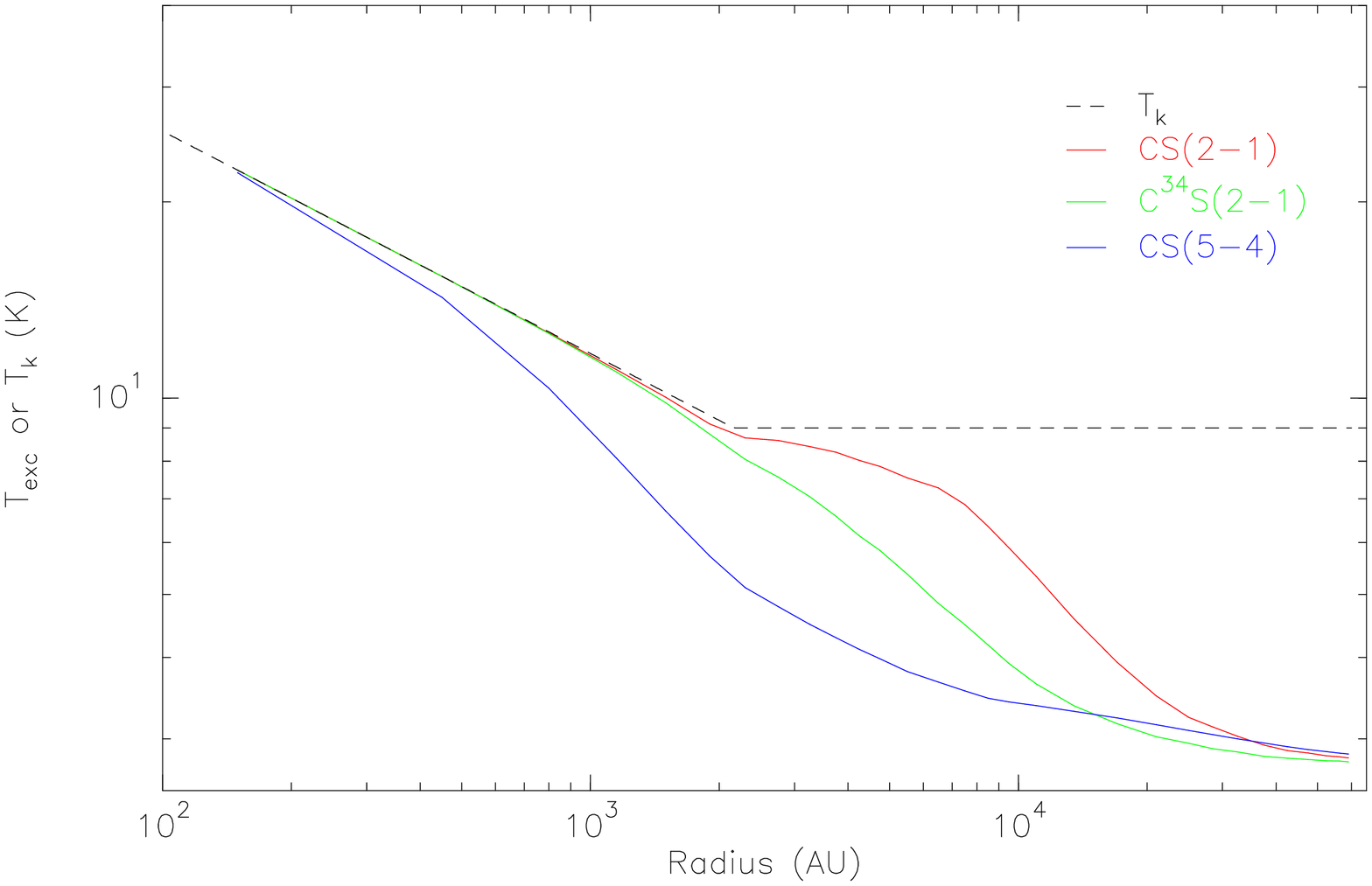}}}
\caption[]{Kinetic (black) and excitation temperature 
(color) for all the transitions of the best fit model 'MCS'.}
\label{fig:CStexc}
\end{figure}
}
%%%%%%%%%%%%%%%%%%%%%%%%%%%%%%%%%%%%%%%%  

\subsubsection{Abundance}

The outer and inner abundances are constrained by the CS~2--1 
absorption dip and the CS~5--4 intensity, respectively. In between, the 
abundance distribution is constrained by the C$^{34}$S 2--1 spectra
which require a CS abundance of 2.5$\times$10$^{-9}$ at a radius of
8000~AU. The depth of the CS 2--1 self-absorption is mostly 
determined by the outer abundance distribution of the 
low-excitation material at radii $\sim 10000$ AU $<r<30000$~AU. We do not constrain the CS abundance at radii larger than 30000 AU. CS 5--4 
probes the innermost parts of the core and sets constraints on the 
abundance at radii $r < 2000$~AU. 
As a result, the CS abundance increases up to 8000~AU and drops by a 
factor of 4 in the outer parts.
As long as the temperature is low enough, CS and other sulphur bearing molecules are expected to be strongly depleted 
towards the centre of dense cores where the density is high \citep[e.g.,][]{tafalla04, bergin01}. 
Depletion occurs due to the freezing-out process onto the dust grains 
\citep[e.g.,][]{tafalla02,stahler10} and observations of starless cores, such 
as L1544, or Class 0 protostars, such as IRAM 04191, also showed CS depletion towards the centre by a factor of $\sim 20$ 
\citep{tafalla02, belloche02}. 

\subsubsection{Turbulence}

We keep the turbulent broadening uniform at 0.3~km~s$^{-1}$ ($FWHM$) up to $\sim12600$ AU as we see no clear variations in our diagrams of turbulent linewidth versus position up to this radius (see Fig.~\ref{fig:posfwhm}). We then let the turbulent linewidth double its value from 12600 AU to the outermost envelope radius at 60000 AU. This is a constraint we derive from both the width of the CS 2--1 absorption dip, and the CO 3--2 modeling (see Sect.~\ref{sec:COmodeling}). 

\subsubsection{Opacities and excitation temperatures}

All transitions of CS and its isotopologue are optically thin on all positions apart from 
CS 2--1 (see Fig.~\ref{fig:CSopac}). The model excitation 
temperatures are shown in Fig.~\ref{fig:CStexc}. CS~2--1, 
C$^{34}$S~2--1, and CS~5--4 are thermalised up to $\sim 2500$ AU, 
$\sim 1500$ AU, and $\sim400$~AU, respectively.

\subsubsection{Discrepancies between model and observations}
\label{sss:CSdiscrepancies}

The model reproduces the absorption dip of CS 2--1 well for all positions. 
However the intensity of the blue peak at the central position and at 
$-17.5^{\prime\prime}$ is higher compared to the observations. The discrepancy 
is not that strong at $+17.5^{\prime\prime}$ and the model 
fits the peak temperature well at the outer positions, i.e., 
$\pm 35^{\prime\prime}$ and $\pm 52.5^{\prime\prime}$. A peculiarity of the 
observed CS 2--1 spectra is the existence of velocity wings at all positions. 
The velocity wings are not reproduced by the model and may partly 
arise due to extended emission stemming from the outflow of the nearby Class I 
object (see Fig.~\ref{fig:CHAmap}). It is however unclear if this can 
explain \textit{both} the redshifted and blueshifted wing emissions.
The C$^{34}$S 2--1 model is weaker at the outermost $\pm 52.5^{\prime\prime}$ positions while matching the intensity of the spectra at the other positions well. CS 5--4, on the other hand, shows one more peculiar feature: the model does not account for the excess of redshifted emission that is prominent toward the central position. 

\subsubsection{Testing the infall velocity field}

 Despite the discrepancies, the 'MCS' model provides a consistent overall 
fit to the data. As we are interested in the infall velocity structure of the core we input different infall velocity profiles to 'MCS' in order to test and constrain its value. The linewidths of the low optical depth lines and the position of the absorption dip of the optically thick lines can be used to place relatively tight constraints on the velocity structure of the source. 
We first test different \emph{uniform} infall velocity profiles and then velocity profiles with a free-fall, power-law ($v\propto r^{-0.5}$) dependence at the inner parts. 

In the case of uniform velocities, the best agreement is found for velocities in the range 0.1~km s$^{-1}$ --~0.2~km~s$^{-1}$. For velocities less than 0.1~km~s$^{-1}$ the peak asymmetry of the CS~2--1 spectra reduces considerably with the red peak becoming too strong in comparison to the observations. In addition, the linewidths of both the CS~2--1 and C$^{34}$S~2--1 lines become narrower than the observed linewidths. On the contrary, when the infall velocities exceed 0.2~km s$^{-1}$ the blue peak of CS~2--1 becomes much broader and the central C$^{34}$S~2--1 spectrum starts showing a double-peaked structure (also seen in the opacity profile), which is inconsistent with the observations. 

We then apply a power-law with an exponent of -0.5 for the inner parts at gradually increasing radii while keeping the velocity constant at the outer parts in order to test whether higher inner velocities are consistent with the observations. From now on, we will refer to the radius at which the velocity profile changes from a power-law to a uniform $r$ dependence as the 'breakpoint'. Our tests suggest that such a power-law velocity distribution matches the observations for radii up to 9000~AU for velocities in the range of $v_{\rm break} = 0.1$~km s$^{-1}$ --~0.2~km~s$^{-1}$, after which we let the velocity remain uniform. For breakpoint radii exceeding 9000~AU, the wide spatial range of high infall velocities produces broader CS~2--1 linewidths and stronger red peak intensities than the observed spectra. The CS~5--4 and C$^{34}$S~2--1 models also start showing a double-peaked structure, which disagrees with the observations. 

Therefore, the CS modeling suggests that the core's envelope is infalling inwards with subsonic to transonic \emph{outer} velocities of 0.1~km s$^{-1}$ --~0.2~km~s$^{-1}$. Inner free-fall power-law velocity distributions are possible with breakpoints at $r \leq 9000$~AU, with infall velocities reaching supersonic values at $r\leq 3500$~AU in this case. We do not constrain the infall velocity structure of the envelope for radii greater than 30000~AU. The infalling motions at the outer parts of the envelope ($< 30000$~AU) contradict the \citet{shu77} assumption of a static envelope in the inside-out collapse model. 

The radial velocity structure of the 'MCS' model shown in Fig.~\ref{fig:ALLprofiles} corresponds to an $r^{-0.5}$ dependence for radii $< 3000$~AU, and a uniform infall velocity of 0.15~km~s$^{-1}$ for $3000$~AU~$\le r \le 60000$~AU (see Fig.~\ref{fig:ALLprofiles}d).

\subsection{HCO$^+$ Modeling}
\label{sec:HCOPmodeling}

\begin{figure*}%[h]
\centerline{\resizebox{1.0\hsize}{!}{\includegraphics[angle=270]{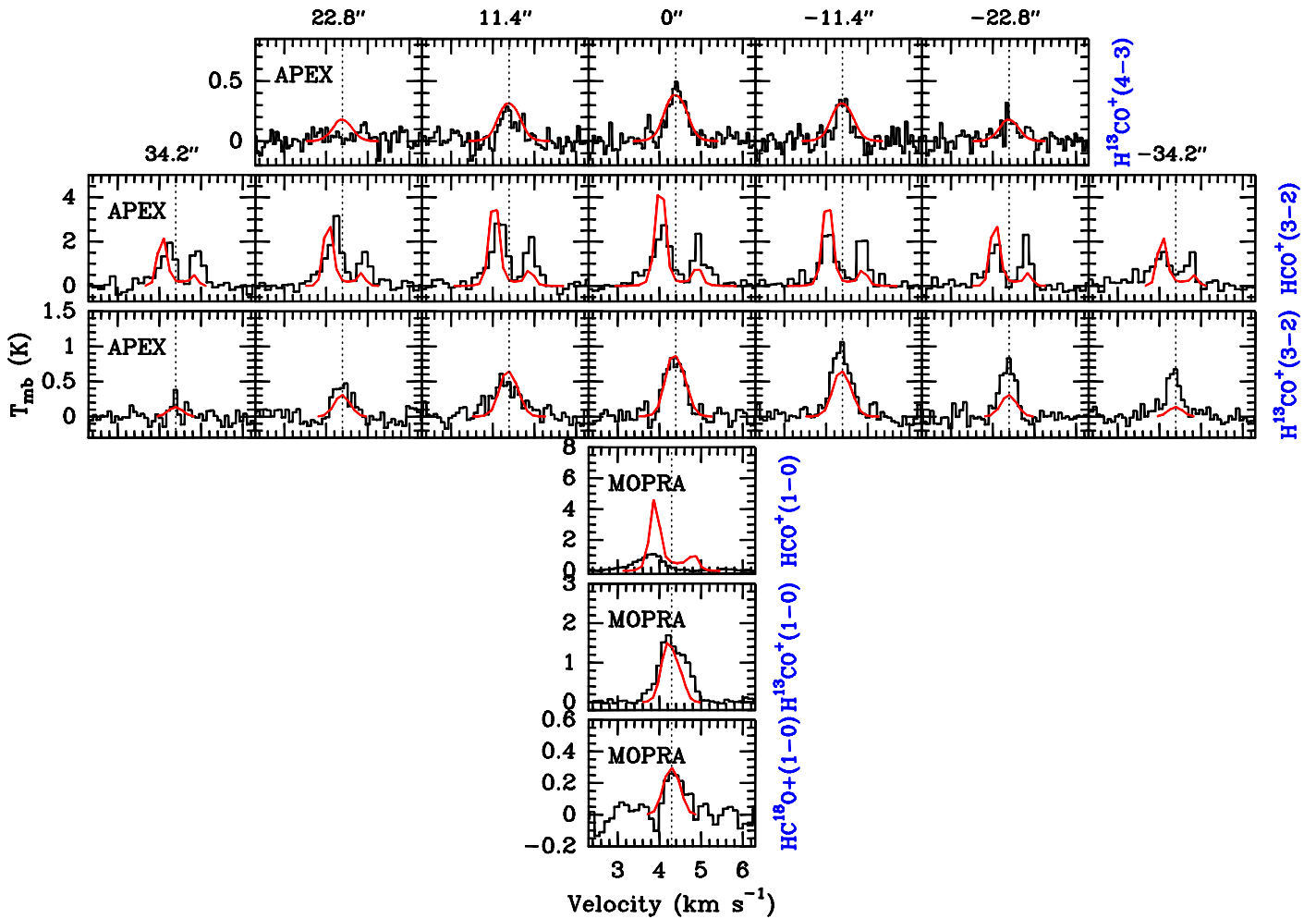}}}
\caption[]{Same as Fig.~\ref{fig:CSmodel} but for the best fit model MHCOP. \label{fig:HCOPmodel}}
\end{figure*}

The following molecular transitions of HCO$^+$ and its isotopologues were modeled: HCO$^+$ 1--0, HCO$^+$ 3--2, H$^{13}$CO$^+$ 1--0, H$^{13}$CO$^+$ 3--2, H$^{13}$CO$^+$ 4--3, and HC$^{18}$O$^+$ 1--0. The 1--0 transitions were observed
with Mopra, the other ones with APEX.
We have central position spectra for HCO$^+$, H$^{13}$CO$^+$, and 
HC$^{18}$O$^+$~1--0, while HCO$^+$ and H$^{13}$CO$^+$~3--2 were also
observed at seven and H$^{13}$CO$^+$ 4--3 at five positions along the direction perpendicular to the filament (see Fig.~\ref{fig:CHAmap}). One of the models that provides a good fit to the data is shown in Fig.~\ref{fig:HCOPmodel} and we will hereafter refer to it as 'MHCOP'. The distributions of density, abundance, kinetic temperature, radial 
velocity, and turbulent broadening characterising 'MHCOP' can be seen in 
Fig.~\ref{fig:ALLprofiles}.

%%%%%%%%%%%%%%% M688 Model %%%%%%%%%%%%%%

\onlfig{
\begin{figure}%[h]
\centerline{\resizebox{1.0\hsize}{!}{\includegraphics[angle=0]{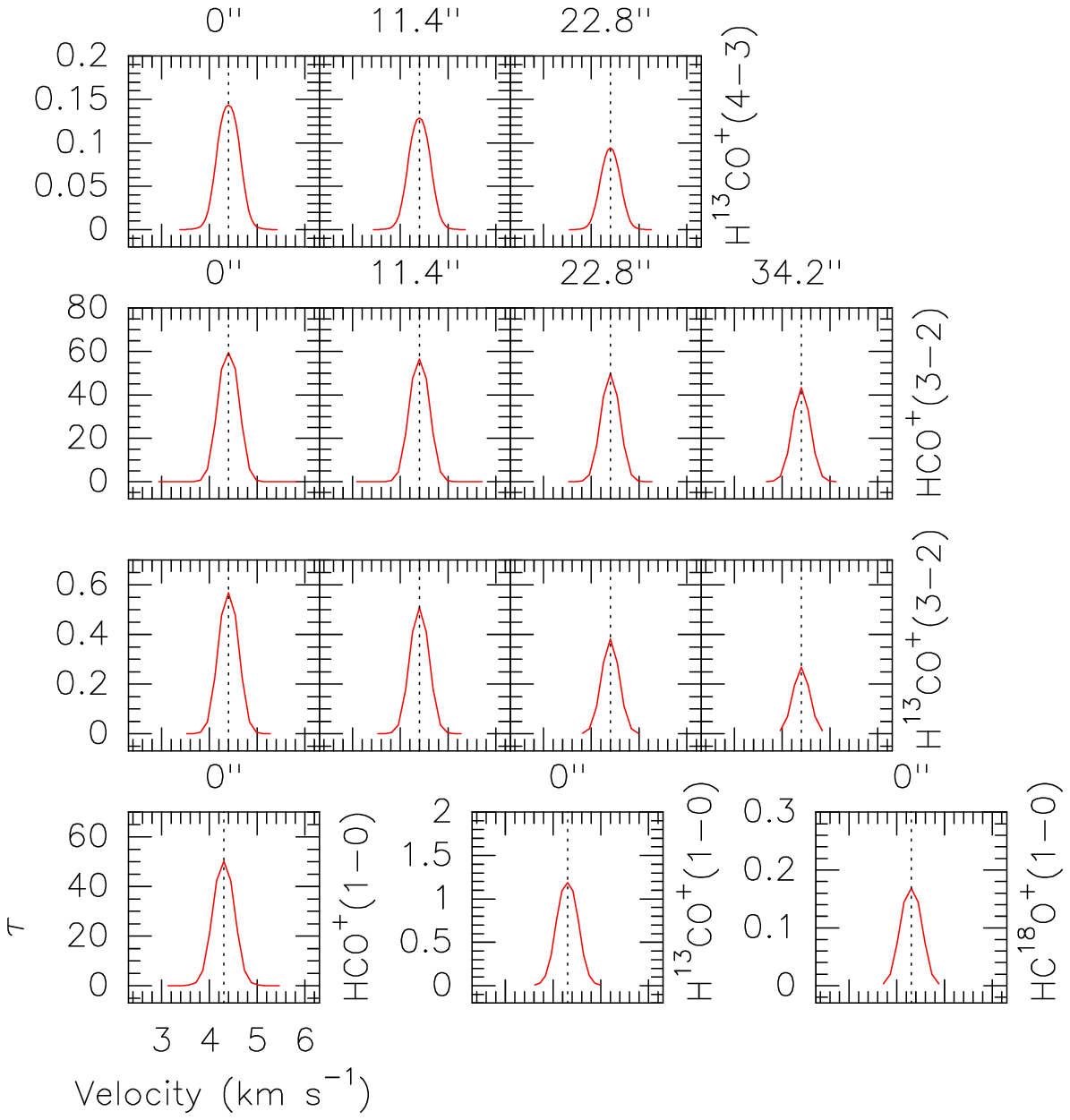}}} 
\caption[]{Transition opacities for the best fit model 'MHCOP'. The dotted line shows the systemic velocity of Cha-MMS1. For the APEX spectra, a correction of 0.1~km~s$^{-1}$ was added (see Sect.~\ref{sec:vlsrvelocities_issue}).
}
\label{fig:HCOPopac}
\end{figure}
}

\onlfig{
\begin{figure}%[!h]
\centerline{\resizebox{1.0\hsize}{!}{\includegraphics[angle=270]{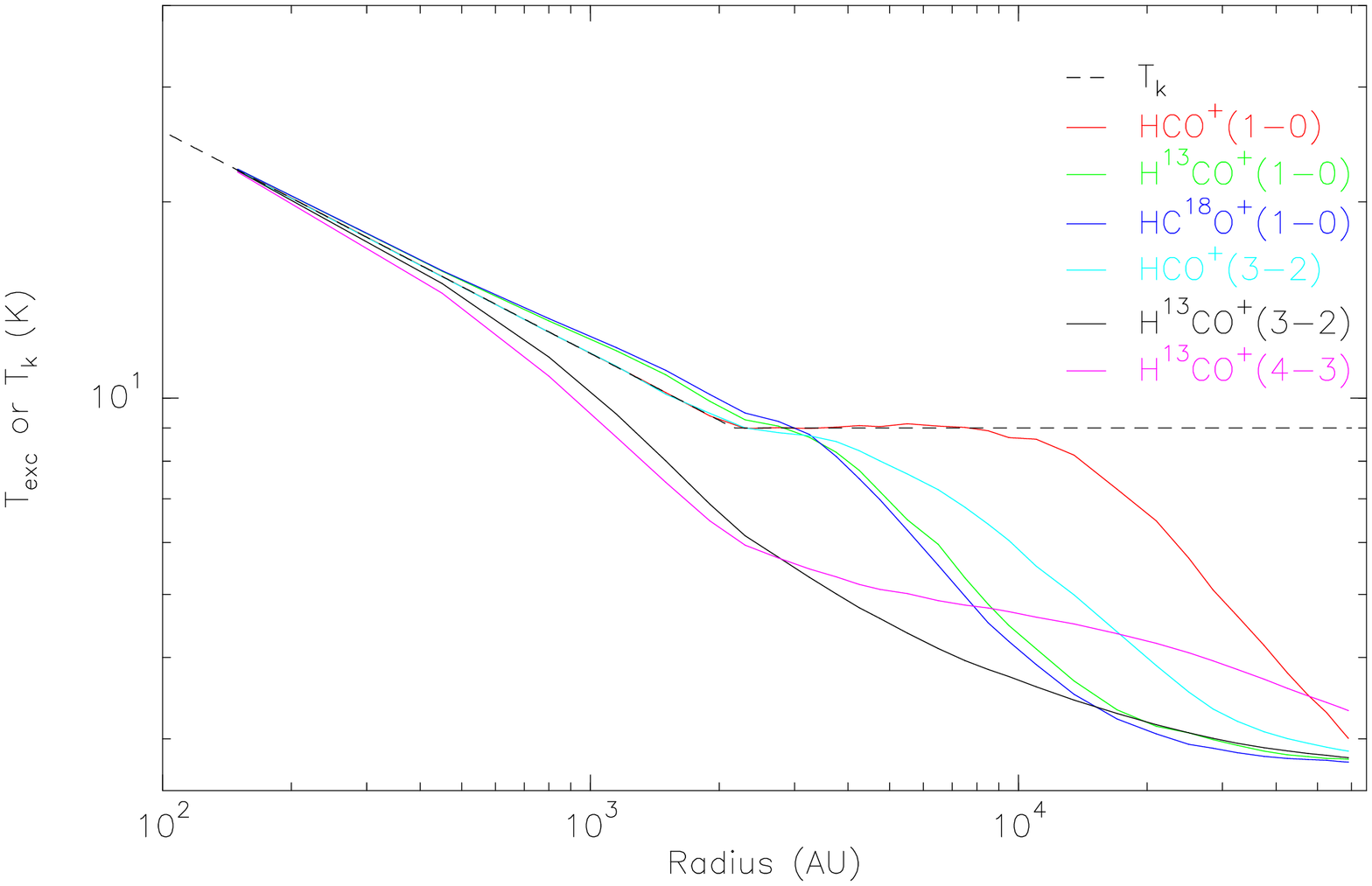}}}
\caption[]{Same as Fig.~\ref{fig:CStexc} for the best fit model 'MHCOP'. \label{fig:HCOPtexc}}
\end{figure}
}

%%%%%%%%%%%%%%%%%%%%%%%%%%%%%%%%%%%%%%%%%%%%%%%%%%%%%%%%%%%%%%%%%%%%%555

\subsubsection{Abundance}

We take into account the following observational constraints to define the input abundance profile (Fig.~\ref{fig:ALLprofiles}b). The inner profile is constrained by the intensities of the optically thin H$^{13}$CO$^+$ 3--2, H$^{13}$CO$^+$ 4--3, and HC$^{18}$O$^+$ 1--0 spectra. 

A decrease of the HCO$^+$ abundance by a factor of $\sim 20$ from a radius of 20000 AU down to the centre fits the intensities of the optically thin lines well. As a test, we extended the plateau of uniform outer abundance towards the centre to radii $r < 20000$ AU, and found that this produced model intensities that were too high for the optically thin spectra. By varying the abundance profile between 20000 AU and 60000~AU we conclude that we do not constrain the outer, $r \ge 20000$~AU, abundance profile with our current HCO$^+$ transitions. We therefore use a uniform outer abundance for $r \ge 20000$~AU as an approximation.

\subsubsection{Opacities and excitation temperatures}

Figures ~\ref{fig:HCOPopac} and ~\ref{fig:HCOPtexc} show the resultant opacity and excitation temperature profiles of the model for each transition. HCO$^+$ 3--2, HCO$^+$ 1--0, and H$^{13}$CO$^+$ 1--0 are optically thick while H$^{13}$CO$^+$ 4--3, H$^{13}$CO$^+$ 3--2, and HC$^{18}$O$^+$ 1--0 are in the optically thin range. HCO$^+$ 1--0 is thermalised up to $\sim 10000$~AU, while HCO$^+$ 3--2, H$^{13}$CO$^+$ 1--0, and HC$^{18}$O$^+$ 1--0  are thermalised up to $\sim 3000$ AU. H$^{13}$CO$^+$ 4--3 and H$^{13}$CO$^+$ 3--2 cease to be thermalised for radii larger than $\sim$~500~AU. HCO$^+$~1--0 is thermalised up to a much larger radius than H$^{13}$CO$^+$ and HC$^{18}$O$^+$ 1--0 because of its much higher optical depth.

\subsubsection{Discrepancies between model and observations}

One of the major discrepancies between model and observations is the failure of the model to reproduce the emission of the red peak of the HCO$^+$~3--2 transition. Raising the outer temperature of the core does not have a significant effect on the red peak of the model spectrum as this transition is not thermally excited at radii larger than 3000~AU. This disagreement might then be a density effect instead, if the true density profile deviates from the spherically symmetric, $r^{-2}$ dependence. The observed H$^{13}$CO$^+$~3--2 spectra show higher intensities in the north-west ($-11.4^{\prime\prime}$, $-22.8^{\prime\prime}$, $-34.2^{\prime\prime}$) direction than the south-east direction. H$^{13}$CO$^+$~3--2 is not the only transition not peaking at the central position.  Higher intensities toward the north-west are also seen in our N$_2$H$^+$ 1--0 spectra. The N$_2$H$^+$ 1--0 emission in the maps presented by \citet{ladd11} is similarly not peaked at Cha-MMS1. 'MHCOP' fits the spectra at the NW side well but it does not fit the excess emission at the SE side. We cannot yet pinpoint the reason for this intensity difference on each side of the central spectrum. H$^{13}$CO$^+$ 4--3 and the central spectrum of HC$^{18}$O$^+$ 1--0 are reproduced rather well, as is H$^{13}$CO$^+$ 1--0 even though the exact lineshape of the observed spectrum is not reproduced. The observed H$^{13}$CO$^+$ 1--0 is broader than the model at the redshifted part of the spectrum, like CS 5--4 (see Sect.~\ref{sec:CSmodeling}). 

The model strongly overestimates the peak temperature of the observed HCO$^+$ 1--0 transition which is even weaker than H$^{13}$CO$^+$ 1--0. HCO$^+$ 1--0 is very optically thick and thermalised up to large radii. It is certainly affected by significant absorption from the low-density material in the ambient cloud at $r \ge 60000$~AU, which is not included in the modeling.

\subsubsection{Testing the infall velocity field}

There is a range of infall velocity profiles that give consistent fits to the data. In the case of uniform, constant velocities up to 60000~AU the range 0.1~km~s$^{-1}$~--~0.2~km~s$^{-1}$ agrees well with the observations. For values higher than 0.2~km~s$^{-1}$ the HCO$^+$~3--2, H$^{13}$CO$^+$~3--2, H$^{13}$CO$^+$~4--3, and HC$^{18}$O$^+$~1--0 spectra become too broad while for velocities lower than 0.1~km~s$^{-1}$ 
the absorption dip is not redshifted enough.
We then tested a velocity profile described by a power-law following an $r^{-0.5}$ dependence at the inner part of the core up to a certain radius after which the infall velocity remains uniform, as we did for CS. For a breakpoint at 3000 AU infall velocities of $v_{\rm break} = 0.1$~km s$^{-1}$ --~0.2~km~s$^{-1}$ give consistent fits. When we increase the breakpoint radius, the range of consistent infall velocities decreases. At 6000 AU and 9000~AU, the breakpoint velocities still consistent with the data are 0.1~km s$^{-1}$ --~0.15~km~s$^{-1}$ and 0.1 km s$^{-1}$ respectively. Overall, inner power-law profiles at radii larger than 9000 AU produce spectra with too broad linewidths for all transitions, apart from H$^{13}$CO$^+$~1--0, and inconsistent double-peaked spectra for H$^{13}$CO$^+$~3--2, H$^{13}$CO$^+$~4--3, and HC$^{18}$O$^+$ 1--0. We also varied the outer infall velocity profile to check whether it is constrained by our data. To accomplish this we let the infall velocity sharply drop to zero at gradually increasing radii and found that the model does not constrain the velocities for radii greater than 35000 AU. Even though the dip of the HCO$^+$3--2 model spectrum in Fig.~\ref{fig:HCOPmodel} is not as redshifted as in the observed spectrum, a velocity drop to zero at radii smaller than $ 35000$~AU makes this discrepancy even stronger by producing model spectra with almost no apparent redshift of the HCO$^+$3--2 absorption dip.
Hence,  we cannot accurately constrain the infall velocity structure of the core for $r \ge 35000$~AU from the HCO$^+$ modeling. 

The velocity distribution corresponding to model 'MHCOP' is the same as the one of 'MCS' (see Fig.~\ref{fig:ALLprofiles}d).

\subsection{CO Modeling}
\label{sec:COmodeling}

\begin{figure*}
\centerline{\resizebox{1.0\hsize}{!}{\includegraphics[angle=270]{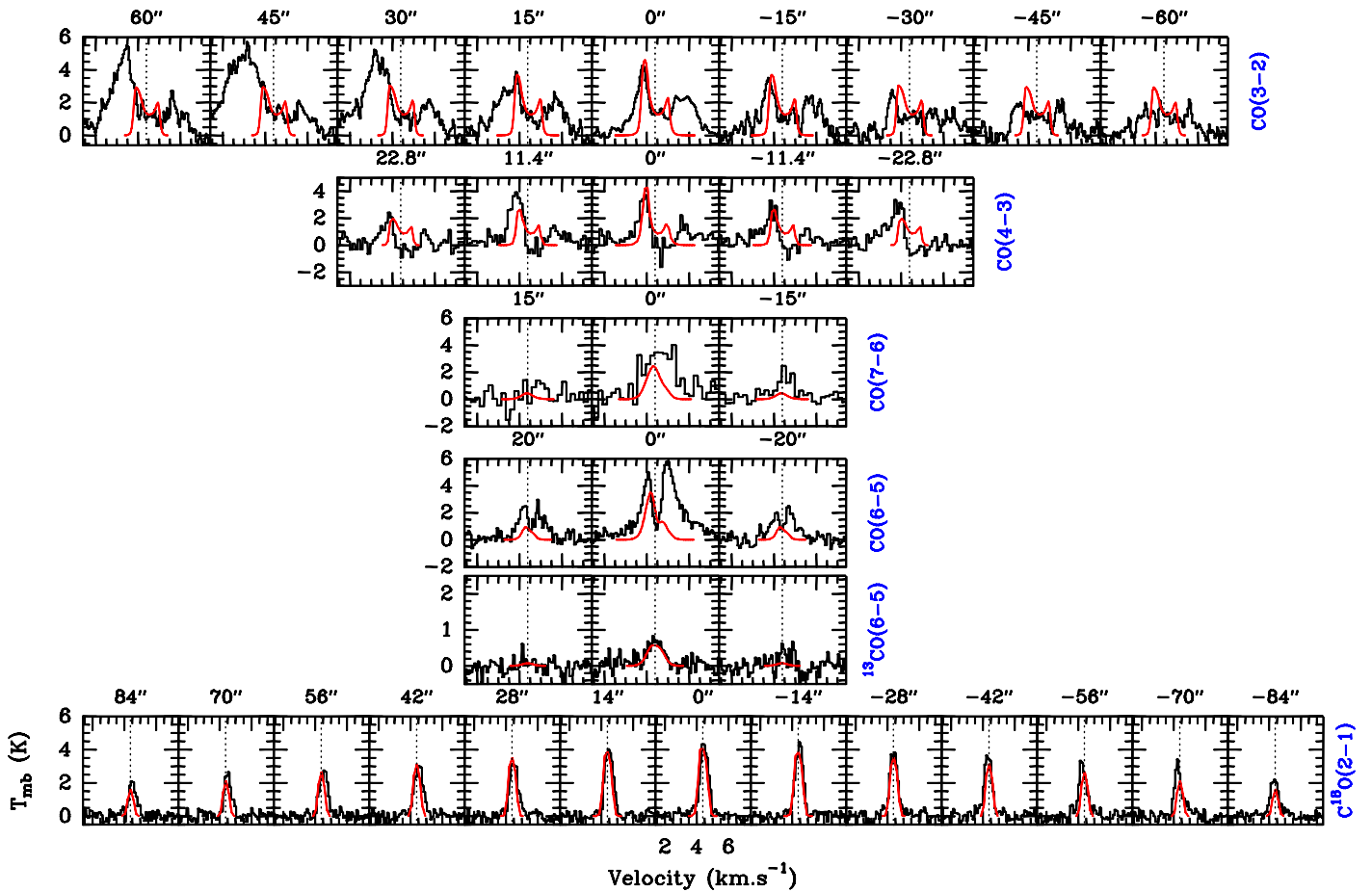}}}
\caption[]{Same as Fig.~\ref{fig:CSmodel} but for the best fit model 'MCO' of the CO transitions.  \label{fig:MCOmodel}}
\end{figure*}

We modeled the following molecular transitions of CO and isotopologues: CO 3--2, CO 4--3, CO 6--5, CO 7--6, $^{13}$CO 6--5, and C$^{18}$O 2--1. 
Figure~\ref{fig:MCOmodel} shows one of the models that fits the data 
relatively well, hereafter 'MCO'. The distributions of density, abundance, kinetic temperature, radial 
velocity, and turbulent broadening characterising 'MCO' can be seen in 
Fig.~\ref{fig:ALLprofiles}.

%%%%%%%%%%%%%%% M760 Model %%%%%%%%%%%%%%

\onlfig{
\begin{figure}
\centerline{\resizebox{1.0\hsize}{!}{\includegraphics[angle=0]{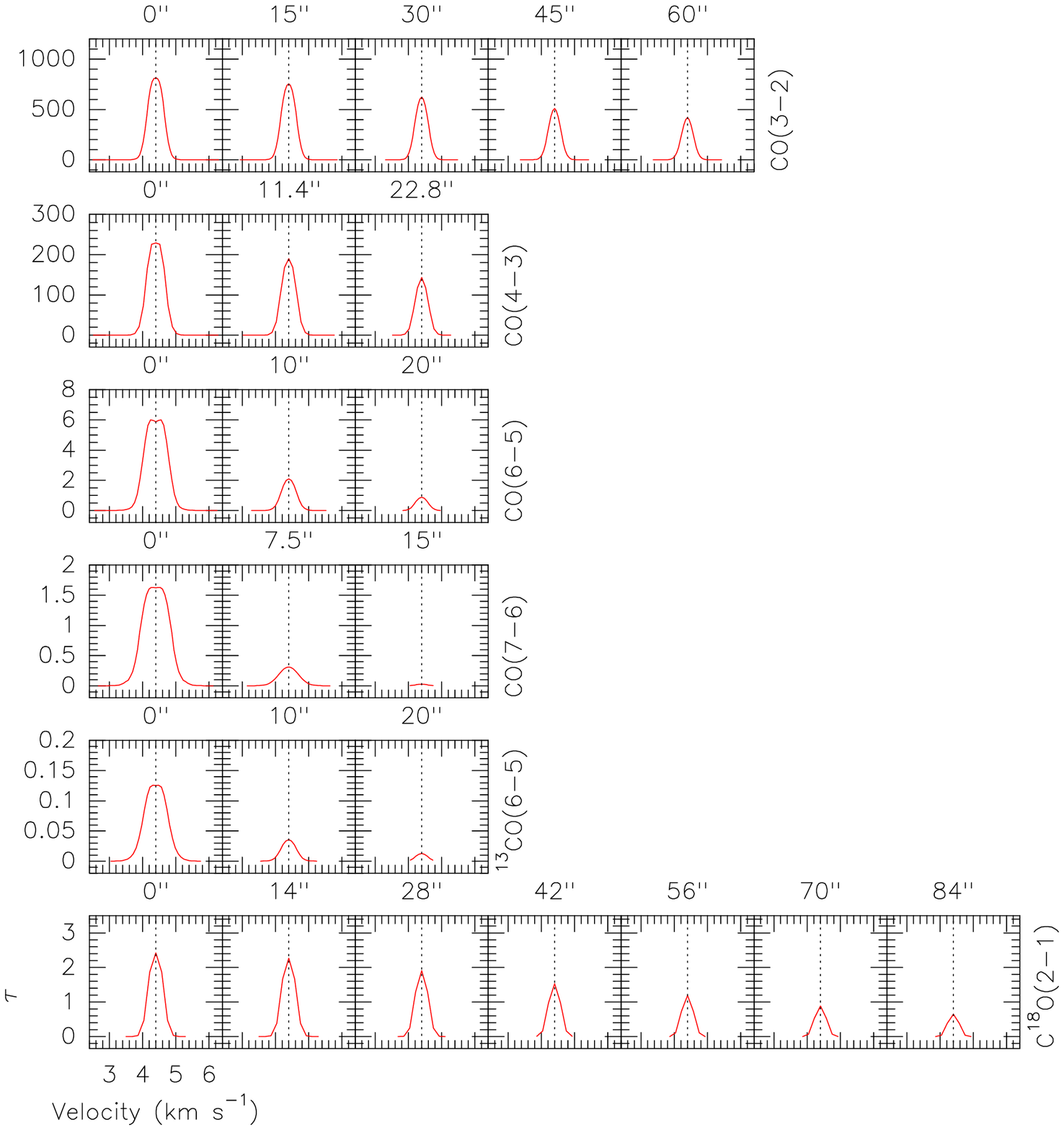}}} 
\caption[]{Transition opacities for the best fit model 'MCO'. The dotted line shows the systemic velocity of Cha-MMS1. For the APEX spectra, a correction of 0.1~km~s$^{-1}$ was added (see Sect.~\ref{sec:vlsrvelocities_issue}).
}
\label{fig:MCOopac}
\end{figure}
}

\onlfig{
\begin{figure}
\centerline{\resizebox{1.0\hsize}{!}{\includegraphics[angle=270]{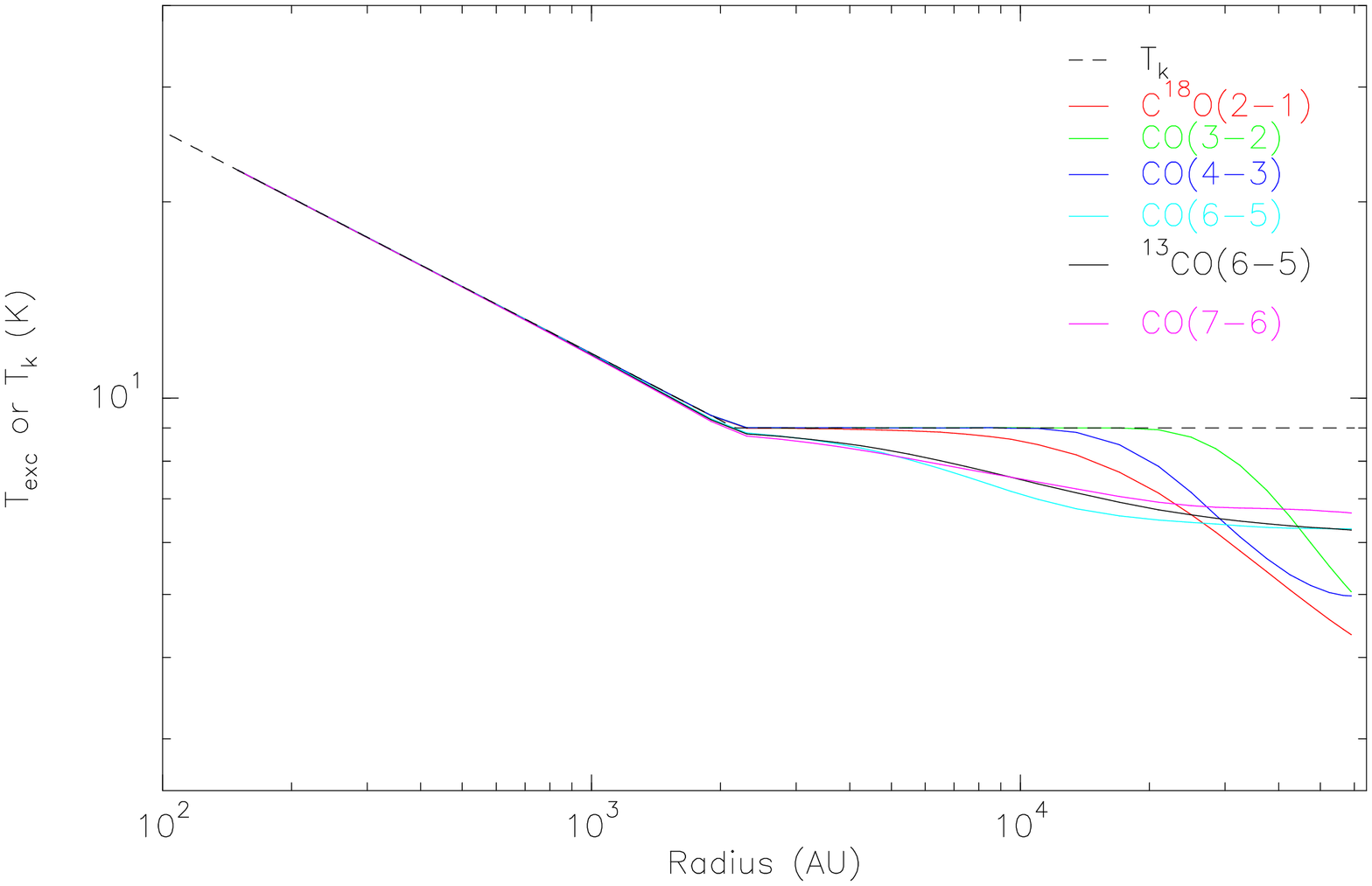}}}
\caption[]{Same as Fig.~\ref{fig:CStexc} for the best-fit model 'MCO'.
}
\label{fig:MCOtexc}
\end{figure}
}

\subsubsection{Abundance}

The abundance profile of CO down to a radius of $\sim 1000$~AU is mainly constrained by the intensity variations of the C$^{18}$O and C$^{17}$O 2--1 transitions.
The ratio of the C$^{17}$O and C$^{18}$O~2--1 integrated intensities yields C$^{18}$O 2--1 opacities ranging from $\sim1.6$ toward the center to less than $\sim0.5$ toward the outer parts (see Appendix~\ref{sec:opacity_c18o}). We take these opacity values into account to further constrain our model. 
Finally, the optically thin $^{13}$CO 6--5 transition, which traces higher densities and is spatially better resolved, sets strong constraints on the abundance in the inner parts of the envelope ($r < 1500$~AU).
Overall, we obtain the abundance profile shown in Fig.~\ref{fig:ALLprofiles}b, with a maximum of $5 \times 10^{-5}$ at 9000~AU, depletion by a factor of $\sim 10$ from $\sim 9000$ AU to $\sim 700$~AU, and also a decrease of the abundance by a factor $\sim7$ from $\sim9000$ AU to $\sim 20000$~AU, in order to match the lower outer C$^{18}$O opacity. We constrain the CO abundance up to 20000~AU.
CO depletion has previously been observed towards the inner parts of various prestellar cores \citep[e.g.,][]{bacmann02, christie12}. 

\citet{furuya12} coupled a gas-grain chemical network to 3D radiation 
hydrodynamic simulations to derive molecular abundances at the first core 
stage. In their model, CO desorbs to the gas phase at a temperature of 25~K, 
somewhat higher than the commonly used 20~K value, due to the 
high density of the first-core envelope. The density at which their
simulation reaches 25~K is $\sim 10^9$~cm$^{-3}$, which is similar in our 
model. We tested the effect of an inner CO desorption region at $r \leq 54$~AU (from Equation~\ref{eq:tdust}), at which radius the temperature reaches 25~K in our model. We used an inner CO abundance of $1 \times 10 ^{-4}$ for the desorption region, which is consistent with the predicted abundance for the first core stage \citep{furuya12}. At radii larger than 54 AU, and therefore at temperatures T $< 25$~K, CO depletion takes place. We found that the model is not sensitive to the abundance within the inner 54 AU and the presence or the absence of a desorption region on such smaller scales has no apparent effect on the spectra. Therefore, we do not account for inner desorption. 

\subsubsection{Turbulence}

The self-absorption of the CO 3--2 spectra is very broad, suggesting a
large turbulent broadening in the outer parts (radii $r \ge 12500$~AU).
We therefore raise it by a factor of $\sim 2$ from 12500 AU to 60000~AU.
In this way, the blueshifted side of the self-absorption is well 
reproduced, but the redshifted side of the model is still too narrow. The 
turbulent broadening is kept uniform for radii $r \le 12500$~AU 
(see Fig.~\ref{fig:ALLprofiles}), as derived in 
Sect.~\ref{sec:turbulence}. 

\subsubsection{Model opacities and excitation temperatures}

Figures~\ref{fig:MCOopac} and ~\ref{fig:MCOtexc} show the model opacities and excitation temperatures for all the transitions. CO 3--2, CO 4--3, and C$^{18}$O 2--1 remain thermalised out to $\sim$~20000 AU, $\sim13000$ AU, and $\sim9000$ AU respectively, while CO 6--5 and CO 7--6 cease to be thermalised at $\sim$~2000~AU. Moreover, all the transitions are optically thick with the exception of $^{13}$CO 6--5 at all positions, CO 7--6 at offset positions, and C$^{18}$O 2--1 at the outermost position (see Fig.~\ref{fig:MCOopac}).  

\subsubsection{Discrepancies between model and observations}

The spectra of CO 3--2 at 15$^{\prime\prime}$, 30$^{\prime\prime}$, 45$^{\prime\prime}$, and 60$^{\prime\prime}$ show strong emission in the blueshifted part of the spectrum while the spectra at -15$^{\prime\prime}$, -30$^{\prime\prime}$, -45$^{\prime\prime}$, and -60$^{\prime\prime}$ show much weaker blueshifted emission. Excess emission in the redshifted part of the spectra is also seen at all positions. We expect the outflow of the neighbouring Class I protostar (see Fig.~\ref{fig:CHAmap}) to contaminate the low density tracers such as CO 3--2, especially at the offset positions, thereby broadening the observed emission. \citet{belloche06} presented a CO 3--2 intensity map of the region toward Cha-MMS1 (see Fig. 1 of their paper). The blueshifted emission they show reaches our offset positions, and especially affects the spectra towards the south-east of Cha-MMS1. They also discuss the possible presence of two separate outflows in this region. The highly broadened blue and red peaks of CO 3--2 might \emph{partly} result due to these outflows if they have a wide opening angle and lie close to the plane of the sky \citep[see][]{cabrit90}.

%\subsubsection{Implications of discrepancies}

The model reproduces well the C$^{18}$O 2--1 emission and its opacity, as well as the $^{13}$CO 6--5 emission. However, the model fails to reproduce the strong redshifted emission of the CO 6--5 and 7--6 transitions, similar to the CS 5--4 emission. As these transitions probe regions of high densities, this extra emission may be an indication of additional warmer, high-velocity material at the inner core (see Fig.~\ref{fig:CSmodel}). This emission may stem from a very compact outflow. To address this question however, we would need to resolve the very inner part of the core at small scales of a few hundreds AU (see Sect.~\ref{sec:FCtheory}).

\subsubsection{Testing the infall velocity field} 

From the CO modeling we draw the following conclusions on the infall velocity structure across the envelope. We find that uniform infall velocities of up to 0.2~km~s$^{-1}$ give fits relatively consistent with the data. Velocities above 0.2~km~s$^{-1}$ produce C$^{18}$O~2--1 linewidths that are too broad compared to observed values. We test an inner power-law velocity dependence ($v \propto r^{-0.5}$), with uniform outer velocities. We find that the power-law profile is consistent with the observations up to $\sim 6000$~AU if the infall velocity is in the range of $v_{\rm break} = 0.1$~km~s$^{-1}$~--~0.15~km~s$^{-1}$ at this radius, after which it remains constant. A radius of 9000~AU is also consistent when $v_{\rm break} = 0.1$~km~s$^{-1}$ at this breakpoint and onwards. For larger radii the model spectra are much broader than the observed spectra due to the extended spatial range of high inner infall velocities. However, the opacity of the optically thick transition CO 3--2 does not allow us to set constraints on the lower limit of the infall velocity. 

The velocity profile that coresponds to the 'MCO' model is shown in Fig.~\ref{fig:ALLprofiles}, and it is identical to the ones used for 'MHCOP' and 'MCS'.

\subsection{Infall velocity distribution of Cha-MMS1: combining modeling results}

\begin{figure}%[h]
\centerline{\resizebox{1.0\hsize}{!}{\includegraphics[angle=0]{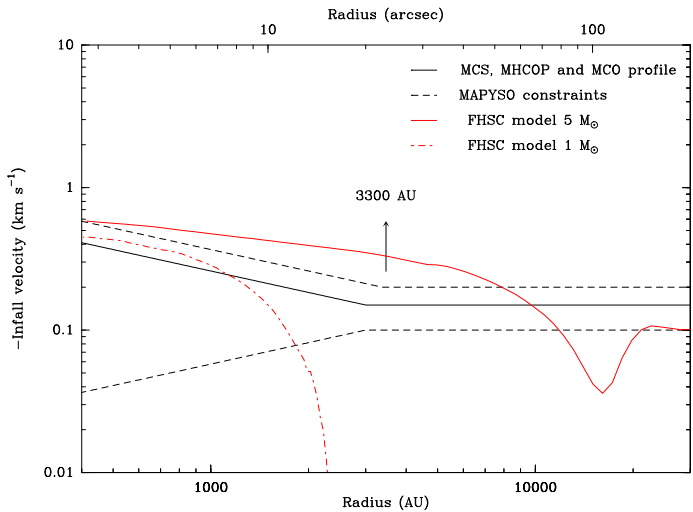}}} 
\caption[]{Infall velocity field of the envelope of Cha-MMS1 based on 
the radiative transfer modeling with MAPYSO. The solid black line 
corresponds to the velocity profile of the 'MCS', 'MHCOP', and 'MCO' models.
The dashed lines show the range of velocity fields that are 
still consistent with the observations. The red solid and dashed curves correspond to the average radial velocity at the equatorial plane of a magnetised FHSC with a core mass of 1 $M_{\odot}$ and 5 $M_{\odot}$, and an age of 850 yr and 2684 yr respectively.  
\citep[][, see details in Sect.~\ref{sec:discussion_models}]{commercon12}.}
\label{fig:vinf}
\end{figure}

We obtain slightly different constraints on the infall velocity structure of Cha-MMS1 from the radiative transfer modeling of the CS, HCO$^+$, and CO molecular transitions. We take the \emph{envelope} of the infall velocity profiles that are consistent with all three datasets as the overall range of possible infall velocities for Cha-MMS1. The consistent velocities are shown as the area enclosed within the dashed lines in Fig.~\ref{fig:vinf}, while the solid line shows the velocity profile of the 'MCS', 'MHCOP', and 'MCO' models.

The infall velocity is relatively well constrained over the range of 
radii 3300 AU to 30000~AU, with subsonic/transonic velocities in the range 0.1~km~s$^{-1}$ to 
0.2~km~s$^{-1}$. Our data do not constrain the velocity field beyond $\sim30000$~AU.
Below 3300~AU, there are more degeneracies and the spectra are consistent with
an increase of the infall velocity as $r^{-0.5}$, but also with a flat velocity
profile, or even a decrease below a radius of $\sim 3000$~AU.

%\clearpage
\section{Discussion}
\label{sec:discussion}

\subsection{Far-infrared emission and luminosity}

Recent radiation-MHD simulations predict that strongly or weakly 
magnetised FHSCs are expected not only to emit at 70~$\mu$m but to 
also produce detectable emission at wavelengths down to 20~$\mu$m, 
especially during the latest phase of their evolution \citep{commercon12}. 
Previous simulations that found no significant first-core emission 
below 30~$\mu$m~--~50~$\mu$m were either spherical 
\citep[e.g.,][]{masunaga98,omukai07} or did not take the magnetic field into
account and used a barotropic equation of state \citep[][]{saigotomisaka11}. 
A first core can therefore also be identified by a Spitzer 24~$\mu$m  (and 
70~$\mu$m) detection at late phases if its inclination is less than 
60$^{\circ}$ and there is no detection at wavelengths smaller than 
20~$\mu$m \citep{commercon12}.
As seen in Sect.~\ref{sec:intlum}, the 24~$\mu$m and 70~$\mu$m 
\textit{Spitzer} fluxes of Cha-MMS1 are consistent with the predictions of the 
RMHD simulation of \citet{commercon12} for a magnetised FHSC with a 
normalised mass-to-magnetic-flux ratio of 2 seen at an inclination lower than 
60$^\circ$. 
The SED of Cha-MMS1 is therefore consistent with Cha-MMS1 
being at the FHSC stage. However, if its actual inclination is higher than 
60$^\circ$ then Cha-MMS1 would have to be in a more advanced stage (Class 0).

\subsection{Outflows}
\label{sec:discussion_outflow} 

Class 0 protostars usually drive fast, extended, and easily detectable 
outflows whereas FHSCs are predicted to drive very compact, slow outflows (see 
Sect.~\ref{sec:FCtheory}). This is a major observational signature 
that can be used to distinguish between the two and break the 
degeneracies that remain in their SEDs when their inclination is not known.

A search for a fast, large-scale outflow driven by Cha-MMS1 in CO 3--2 with 
APEX was negative \citep{belloche06}. We observed the CO~6--5, 
CO~7--6, and $^{13}$CO~6--5 transitions in order to search for signs of a 
compact, unresolved outflow. The modeling of these transitions 
gave ambiguous, yet maybe promising results (see 
Sect.~\ref{sec:COmodeling}). Our model does not reproduce the 
redshifted part of the CO~6--5 and 7--6 emission while it fits well 
the C$^{18}$O~2--1 transition. This excess emission might point to the 
presence of unresolved, higher-velocity material at the inner core.
In addition to this, the CS 5--4 model in Sect.~\ref{sec:CSmodeling} shows a 
similar discrepancy. Its observed spectrum has an excess of redshifted 
emission that the model fails to reproduce while it fits well the 
lower-$J$ CS and C$^{34}$S transitions.
CS~5--4 probes material at higher densities and hence, its broad 
spectrum indicates higher-velocity, dense material close to the 
centre of the core which cannot be seen with the other, lower-density 
transitions. 

\begin{figure}[h]
\begin{tabular}{c}
\centerline{\resizebox{1.0\hsize}{!}{\includegraphics[angle=270]{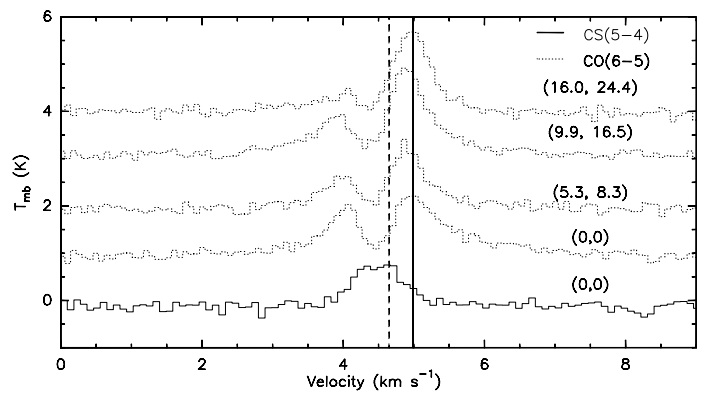}}} 
\end{tabular}
\caption[]{Comparison of the CS 5--4 spectrum (solid) toward the 
central position of Cha-MMS1 and CO 6--5 spectra (dotted) taken at the central 
position and along the north-east direction toward the nearby Class~I 
protostar. The offset position of each spectrum is given in 
arcseconds relative to the central position. The CO~6--5 spectra are 
shifted vertically by a step of 1~K for clarity. The solid vertical line roughly 
corresponds to the velocity of the CO 6--5 red peak, while the dashed 
line marks the velocity of the peak of the CS~5--4 excess emission.
}
\label{fig:excessemission}
\end{figure}

Despite the hints for the presence of dense, higher-velocity 
material confined to the centre of the core, there is also an 
alternative explanation. Figure~\ref{fig:excessemission} compares the 
CO~6--5 spectra at the centre and toward the north-east direction of the 
filament (hence toward the nearby Class I outflow) to the central spectrum of 
CS~5--4. The CO~6--5 red peak emission becomes stronger as we move up along
the filament. It peaks at a velocity (solid line) where CS 5--4 has 
some wing emission. This suggests that at least part of the excess emission 
in the CS~5--4 spectrum is not confined to the inner parts of Cha-MMS1 but
extends toward the nearby Class I outflow lobe. The bulk of the CS~5--4 excess 
emission, however, peaks at a lower velocity (dashed line) and it is unclear
if it represents a similar extended component. 

Higher-angular resolution observations are certainly needed to make any reliable conclusions about the presence of a compact, slow outflow driven by Cha-MMS1.

As mentioned in Sect.~\ref{sss:CSdiscrepancies}, all CS 2--1 spectra 
in Fig.~\ref{fig:CSmodel} show blueshifted and redshifted wing emissions that 
are not accounted for by 
our model. Such wing emissions are even more pronounced in CO~3--2 
(Fig.~\ref{fig:MCOmodel}), the blueshifted one being clearly associated with
emission from the outflow driven by the nearby Class~I protostar 
\citep[][]{belloche06}. \citet{hiramatsu07} and \citet{ladd11} proposed
that the curved shape of the blue lobe of this outflow is due to its partial
interaction with the Cha-MMS1 dense core. This would explain the presence of 
the extended blueshifted CS 2--1 emission around Cha-MMS1. As recalled by 
\citet{ladd11}, the HH objects
49 and 50 which are thought to be associated with this outflow are 
\textit{redshifted}. This suggests that the outflow axis lies close to the 
plane of the sky and, provided its opening angle is large enough, it could 
also explain the presence of extended, redshifted, wing emission in CS~2--1 in
the vicinity of Cha-MMS1. This emission is however difficult to disentangle 
from the ambient cloud emission in CO 3--2 and it was excluded from the 
range of velocities used by \citet{belloche06} to produce their map of the 
redshifted outflow lobe.

\subsection{Interpretation of the P-V diagrams}
\label{sec:discussion_rotation}

Determining the nature of the velocity gradients in position-velocity diagrams 
is not straightforward because many processes such as rotation, 
infall, outflow, and turbulence can influence their shape
\citep[e.g.,][]{tobin12, burkertbodenheimer00}. 
\citet{tobin12} suggested that projected infall in filamentary 
protostellar envelopes can dominate the velocity gradients for scales 
larger than 1000 AU. \citet{burkertbodenheimer00} showed that 
turbulence can produce velocity gradients that can be mistaken for 
rotation. In Sect.~\ref{sec:rotation} we found no significant velocity
gradient along the direction parallel to the filament in which Cha-MMS1 is 
embedded, but a significant one in the perpendicular direction. If we 
can approximate the filamentary geometry of the continuum emission as 
axisymmetric, and if infall plays a significant role in producing these 
velocity shifts, we would expect to see its contribution in the P-V diagrams 
along the filament direction. That this is not the case enhances the probability that 
the velocity gradients we observe perpendicular to the filament are due to 
rotation. 

The P-V diagrams of C$^{18}$O 2--1 and C$^{17}$O 2--1 cover almost the same extent as the width of the filament ($\sim 0.1$ pc). As the outermost positions of C$^{18}$O 2--1 and C$^{17}$O 2--1 reach the edges of the filament (see Fig.~\ref{fig:CHAmap}d), we cannot exclude the possibility that turbulent motions influence the shape of the velocity gradients at these scales. However, the ``S'' shape of the C$^{18}$O 2--1 and C$^{17}$O 2--1 P-V curve at the outermost positions on either side of the central position is relatively symmetric (see Fig.~\ref{fig:pvapex}c), and therefore probably indicative of rotation rather than random turbulent motions. 
If the interpretation in terms of rotation is valid, then the ``S'' 
shape of the P-V curve indicates that the filament is rotating in a 
differential manner, the outer parts ($r > 8000$~AU) rotating more slowly than 
the inner parts. The turn-over of the C$^{18}$O and C$^{17}$O P-V curves occurs 
very close to the edge of the filament as traced with LABOCA. We speculate that
this behaviour is related to the formation process of the filament, but the
physics of this process would have to be investigated.

We obtained a velocity gradient of $3.1 \pm 0.1$~km~s$^{-1}$~pc$^{-1}$ over 
$r < 8000$~AU for Cha-MMS1. This is similar to the velocity 
gradients often found in dense cores and attributed to rotation, with magnitude 
typically ranging from $\sim 0.3$~km~s$^{-1}$~pc$^{-1}$ to $\sim 6$~km~s$^{-1}$~pc$^{-1}$ on typical 
scales of $\sim 0.1$~pc 
\citep[e.g.,][]{goodman93,caselli02,belloche02, tafalla04}.
The ``S'' shape of the C$^{17}$O 2--1 and C$^{18}$O 2--1 P-V diagrams 
of Cha-MMS1 is very reminiscent of the P-V diagrams of the young 
Class~0 protostar IRAM~04191 located in Taurus. \citet{belloche02} 
derived a rotational angular velocity of $9 \pm 3$~km~s$^{-1}$~pc$^{-1}$ 
at a 
radius of 2800~AU and $1.9 \pm 0.2$~km~s$^{-1}$~pc$^{-1}$ at 7000 AU. 
They concluded that IRAM 04191 shows clear signs of differential 
rotation in the envelope. At scales of $\sim 1000$~AU, the envelope 
rotates even faster \citep[][]{belloche04}.
As mentioned in the previous paragraph, the P-V diagram of Cha-MMS1 is
also consistent with differential rotation but a major difference compared to
IRAM~04191 is that this concerns the outer parts of the envelope only 
($r > 8000$~AU). At smaller radii down to $\sim 4000$~AU, the velocity 
profile is consistent with solid-body rotation. 
However, a puzzling feature of the P-V diagrams of Cha-MMS1 is the 
even flatter velocity gradient in the inner parts of the envelope below 
4000~AU, with an amplitude lower than 2 km~s$^{-1}$~pc$^{-1}$.

If the velocity gradients really trace rotation, then the envelope of
Cha-MMS1 has a very peculiar rotational structure: the inner parts rotate more 
slowly than the bulk of the envelope, and the outer parts rotate also more
slowly. A collapsing, magnetised core is expected to have an angular velocity 
increasing toward the center \citep[][]{basu95a}, the exact shape of the 
profile depending on the initial angular momentum distribution. Since 
we see evidence for infall motions in
the envelope of Cha-MMS1, its peculiar rotational structure suggests that
an efficient mechanism removing angular momentum during the collapse 
is at work over the range of radii 2000 AU to 8000~AU. It would be interesting
to investigate if magnetic braking is efficient enough to account for this
angular momentum removal in Cha-MMS1. Measuring the magnetic field structures 
and ionisation levels of the Cha-MMS1 and IRAM~04191 envelopes would then be 
necessary to compare the two sources and understand why they behave so 
differently in terms of rotation.

\subsubsection{Centrifugal acceleration and rotational energy}
\label{sec:discussion_cent_accel}

If we interpret the velocity gradients in the P-V diagrams as rotation, then 
we can estimate the dynamical importance of rotation for Cha-MMS1. 
We assume solid-body rotation for the inner envelope at 
$r<8000$~AU and an inclination of the rotation axis in the range 
45$^\circ$~--~60$^\circ$, as derived in Sect.~\ref{sec:intlum}. The angular velocity then
ranges from 4.4~km~s$^{-1}$~pc$^{-1}$ to $3.6 \pm 0.1$~km~s$^{-1}$~pc$^{-1}$.
We compute the centrifugal acceleration and the local gravitational field as 
follows:
\begin{equation}
a_{\rm cen} = \frac{v_{\rm rot}^2}{r} = \Omega^2 r ,
\end{equation} 
\begin{equation}
g = G \times \frac{M_{\rm env}+M_{\rm obj}}{r^2} ,
\end{equation}
where $a_{\rm cen}$, $v_{\rm rot}$, $\Omega$, $g$, $G$, $M_{\rm env}$, 
$M_{\rm obj}$, and $r$ are the centrifugal acceleration, the rotational 
velocity, the angular velocity, the gravitational acceleration, the 
gravitational constant, the envelope mass, the mass of the central object, and 
the radius, respectively.  We assume that the envelope mass is proportional to the radius (density proportional to $r^{-2}$) and that a mass of 1.44~$M_{\odot}$ is enclosed within a radius of 3750~AU, as derived from the LABOCA 870~$\mu$m dust continuum map \citep{belloche11a}. 

\begin{figure}[h]
\begin{tabular}{c}
 \centerline{\resizebox{0.9\hsize}{!}{\includegraphics[angle=0]{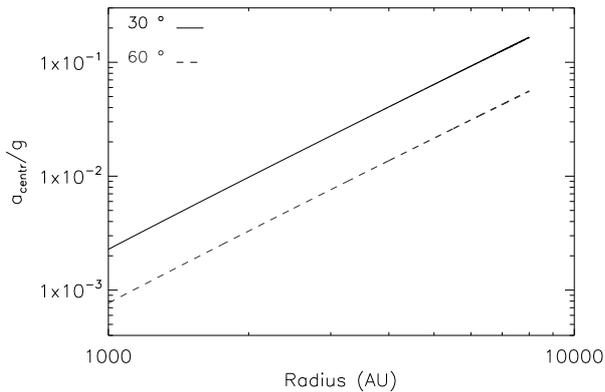}}}
\end{tabular}
\caption[]{\small Ratio of centrifugal to gravitational accelerations
    for the inner part of Cha-MMS1 that may be undergoing solid-body 
rotation. The upper and lower curves are for an inclination of 30$^\circ$ and 
60$^\circ$, respectively. 
}
\label{fig:acen_g}
\end{figure}

Figure~\ref{fig:acen_g} shows the variation of $a_{\rm cen}/g$ as a
function of radius. Within the framework of our assumptions, 
the centrifugal acceleration represents at most $20\%$ of the 
gravitational acceleration. Thus rotation does not provide significant support 
to the envelope on scales of a few thousand AU. A similar conclusion was drawn 
by, e.g., \citet{caselli02} for their sample of dense cores, while 
\citet{belloche02} found that the centrifugal acceleration was a sizeable 
fraction of the gravitational acceleration on such scales in IRAM~04191
(up to $\sim 40\%$).

The ratio of rotational kinetic energy to the core's gravitational energy for a centrally peaked $r^{-2}$ density profile, $\beta_{\rm rot}$, is given by \citep{goodman93}:
\begin{equation}
\beta_{\rm rot} = \frac{\Omega^2 R^3}{9 GM} .
\end{equation}
We obtain values of $\beta_{\rm rot}$ $\sim$~0.02 and $\sim$~0.006 at 8000 AU for inclinations of 30$^{\circ}$ and 60$^{\circ}$ degrees, respectively. 

\subsection{Implications of the infall velocity structure of Cha-MMS1}

%%non-magnetized

In this section we compare the infall velocity distribution of Cha-MMS1 with predictions from theory and observations of other collapsing cores. The number of uncertainties involved is large, both in constraining the velocity profile of Cha-MMS1 from the radiative transfer modeling, as well as in the various existent theoretical collapse models in terms of initial and boundary conditions,
and the inclusion or omission of either magnetic fields or rotation.
Therefore, a comparison of our source with theoretical models is not straightforward and we only aim to qualitatively discuss which collapse models provide better agreement with the infall profile we derived for Cha-MMS1. 

\subsubsection{Infall in Cha-MMS1 and in other observed cores}

Figure~\ref{fig:vinf} shows the infall velocities consistent with the envelope of Cha-MMS1 as derived from the radiative transfer modeling in Sect.~\ref{sec:mapyso}. We can discern two distinct regimes in terms of the envelope motions of Cha-MMS1, one below 3300 AU and one within $3300$~AU~$\leq r \leq30000$~AU. At radii larger than 3300 AU we better (more tightly) constrain the velocity profile of Cha-MMS1. The velocities are sub- to transonic within the range of 0.1~km s$^{-1}$ --~0.2 km s$^{-1}$. At radii below 3300 AU, where the model degeneracies are greater,  Cha-MMS1 is consistent both with increasing \emph{and} decreasing central velocities, ranging from 0.04 km s$^{-1}$ to 0.6 km s$^{-1}$. Therefore, the inner ($r<3300$ AU) envelope is characterised by velocities that can reach supersonic values compared to the outer ($r>3300$~AU), more quiescent regions. 

We also found that an inner, free-fall velocity distribution proportional to $r^{-0.5}$ is consistent with the envelope of Cha-MMS1 with an upper limit of $r\sim9000$ AU. If we assume that the latter radius signifies the distance that the reflected expansion wave has so far covered while travelling outwards at the sound speed, as described in the inside-out collapse scenario \citep{shu77}, then we can estimate an upper limit to the age of Cha-MMS1 as $t\sim\frac{r_s}{a_s}$ = 2$\times$10$^5$ years or 0.2 Myr. The sound speed, $a_s\sim0.2$ km s$^{-1}$, is computed for a temperature of 9 K and assumes isothermality. In our models we have an isothermal temperature of 9 K only for $r>2000$ AU and hence, this condition does not hold for the very inner radii. However, we only use the derived value as a crude estimate. Since this lifetime is an upper limit, it is consistent with both a first core (see Sect.~\ref{sec:FCtheory} for FHSC lifetimes) or a Class 0 protostar. The prestellar phase lifetime is estimated to be $\sim 0.5$ Myr \citep{evans09} and Class 0 objects have estimated lifetimes ranging from $\sim4$~--~$9\times$10$^{4}$~yr \citep{maury11} to $\sim1.6\times$10$^{5}$~yr \citep{evans09}. 

\citet{belloche02} similarly performed radiative transfer modeling of their molecular transitions towards the young Class~0 IRAM 04191 protostar using the MAPYSO code. They obtained an infall velocity of 0.15~km~s$^{-1}$ at a radius of 1500~AU and 0.1~km~s$^{-1}$ up to $\sim11000$~AU for their ``best'' fit. Their range of consistent infall velocities is similar to ours, albeit it is better constrained at small radii. For $r\ge2000$~AU they find infall velocities in the range of $\sim0.07$ and $\sim0.15$~km~s$^{-1}$, similar to the range 0.1 km s$^{-1}$~--~0.2 km s$^{-1}$ that we obtain for Cha-MMS1 at $r\ge3300$ AU. Nevertheless, they constrain the inner $r\le3000$ AU velocity profile as to that of a power-law, whereas in the case of Cha-MMS1 power-law, uniform \emph{and} decreasing velocity distributions are consistent with our data below 3300 AU. Therefore, the infall velocity structures of both Cha-MMS1 and IRAM~04191 cover a similar range of velocities for $r>3000$ AU, with Cha-MMS1 being additionally consistent with a larger range of velocity profiles at small radii. The flat, or decreasing velocity distributions at the inner core of Cha-MMS1 could represent an evolutionary phase younger than that of a protostar, if central free-fall has not yet taken place at these scales. The age of IRAM 04191 was estimated to be $\le 3 \times$10$^4$ yr, assuming 2000 AU is the radius below which there is a free-fall velocity field structure, while we previously estimated an age of $\le 2 \times$10$^5$ yr for Cha-MMS1.

The infall velocities we derive are mostly within the sub- to transonic range for $r\ge 3300$ AU ($\ge0.02$ pc). \citet{lee99} found subsonic inward velocities of 0.04 km s$^{-1}$~--~0.1 km s$^{-1}$ for a sample of 17 starless cores showing evidence for infall motions. L1544 is one of the starless cores that has been extensively studied \citep[e.g.,][]{tafalla98, williams99}. These studies have shown that L1544 is undergoing inward motions with velocities up to $\sim0.1$ km s$^{-1}$ over $\sim0.1$~pc. Such extended inward velocities imply that the velocity fields of Cha-MMS1, like that of L1544, are consistent with each other but inconsistent with the static SIS \citet{shu77} model. 

\subsubsection{Comparison to theoretical models}
\label{sec:discussion_models}

In this section, we compare the infall velocity field derived for Cha-MMS1 with the velocities predicted by theoretical models. We resolve radii $r\ge700$ AU from our observations, for which radius we find an infall velocity range of $\sim0.05$ km s$^{-1}$~--~0.5 km s$^{-1}$. 

We used 3D radiation-magneto-hydrodynamic models to extract the radial velocity structure for a 1~$M_{\odot}$ \citep[MU2 model;][]{commercon12} and 5~$M_{\odot}$  dense core (Fig.~\ref{fig:vinf}). The mass estimate of Cha-MMS1 from continuum observations yields 1.44~$M_{\odot}$ within a radius of 3750~AU \citep{belloche11a}.  

Both models have a normalised mass-to-magnetic-flux ratio of 2. They employ an azimuthal density perpurbation to assist fragmentation. The ratio of thermal to gravitational energy is 0.37 for the 1 $M_{\odot}$ model and 0.35 for the 5 $M_{\odot}$ model, the ratio of rotational to gravitational energies 0.045 and 0.03 ($\sim2$ and $\sim1.5$ times larger than the upper limit of Cha-MMS1's $\beta_{\rm rot}$, respectively), and the initial temperatures are 11 K and 10 K respectively. Their corresponding initial radii are 3300 AU and 20000 AU. The density profile of the 1 $M_{\odot}$ case is initially uniform, whereas it scales as $r^{-2}$ at the outer radii in the 5 $M_{\odot}$ case. The 1 $M_{\odot}$ and 5 $M_{\odot}$ models are denser at the equatorial plane of the cores by a factor of $\sim3$ and $\sim4$, respectively, compared to the density profile we used for Cha-MMS1 in Sect.~\ref{sec:mapyso}.

The models predict the evolution of the 24 $\mu$m and 70 $\mu$m flux densities during the first core phase. The 3D radiative transfer code RADMC-3D by Dullemond\footnote{http://www.ita.uni-heidelberg.de/~dullemond/software/radmc-3d/} is used for the post-processing of the RMHD calculations \citep[see][ for more details]{commercon12}.  The flux densities of Cha-MMS1 are, within a factor of $\sim$~2, consistent with a first core age of $\sim$850 yr based on the 1 $M_{\odot}$ model, and 2680 yr based on the 5 $M_{\odot}$ model (see Sect.~\ref{sec:intlum_method2}). The average radial velocity distribution of both first core models close to the equatorial plane (in order to avoid outflow contamination) are shown in Fig.~\ref{fig:vinf} as red solid and dashed lines. We can only explore the velocity profile of the 1 $M_{\odot}$ model for $r<2500$ AU due to its smaller initial radius (3300 AU). The radial velocities it predicts up to $\sim2000$ AU are within the range of velocities we constrain for Cha-MMS1. For $r>2000$ AU the velocities decrease to zero as we approach the boundary of the core. The average velocity extracted from the 5 $M_{\odot}$ model exhibits a velocity profile that follows the \emph{upper limit} of Cha-MMS1's velocity range up to $\sim8000$ AU, and is \emph{within} this range for $8000$~AU~$<r<12000$~AU. The slightly larger velocities predicted by the model compared to our upper limit for Cha-MMS1 may be due to the higher densities of the model.The velocity dip that peaks at $\sim18000$ AU in Fig.~\ref{fig:vinf} is a remnant of a small initial expansion at the initial core radius of $\sim20000$ AU that preceded the first core collapse, and therefore is of no physical importance for the interpretation of the collapse process in the model compared to the observational constraints.  

We now compare our results with other first core models that show reasonable agreement with the properties of Cha-MMS1. \citet{masunaga98} explored the protostellar collapse of a cloud core via radiation hydrodynamic simulations assuming spherical symmetry, and specifically focused on the formation of the first hydrostatic core. For the typical case of 1 $M_{\odot}$ and 10 K they find a homogeneous density structure inside the first core and an outer density distribution proportional to $r^{-2}$. The velocity field structure after the first core formation ($\rho\sim10^{-13}$~g~cm$^{-3}$) does not change for scales larger than $\sim1000$ AU for the different evolutionary times they present. They find supersonic velocities out to radii of $\sim3000$ AU, in very good agreement with the radius of $\sim$3300 AU below which supersonic motions are compatible with the envelope of Cha-MMS1. The infall velocities at 1000 AU, 3000 AU, and 9000 AU they predict are $\sim0.25$~km~s$^{-1}$, $\sim0.2$~km~s$^{-1}$, and $\sim0.1$~km~s$^{-1}$. At the same radii we obtain velocity ranges of 0.06~--~0.5~km~s$^{-1}$, 0.1~--~0.25~km~s$^{-1}$, and 0.1~--~0.2~km~s$^{-1}$. The velocity structure of Cha-MMS1 is therefore also consistent with the predictions of \citet{masunaga98} for a first core resulting from the collapse of a 1 $M_{\odot}$ nonrotating, non-magnetised cloud.

\citet{saigo08} investigated the effect of different initial rotation speeds during protostellar collapse and the formation of the first and second hydrostatic cores by performing 3D hydrodynamical simulations of cores with slow, moderate, and fast initial rotation. For a slowly rotating cloud ($\beta_{\rm rot}\sim 0.015$~--~$0.09$, ie., consistent with the respective ratio of Cha-MMS1), the infall velocity structure during the first core phase is described by supersonic motions below a radius of $\sim3000$ AU, and infall velocities of $\sim0.4$~km~s$^{-1}$, $\sim0.3$~km~s$^{-1}$, and $\sim0.15$~km~s$^{-1}$ at radii of 1000 AU, 3000 AU, and 9000 AU. 
The infall velocity predictions of \citet{saigo08} for the model with slow initial rotation is therefore consistent with the velocity ranges of 0.06~--~0.5~km~s$^{-1}$, 0.1~--~0.25~km~s$^{-1}$, and 0.1~--~0.2~km~s$^{-1}$ that we obtain for Cha-MMS1 at the same radial distances.

Finally, \citet{ciolekbasu00} presented an ambipolar diffusion model incorporating the observational constraints and physical parameters previously derived for the protostellar core L1544. They presented a nonrotating, ambipolar diffusion model with a lower background magnetic field strength (initial mass-to-flux ratio 0.8 in units of the critical value), that could reproduce the extended observed infall velocities of L1544 and concluded that L1544 might be a supercritical core undergoing magnetically diluted collapse. Their model predicts infall velocities of $\sim0.2$~km~s$^{-1}$ and $\sim0.15$~km~s$^{-1}$ at late times (approaching the first core densities), at 3000 AU and 9000 AU from the supercritical core, respectively, which are consistent with the infall velocity field structure we derive for Cha-MMS1 (Fig.~\ref{fig:vinf}). The spatial scale for supersonic infall velocities they predict at late times is $\sim2000$ AU. The agreement between the velocities of the two cores, especially for the late time evolution models (approaching first core densities), might be indicating that the initial conditions of the model for L1544 adequately describes the initial conditions of Cha-MMS1. Cha-MMS1 might therefore be undergoing collapse physically similar to that of the prestellar dense core L1544.

In general, the velocities of the RMHD model are in agreement with the inner envelope ($r<2000$ AU) of Cha-MMS1 for a 1 $M_{\odot}$ collapsing dense core and for $r<12000$ AU for a 5 $M_{\odot}$ dense core at the FHSC stage. In the innermost 2000 AU the velocity predictions of both models ``bracket'' the upper limit of the observationally constrained velocities for Cha-MMS1, and for larger radii the 5 $M_{\odot}$ predictions closely follow this upper limit. Non-magnetised, rotating as well as nonrotating models for the first core phase \citep{saigo08, masunaga98} also produce infall velocity structures that are consistent with the infall motions of Cha-MMS1 within scales of $\sim10000$ AU ($\sim0.05$ pc). Consequently, the collapse motions in the envelope of Cha-MMS1 are consistent with first core predictions.

As our observations are not sensitive to the very inner part of Cha-MMS1, where rapid infall velocity changes are expected, we cannot distinguish between an evolved prestellar core, a first core, and a young Class 0 object based on the kinematics alone. The 24 $\mu$m and 70 $\mu$m detections of Cha-MMS1 rule out its prestellar nature, and thus shifts the dilemma between a first core and a young Class 0 protostar. This dilemma can be solved via the detection of a slow, compact outflow stemming from Cha-MMS1 via interferometric studies, which would differentiate between the two evolutionary phases. In any case, our kinematic and dynamical study has so far shown that its properties do \emph{not} contradict the first core predictions and it is an interesting target for exploring the early protostellar stages of star-formation.  

\section{Summary and conclusions}
\label{sec:conclusions}

We performed observations of the dense core Cha-MMS1 in various molecular transitions and conducted an analysis of the kinematics within the core in order to investigate its physical properties and dynamical state. We utilised a 1D radiative transfer code to constrain the infall velocity structure of the core. Our conclusions can be summarised as follows:

\begin{enumerate}

\item The internal luminosity of Cha-MMS1 is estimated from the predicted inclination-dependent time evolution of SEDs for the first core phase for a 1 $M_{\odot}$ \citep{commercon12} and 5 $M_{\odot}$ model. The 24 $\mu$m and 70 $\mu$m flux densities imply inclinations within the range of $30^{\circ} \leq i < 60^{\circ}$ and an internal luminosity range of 0.08 $L_{\odot}$~--~0.18 $L_{\odot}$.  

\item The classical infall signature is detected in optically thick transitions, suggesting that the envelope of Cha-MMS1 is undergoing inward motions.

\item The position-velocity diagrams of optically thin transitions show velocity gradients perpendicular to the filament in which Cha-MMS1 is embedded. The average gradient over an extent of $\sim16000$ AU in diameter is $3.1\pm0.1$~km~s$^{-1}$~pc$^{-1}$ while we found no significant gradient along the filament. Interpreted in terms of rotation, these velocity variations imply solid-body rotation in the envelope up to a radius of $\sim8000$ AU, and slower, differential rotation from $\sim8000$ AU to $\sim12500$ AU. The average velocity gradient in the range 2000 AU~--~4000~AU is surprisingly flatter, which is difficult to understand in terms of rotation.

\item The turbulent velocity dispersion in the core is uniform within a radius of $r\sim5000$ AU parallel to the filament and $\sim12500$ AU perpendicular to the filament. The non-thermal dispersion is of the same order as the mean thermal dispersion at a temperature of 9 K, therefore implying an equipartition between thermal and non-thermal motions.  

\item Our radiative transfer modelling yields subsonic to transonic infall velocities in the range 0.1~km~s$^{-1}$~--~0.2~km~s$^{-1}$ for $3300$ AU $ < r < 30000$~AU. The velocity field is less well constrained in the inner parts for $r < 3300$~AU. A velocity increasing as $r^{-0.5}$ toward the center is consistent with the data, but we cannot exclude a decrease either. We find subsonic to supersonic velocities in the range 0.04~km~s$^{-1}$~--~0.6~km~s$^{-1}$ for $r\leq3300$ AU.

\item Part of the redshifted emission of the high density tracers CS 5--4, CO 7--6, and CO 6--5 is not reproduced by the radiative transfer model. This excess emission may indicate the presence of unresolved, higher velocity material at the inner core originating from a compact outflow driven by Cha-MMS1, or alternatively, it could arise due to contamination from the outflow of the nearby Class I protostar.

\item We find a relatively good agreement between the infall velocity profile derived for Cha-MMS1 and predictions of 3D RMHD simulations for the first hydrostatic core phase.

\end{enumerate}
Both the kinematical agreement with the predictions of RMHD simulations and the possible presence of a compact outflow suggested above are consistent with Cha-MMS1 being at the stage of the first hydrostatic core. However, we cannot affirm the object's nature without high resolution interferometric observations to search for and image a compact, slow, outflow. With the early prestellar core phase ruled out due to the object's 24 $\mu$m and 70 $\mu$m detection, Cha-MMS1 is either a first core or a young Class 0 protostar and our kinematical study cannot exclude either possibility.   \\

\noindent
\emph{Acknowledgments.} We thank B{\'e}reng{\`e}re Parise, Philippe Andr{\'e}, and Tyler Bourke for their insightful comments and suggestions, the APEX and Mopra staff for their support during the observations, and the referee for his/her feedback that helped in improving the quality of this paper. AET was supported for this research through a stipend from the International Max Planck
Research School (IMPRS) for Astronomy and Astrophysics at the Universities of Bonn and Cologne.

\bibliographystyle{aa} % style aa.bst
\bibliography{manuscript_astroph} % your references Yourfile.bib

\begin{appendix}
\section{Mopra calibration and efficiency}
\label{sec:mopracal}

\subsection{Calibration ambiguities}

Some of our Mopra data show differences in the peak temperatures of the two polarisations, POL0 and POL1. The differences are not systematic for all the transitions observed. The two polarisations in some cases differ up to $\sim10$\%. The CS 2--1 POL1 intensity for position P6 is stronger by $\sim5$\% within the uncertainties, while C$^{34}$S 2--1 has a stronger POL0 intensity by $\sim9$\%. POL0 also shows higher intensities for positions P1 and P6 of the N$_2$H$^+$~1--0 transition by $\sim7$\% and $\sim6$\% respectively. Most of the other transitions (and positions) do not show significant differences. As the observed discrepancies are not systematic and are only seen in very few cases, we use the average of both polarisations per transition for the analysis in this paper. 

Pronounced differences in intensity are seen when comparing a pair of spectra belonging to the same transition that was observed with different tuning frequencies. HC$_3$N 10--9 and HNC 1--0 are the two transitions observed in both setups and show intensity differences of $\sim38$\% and $\sim33$\%. Unfortunately, we have not yet found a satisfactory explanation for these discrepancies.

%Line    & POL0/POL1$^a$  
%CS(2-1) & 

%Notes $^a$ The ratios correspond to the average weighted ratios of all nine offset positions observed with Mopra for each transition.

\subsection{Efficiency}

Even though the expected Mopra beam efficiency at the 90 GHz band is $\sim$~0.5\footnote{see http://www.narrabri.atnf.csiro.au/mopra/obsinfo.html.}, we derive a value that is lower by $\sim$~30\% after performing independent calibration tests. IRAM 04191 in Taurus and OPH A SM1N in Ophiuchus were observed with both the Mopra and the IRAM 30 m telescopes \citep{belloche02, andre07}. IRAM 04191 was observed in the molecular transitions CS~2--1, N$_2$H$^+$~1--0, C$^{34}$S~2--1, and H$^{13}$CO$^+$~1--0 while OPH A SM1N in N$_2$H$^+$~1--0. After smoothing the 30 m data to the Mopra angular resolution we directly compared their integrated intensities with the equivalent Mopra intensities. We derive lower efficiencies for most of the observed transitions (see Table~\ref{table:eff}) with the overall weighted average efficiency being $\sim$~0.34. 

\begin{table}[h]
\caption{Mopra beam efficiencies$^a$} 
\vspace*{-1ex}
\begin{tabular}{lll} \hline\hline
Source     &  Line               &  Mopra Efficiency \\ \hline
IRAM 04191 & N$_2$H$^+$ 1--0     &  0.25 $\pm$ 0.01 \\
IRAM 04191 & CS 2--1             &  0.35 $\pm$ 0.02 \\
IRAM 04191 & C$^{34}$S 2--1       &  0.52 $\pm$ 0.13 \\
IRAM 04191 & H$^{13}$CO$^+$ 1--0  &  0.36 $\pm$ 0.03 \\ 
OPH A SM1N   & N$_2$H$^+$ 1--0     &  0.36  $\pm$ 0.003 \\ 
\hline
\end{tabular}\\[0.5ex] 
\label{table:eff} 
Notes 
$^a$ Mopra efficiencies derived after comparing independent observations from the IRAM 30 m and Mopra telescopes. The IRAM 30 m spectra were smoothed to the Mopra angular resolution.
\end{table}

\section{Calculation of opacities}

\subsection{Opacity of the C$^{18}$O 2--1 line}
\label{sec:opacity_c18o}

We observed both C$^{18}$O 2--1 and C$^{17}$O 2--1 at 13 positions perpendicular to the filament (Fig.~\ref{fig:CHAmap}). An isotopic ratio of [$^{18}$O]/[$^{17}$O] $\sim$ 4.11 was found for the nearby (140~pc) low-mass cloud $\rho$ Ophiuchus \citep{wouterloot05}. We use this value to derive the opacity of C$^{18}$O 2--1 using the following relation: 

\begin{equation}
\frac{I_{C^{18}O}}{I_{C^{17}O}} = \frac{1-e^{-\tau_{C^{18}O}}}{1-e^{-\tau_{C^{17}O}}}
\end{equation}
where I$_{C^{18}O}$ and I$_{C^{17}O}$  are the intensities of the two transitions, $\tau_{C^{18}O}$ and $\tau_{C^{17}O}$ their opacities and $\tau_{C^{17}O}$=$\tau_{C^{18}O}$/4.11. We find opacities that gradually increase from  $\le 0.5$ at the outermost position to $\sim1.6$ at the centre of the core for the C$^{18}$O 2--1 transition. 

\end{appendix}

\end{document}